\documentclass[varenna]{hepph_cimento}
\usepackage{graphicx}
\usepackage{epsfig}
\usepackage{amssymb}
\input{def_panda.tex}
\input{def_pwa.tex}
\begin{document}
\title{A Primer on Partial Wave Analysis}
\author{K.~Peters}
\institute{Institut f\"ur Experimentalphysik,\\ Ruhr-Universit\"at Bochum,\\ D-44780 Bochum, Germany}
\PACSes{\PACSit{11.80}{Partial wave analysis}}
\maketitle
\begin{abstract}
In the 90s of the last century high statistics experiments with fully equipped
4$\pi$ detectors have lead to a better insight in the spectrum of hadrons. In particular
the finding of crypto-exotic and $\jpc$ exotic states tremendously improved the
experimental situation in meson spectroscopy. All this was possible only with sophisticated
analysis methods like the decomposition of measured phase-space distribution into partial waves
and to express the partial waves in terms of complicated dynamical functions. This paper
gives an introduction about the concepts and formalisms involved.
\end{abstract}
%
%
\section{Introduction}
\subsection{Goals}
%
%
The spectrum of hadrons, mesons and baryons is the result of bound states of quarks like
$\qqbar$ and $qqq$ respectively. Although the quark-model was motivated partly
by the structure of the spectra and the symmetry and decay properties of the
hadrons, the gauge field theory of strong interactions, Quantum Chromodynamics (QCD), has
a very strong coupling constant at very low momentum transfer (e.g. low masses) so that
there are no firm predictions in the classical field of light quark spectroscopy and medium energy 
physics. Therefore there is no possibility to map out the states from first principles
(if we leave lattice gauge theory out for the moment). To understand the spectrum of
light hadrons it is important to collect the properties to derive effective models to be used
in that field, which may tie up to the very high energies where perturbative QCD works well.
In order to uncover the spectrum it is necessary to investigate the scattering and the decay
of hadrons and to identify the intermediate states throughout the whole reaction.
\par
Lattice gauge theory could be a way out of the problem, but since the results reflect a
measurement on the lattice (with very limited precision and a lot of assumptions so far)
it is not suited to actually understand the principles and the physics behind the measurements,
e.g. an actual state and/or it's properties. 
\par
The basic task is to find all resonances, with their static properties like mass, width, spin and
parities. Since $SU_F(3)$~\cite{Peters95} provides a lot of relations among the decays of pure states (without
Fock states) without complicated final state effects it is important to measure the partial decays widths as well.
This is a very demanding task, since a lot of resonances overlap. In addition complicated production
processes or scattering with many waves in the intermediate state complicate the situation.
To disentangle the waves and to identify resonances and their actual yield.
\subsection{Technique}
%
%
The techniques are manifold and we should to concentrate only on two basic approaches.
The first one is scattering of hadrons, usually on a proton or a deuterium target,
depending on the desired process. Typical examples are
\bi
\item $\pi$N scattering with or without charge exchange (GAMS at CERN, E852 at AGS),
\item $\gamma$N scattering (CEBAF, MAMI, Elsa, Graal),
\item $\pi$p or pp in the central region (WA76, WA92, WA102 at CERN, E690 at FNAL),
\item pp near meson production thresholds (WASA at Celsius, Anke and TOF at COSY),
\item $\pbarp$ in flight (Crystal barrel, Jetset and PS185 at CERN, Panda at GSI).
\ei
These reactions involve usually a lot of partial waves. Single or double polarization experiments, low 
excitation energies and/or a selection of specific exclusive channels which impose reasonable constraints on the reaction
are required to enable an unambiguous decomposition of the system.
\par
The second one is where the initial system is at rest
and/or the remaining part of the reaction is not of interest like
\bi
\item $\pbarN$ reactions at rest into many body final states (Asterix, Crystal barrel and Obelix at LEAR),
\item $K^0$ and $K^\pm$ decays (NA48 at CERN, Kloe at Da$\phi$ne, kTev at FNAL),
\item $\phi(1020)$ decays (Kloe at Da$\phi$ne, VEPP at Novosibirsk),
\item $D$ and $D_s$ decays in high energy reactions (photo-production at FNAL, Babar/Belle at PEP-2/KEK-B, CLEO-c at CESR),
\item $\jpsi$ decays (MarkIII at SLAC, DM2, CLEO-c at CESR, BES at BEPC).
\ei
For all experiments of this type it is important that spin densities are well known.
Otherwise a full decomposition is not possible. In that case additional assumptions have to made to relate the different parameters.
\subsection{Methods}
\subsubsection{The Partial Wave Approach}
%
%
\paragraph{Motivation}
Partial waves are easily introduced in a scattering process.  To get a first impression
we start with Schr\"odinger's equation
\be
-\frac{\hbar}{2\mu}\Nabla^2\Psi(\rvec)+V(\rvec)\Psi(\rvec)=E\Psi(\rvec)
\ee
with  $\dsty\vec{k}=\frac{\pvec}{\hbar}=\mu\frac{\vec{v}}{\hbar}$ and the reduced mass $\dsty\mu=\frac{m_1m_2}{m_1+m_2}$.
The incident wave can be expressed like $\Psi_i(r,\tht,\ph)=\Exp{\imath kz}$ and we assume a vanishing potential $V(\rvec)=0$.
Then we can expand the initial state $\ket{i}$ in terms of Legendre polynomials $P_l$, thus separating
angular and radial wave function
\be
\ket{i}=\Psi_i=\sum_{l=0}^\infty U_l(r)P_l(\cth).
\ee
The scattering wave function $\Psi_S$ is the difference between incoming and outgoing wave. We parametrize $U_l$ 
in terms of a phase $\dll$ and an inelasticity $\etl$, which are motivated from our knowledge of resonance curves,
where the phase moves from $-\frac{\pi}{2}$ to $\frac{\pi}{2}$ and the inelasticity carries all dissipated probability
(e.g. into other channels):
\be
\Psi_S=\Psi_f-\Psi_i=\frac{1}{k}\sum_{l=0}^\infty(2l+1)\frac{\etl\Exp{2\imath\dll}-1}
{2\imath}P_l(\cth)\frac{\Exp{\imath kr}}{r}
\ee
The total cross section can then be written
\bn
\frac{d\sigma}{d\ph d\cth}&=&\frac{1}{k^2}\left|\sum_{l=0}^\infty(2l+1)\frac{\etl\Exp{2\imath\dll}-1}
{2\imath}P_l(\cth)\right|^2\nonumber \\
&=& \frac{1}{k^2}\left|\sum_{l=0}^\infty(2l+1)\Tl P_l(\cth)\right|^2\label{eq:tm_fact}
\en
with
\be
T_l=\frac{\etl\Exp{2\imath\dll}-1}{2\imath}
\ee
the so called $T$-matrix. We will discuss it's properties later in more detail (see \refsec{sec:tm_def}).
To visualize the properties of $T$ Argand
plots are quite useful, since they provide a check on the unitarity of $T$ (which ensures that the probability of 
a reaction does not exceed unity). The Argand plot is a graph of the imaginary part of $T$ versus the real part of $T$
(see \reffig{fig:argand_def}). It is seen easily from this plot that the phase $\dll$ is not the same as the angle $\phi$ which
apears in the polar representation of complex numbers.
\bfg[hbt]\bc
\psfig{figure=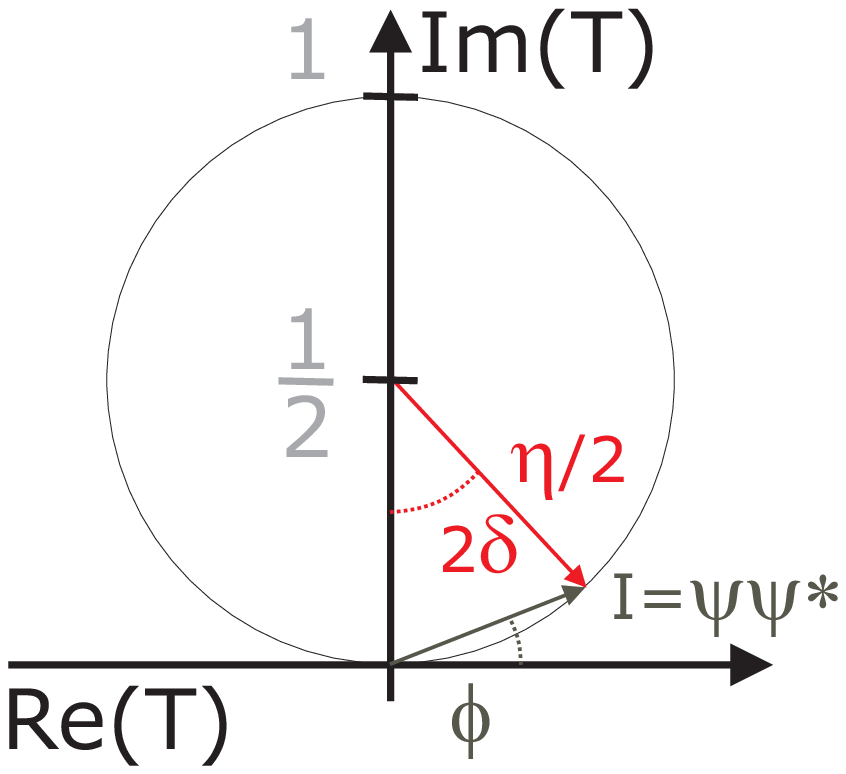,scale=0.5}
\ec
\caption[Argand plot definitions]{Argand plot definitions.}
\label{fig:argand_def}
\efg
As an example, \reffig{fig:argand_example_bw} shows the behavior of $T$ for a relativistic Breit-Wigner function. If the inelasticity is zero,
e.g. there is only one elastic channel, the Argand circle has a constant radius of $\half$ and crosses the imaginary axis at one.
\bfg[hbt]\bc
(a)\vps\psfig{figure=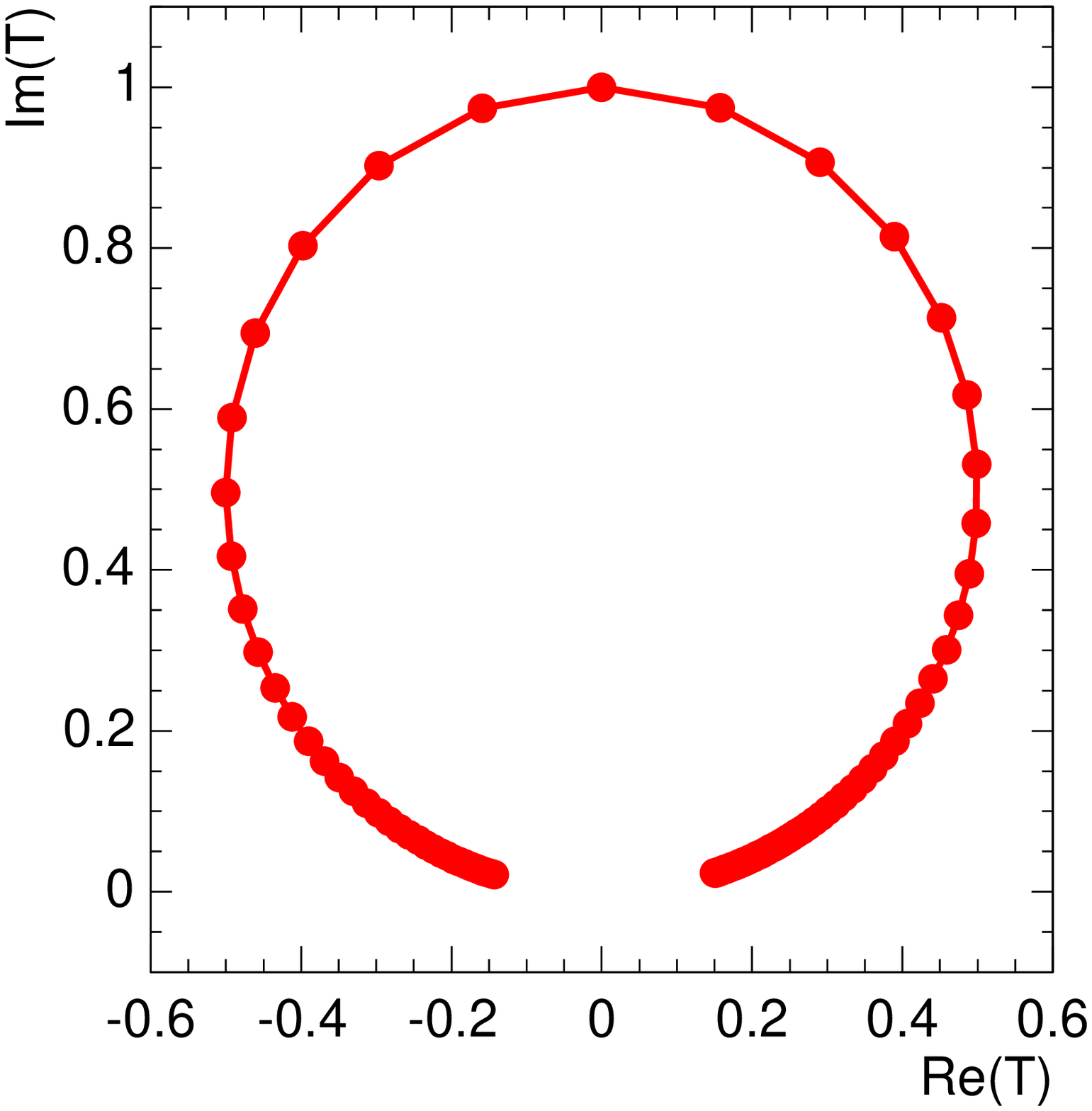,scale=0.2}
\vps(b)\vps\psfig{figure=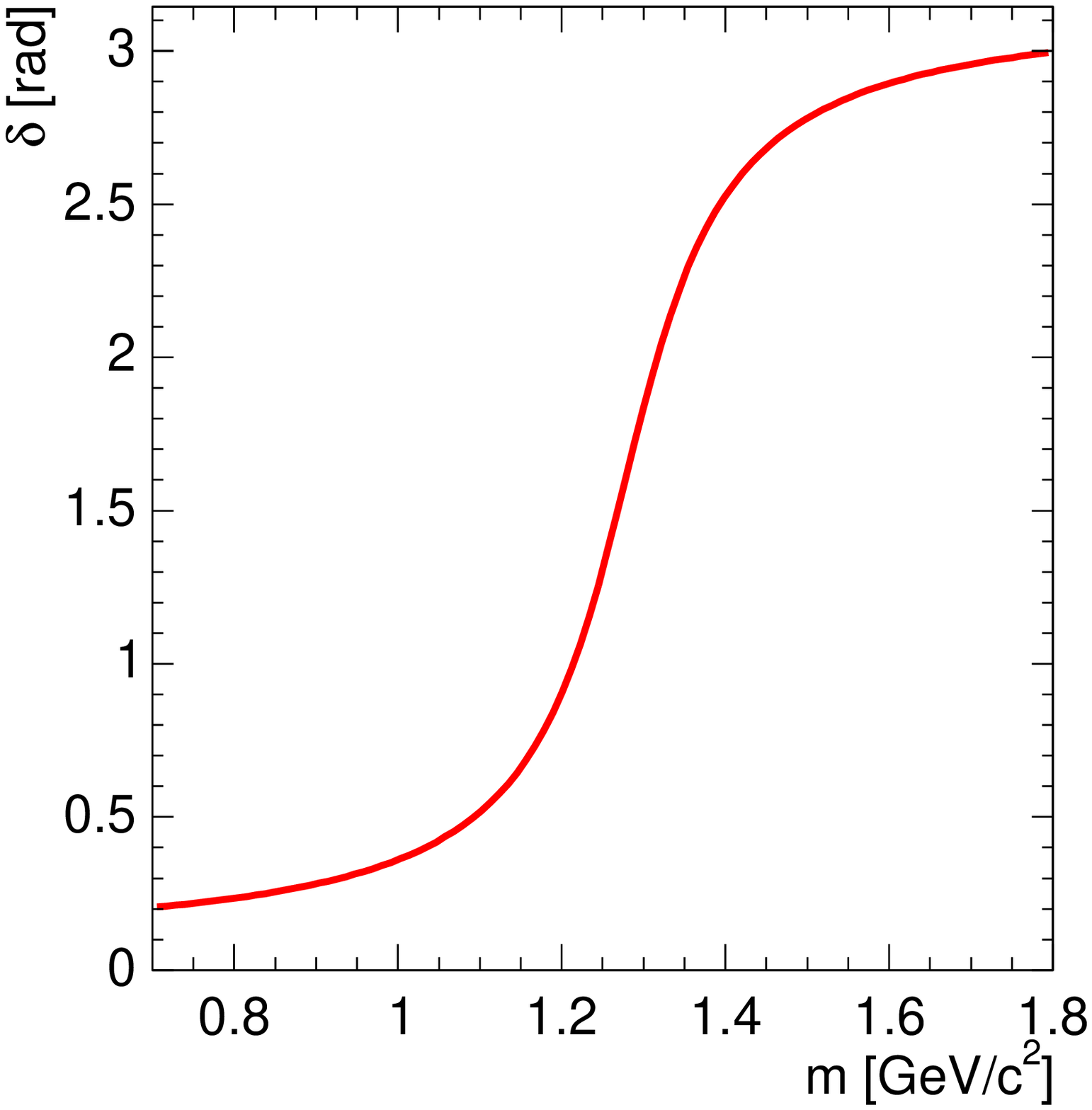,scale=0.2}
\vps(c)\vps\psfig{figure=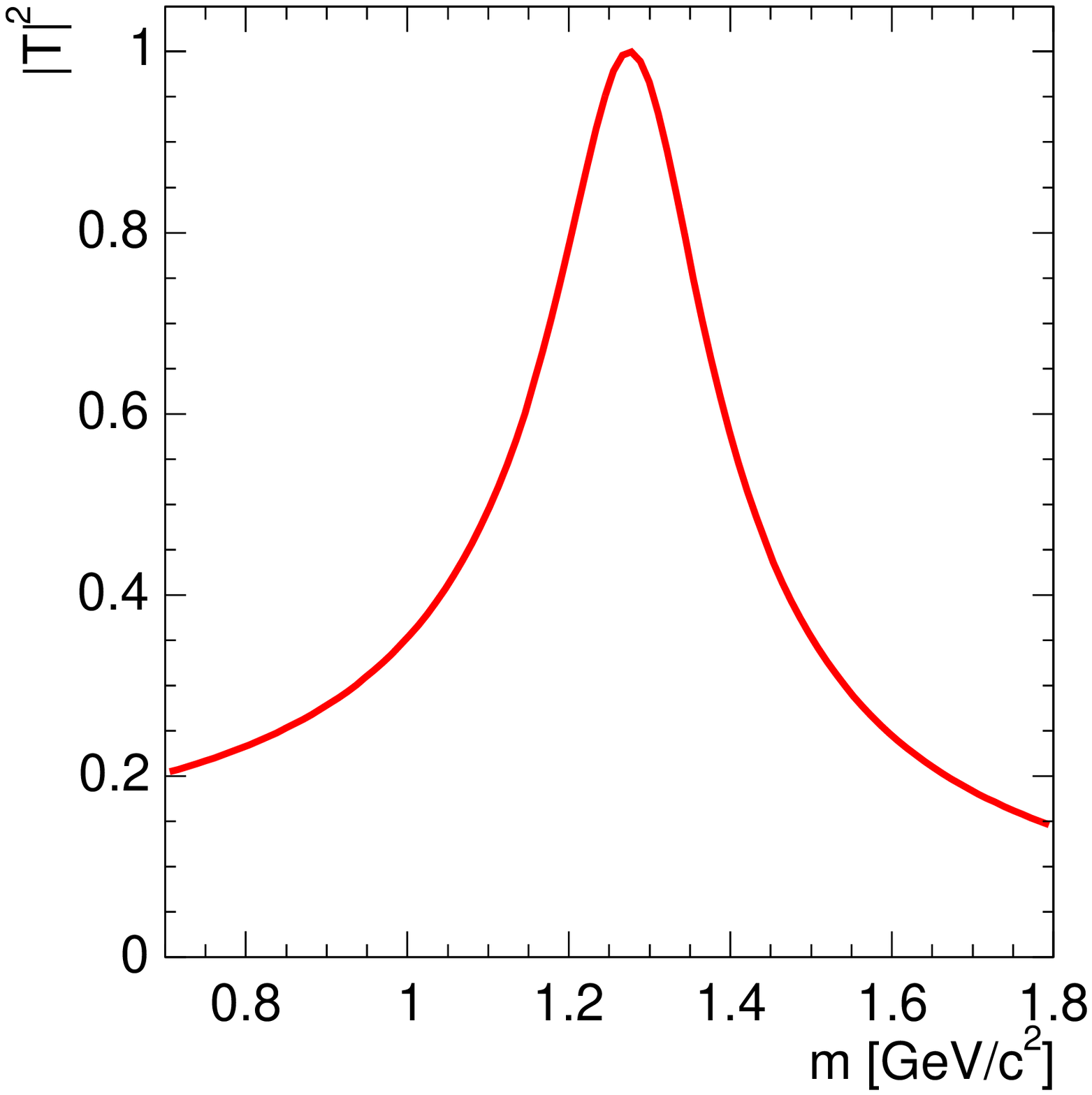,scale=0.2}\ec
\caption[Diagrams of a simple Breit-Wigner]{Simple relativistic Breit-Wigner. (a) Argand, (b) phase $\delta$ and (c) intensity plot.}
\label{fig:argand_example_bw}
\efg
From \refeq{eq:tm_fact} we see that the angular amplitudes ($P_l$) and the dynamic amplitude ($\Tl$) factorize.
\subsubsection{The Isobar Model}
%
%
The last preparative step to write down an amplitude is to make assumptions how particles are grouped to
construct a decay/reaction chain. An empirical approach is the isobar model. It assumes
that all subsequent decays appear to be two-body reactions. This model seems to work extremely well
in very different environments and for most hadrons (exceptions may be $\omega\to\pip\pim\piz$ and $\eta\to3\pi$).
The type of valid reactions is sketched in \reffig{fig:reaction_types}a while \reffig{fig:reaction_types}b shows a
more complicated reaction involving rescattering. The isobar approach will not work easily in those circumstances.
In some cases it can be retained if the rescattering process is refactorizable 
in terms of many isobar reactions. This involves usually
very many parameters and may not lead to a sufficient description. In those cases model assumptions have to be
made.
\bfg[hbt]\bc
(a)\vps\psfig{figure=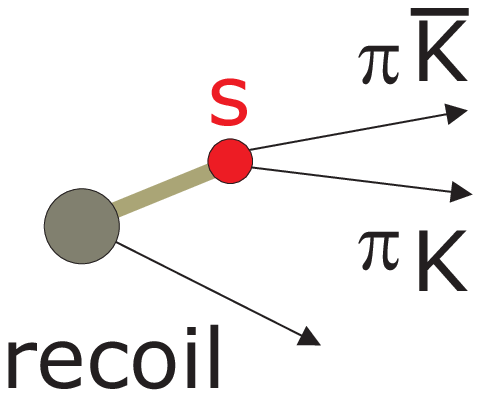,scale=0.5}
\vps(b)\vps\psfig{figure=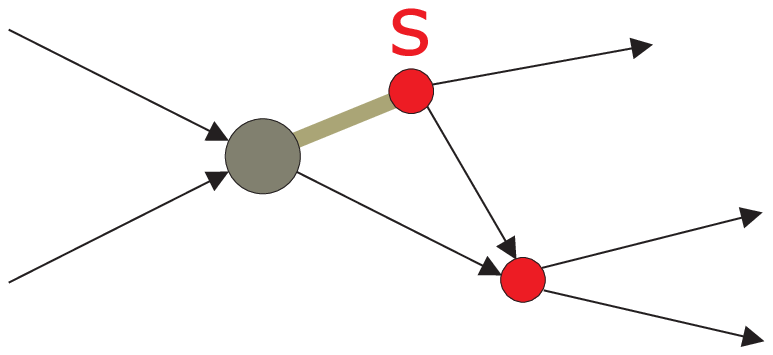,scale=0.5}\ec
\caption[Production process with propagation and rescattering]{Production process with propagation (a) and rescattering (b).
(a) can be treated easily with an isobar method, while (b) can not.}
\label{fig:reaction_types}
\efg
In the isobar model, the two-body decay into particle 1 and 2 factorizes (\reffig{fig:isobar}) completely from the recoil system 3,
which might decay as well. Any node in the decay tree is represented by the same isobar-like amplitude. The recoil systems
are only involved in maintaining the conservation of angular momentum and spin-projections, which imply summation over
unobservable quantities.
\bfg[hbt]\bc
\psfig{figure=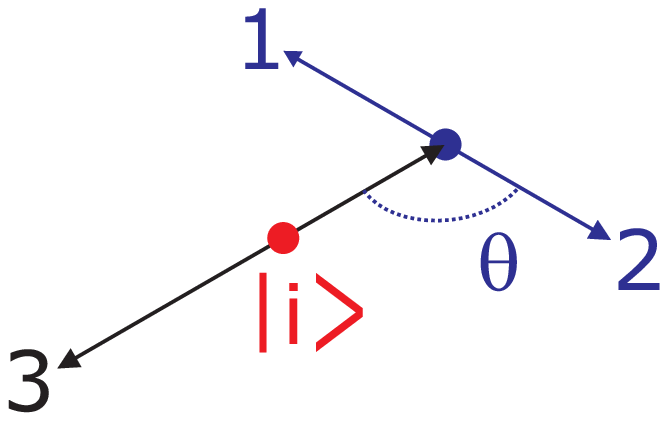,scale=0.5}
\ec
\caption[Isobar definition.]{Isobar definition in a chain of subsequent decays. $\theta$ defines the angle for angular decay distributions.}
\label{fig:isobar}
\efg
\subsubsection{Construction of the Amplitude}
%
%
The full amplitude for each node of the decay tree consists of a dynamical part $\Tl$ and an angular part which we will
call $R_l$. If we deal with strong or electromagnetic interactions also isospin is conserved, which means that we
have to include an isospin related part $I_l$, so that decays with different isospins $\ket{II_3}$ can be related to each
other. This is important if the same charged particle occur with different charge in various parts of the decay tree.
With these ingredients, the full amplitude is defined via
\be
f(I,I_3,s,\Om)= I_l(I,I_3) \Tl(s) R_l(\Om)\label{eq:theamplitude}
\ee
One should note, that after combining the amplitude for all nodes in the decay tree, all conservation laws (like $J$, $I$, and
their respective projections) have to be taken into account, which results in a sum over all unobservable values.
\paragraph{Example: Isospin Relations in $\pbarp (\jpczeromp,\jpconemm)\to\rho\pi$}
The initial $\pbarp$ state has either $\ig(\jpc) =1^-(\jpczeromp)$ (called $^1S_0$) or $\ig(\jpc) =0^-(\jpconemm)$ (called $^3S_1$).
The two final state gammas have $\ig(\jpc)=0^+(\jpconemm)$. Both initial states can decay to $\rho\pi$ but this case illustrates how important
isospin Clebsch-Gordan coefficients are, since 
\bn
\rho^0\pi^0 & \to& (1010|00)=-\sqrt{\frac{1}{3}}\nonumber\\
\rho^0\pi^0 & \to& (1010|10)=0\nonumber\\
\rho^\pm\pimp&\to&(1(\pm1)\ 1(\mp1)|00)=\sqrt{\frac{1}{3}}\\
\rho^\pm\pimp&\to&(1(\pm1)\ 1(\mp1)|10)=\pm\sqrt{\frac{1}{2}}\nonumber
\en
it is evident that $^1S_0$ has destructive $\rho^\pm/\rho^0\pi^0$ interference, while for $^3S_1$ only $\rho^\pm\pi^\mp$ contributes and 
$\rho^0\pi^0$ does not exist.
%

%

%
\section{Spin formalisms}
\subsection{Preface}
%
%
There are various spin formalisms available. In principle there are three basic types:
\bi
\item tensor formalisms, in non-relativistic (Zemach) or covariant form,
\item spin-projection formalisms, where a quantization axis is chosen and proper 
rotations are used to define a two-body decay
\item formalisms based on Lorentz invariants (Rarita-Schwinger) where each 
operator is constructed from Mandelstam variables only.
\ei
The tensor formalisms are very fast algorithms if waves with small angular momentum 
are involved but is getting very complicated if higher waves are present and/or 
a lot of subsequent decays occur. Using the Lorentz invariants is not really a formalism,
since one has to construct an amplitude with the proper transformation and symmetry properties
as the actual particle and wave to be represented. As for the tensor formalism this might be simple
and extremely elegant for waves with low angular momentum but is virtually impossible for a complicated
decay cascade. The Zemach formalism as a representative of tensor formalisms is discussed in~\refsec{sec:zemach}.
\par
If we restrict ourselves now to the other formalisms the key steps in the specification of a scattering formalism are
as follows:
\bi
\item the definition of single particle states of given momentum and spin component ($\pvec$-states),
\item the definition of two-particle $\pvec$-states in the $s$-channel center-of-mass system and of
amplitudes between them,
\item transformation to states and amplitudes of given total angular momentum ($J$-states),
\item symmetry restrictions on the amplitudes,
\item formulae for observable quantities,
\item specification of kinematic constraints.
\ei
The basic three formalisms are known as helicity, transversity and canonical (historically known as orbital)
formalisms. Their basic properties are summarized in~\reftbl{tbl:formalism}. Since they have different
symmetry relations they can be used in different fields. While the helicity formalism is applicable to most
spectroscopy experiment, it is the transversity formalism which conserves parity and can therefore be applied
for partial wave decomposition in $CP$ measurements.
\begin{table}[ht]
\begin{tabular}{rccc}\hline
& Helicity & Transversity & Canonical (Orbital)\\ 
property & \multicolumn{3}{c}{possible/simplicity}\\ \hline
partial wave expansion & simple & complicated & complicated \\
parity conservation & no & yes & yes \\
crossing relation & no & good & bad \\
specification of kinematic constraints & no & yes & yes \\ \hline
\end{tabular}
\caption[Comparison of various spin formalisms]{Comparison of various spin formalisms. 
Depending on the choice of the quantization axis, the amplitudes have different symmetry properties.}
\label{tbl:formalism}
\end{table}
The three formalisms differ mainly in the choice of the spin quantization direction
for each particle.
\par
In the helicity formalism each particle spin is quantized parallel to its own direction of motion
so that its projection, the helicity $\lambda$ is diagonal.
\par
In the transversity approach, the component $\tau$ normal to the scattering plane is used.
\par
In the orbital (or canonical approach) the component $m$ in the incident $z$-direction is diagonal.
\par
The merits of each approach arise largely from the invariance of the projections. These properties are
made precise by the definition of the particle states. It is convenient to define them by explicit
Lorentz transformation from the spin states $m$ of a particle at rest. Detailed discussions
are found in the original papers \cite{Jacob59,Kotanski66,Choushirokov58}. The essential principle in
all cases is, that the spin projection of a particle is defined in its rest frame.
The general single particle state is then defined by applying a definite sequence of boosts
$L(p_x,p_y,p_z)$ and rotation $R(\alpha,\beta,\gamma)$ to the system. We define
\bn
\Psi_\lambda &=& |\pfr,\lambda\rangle=
\wht{R}(\phi,\theta,-\phi)\wht{B}(0,0,p)\ket{m}\equiv \wht{H}(\pfr)\ket{\lambda}\\
\Psi_\tau &=& |\pfr,\tau\rangle= \sum_{\lambda}|\pfr\lambda\rangle\Delta_{\lambda\tau}^J
=\wht{\Delta}\wht{H}(\pfr)\wht{\Delta}\inv\ket{\tau}=\wht{T}\ket{\tau}\\
\Psi_m &=& |\pfr m\rangle=
\sum_{\lambda}|\pfr,\lambda\rangle D_{\lambda\tau}^{J\ast}\wht{R}(\phi,\theta,-\phi)
=\wht{R}\inv(\phi,\theta,-\phi)\wht{H}(\pfr)\ket{m}=\wht{O}\ket{m}.
\en
It is clear that it is necessary to calculate the effect of a general Lorentz transformation
on the states $\ket{\pfr\xi}$. While $L\pfr=\pfr\prm$ the corresponding state is not
$\ket{\pfr\prm,\xi}$ since the states are being always obtained from the rest state by a definite
sequence $X(\pfr)$ and these do not in general commute. However, since both have the same value
for the momentum in a particular system they can both be transformed back to the rest frame by $X\inv$
can there they can differ at most by a rotation
\be
L\ket{\pfr\xi}=X(L\pfr)R(L,\pfr)\ket{0\xi}=\sum_{\xi\prm}\ket{L\pfr,\xi\prm}
D_{\xi\prm\xi}^{J}(r)
\ee
with the Wigner rotation given by
\be
R=X\inv(L\pfr)L X(\pfr).
\ee
The two-particle states are then defined essentially as direct products of single particle states, but
phase space factors may be included to make the form in the center-of-mass system convenient for various
purposes. More details are discussed in the particular sections about the different formalisms (see~\refsec{sec:spin_formalism}).
\subsection{Zemach Formalism}
%
%
\label{sec:zemach}
\subsubsection{Formalism}
The Zemach formalism was originally developed for the investigation of the $K^0$ decays 
into 3$\pi$~\cite{Zemach64}. The basic concept is that every angular momentum involved in 
the reaction is represented by a symmetric and traceless tensor of rank $l$in 3-dim. phase-space.
The tensors $A$ for spins up to two are
\bn
l&=0& \qquad A^0=1\nonumber\\
l&=1& \qquad A^1(\qvec)=\qvec \\
l&=2& \qquad A^2(\qvec)=\threehalf\left[\qvec\cdot \qvec^T-\ub{\frac{1}{3}|\qvec|^2}_{\mbox{\tiny for tracelessness}}\right]\nonumber
\en
with
\be
\qvec\cdot\pvec^T=\ba{c}q_1\\q_2\\q_3\ea\ba{ccc}p_1 & p_2 & p_3\ea
=\ba{ccc}q_1p_1 & q_1p_2 & q_1p_3 \\q_2p_1 & q_2p_2 & q_2p_3\\q_3p_1 & q_3p_2 & q_3p_3\ea
\ee
or with all indices
\bn
l&=0& \qquad A^0=1\nonumber\\
l&=1& \qquad A_{i}^1=q_i \\
l&=2& \qquad A_{ij}^2=\threehalf q_iq_j-\half |q_i|^2\delta_{ij}.\nonumber
\en
The coupling of spins and/or angular momenta, like spin $s$ of a particle and angular momentum $l$
relative to another spinless particle is then done by multiplying the tensors and contraction
of the resulting tensor.
\subsubsection{Example: $\pbarp(\jpczeromp)\to f_2\piz\to \pip\pim\piz$}
The two step process $\pbarp\to f_2\piz$ and $f_2\to\pip\pim$ may serve as an example to illustrate the method.
For $\pvec$ being the momentum between $\piz$ and $f_2$ and $\qvec$ the momentum
in the subsequent $f_2\to\pip\pim$ two-body decay the construction of the amplitude leads to
\bn
A^0 &=& A_{f_2\piz,ij}^2 A_{\pip\pim,kl}^2\underbrace{ \delta_{ik}\delta_{jl}}_{\mbox{\tiny unpolarized}}\nonumber\\
&=& \sum_{i,j,k,l} A_{f_2\piz,ij}^2 A_{\pip\pim,kl}^2\delta_{ik}\delta_{jl}\nonumber\\
&=& \sum_{i,j} A_{f_2\piz,ij}^2 A_{\pip\pim,ij}^2\label{eq:zemach_ex_fullamp}
\en
with
\bn
A_{f_2\piz,ij}^2&=& \threehalf p_ip_j-\half |p_i|^2\delta_{ij}\label{eq:zemach_f2}\\
A_{\pip\pim,kl}^2&=& \threehalf q_kq_l-\half |q_l|^2\delta_{kl}\label{eq:zemach_pp}.
\en
Combining \refeq{eq:zemach_pp} and \refeq{eq:zemach_f2} with \refeq{eq:zemach_ex_fullamp} gives
\bn
A^0 &=& \left(\threehalf p_ip_j-\half |p_i|^2\delta_{ij}\right)\left(\threehalf q_iq_j-\half |q_i|^2\delta_{ij}\right)\nonumber\\
&=& \frac{9}{4}(\qvec\cdot\pvec)^2\nonumber 
-\frac{3}{4}\qvec^2\pvec^2-\frac{3}{4}\qvec^2\pvec^2
+ 3\frac{1}{4}|\qvec|^2|\pvec|^2\\
&=& \frac{9}{4}(\qvec\cdot\pvec)^2-\frac{3}{4}\qvec^2\pvec^2
\en
and finally the angular distribution is calculated by the squared amplitude
\bn
\frac{9}{4}\left[(\qvec\cdot\pvec)^2-\frac{1}{3}\qvec^2\pvec^2\right]&=&
\frac{9}{4}\left[(qp \cth)^2-\frac{1}{3}q^2p^2\right]\nonumber\\
&=& \frac{9}{4}(\cths-\frac{1}{3})^2= P_2^0(\tht)^2.
\en
\subsection{Canonical and Helicity Formalism}
\subsubsection{Few Particle States}
%
%
\label{sec:spin_formalism}
\paragraph{Canonical (Orbital) Formalism}
\subparagraph{Single Particle States}
May $\ket{jm}$ be an at-rest state with spin $j$ and spin projection $m$ on to the $z$-axis
in an euclidean system with coordinates $(x,y,z)$. The single particle state with a momentum
$\pvec$ is constructed through a pure Lorentz transformation $L\pfr$
\bn
\ket{\pfr,jm}&\stackrel{def}{=}& L\pfr \ket{jm}\label{eq:cano_lordef}\\
&=& \wht{R}_0 L_zp \wht{R}_0\inv\ket{jm}\nonumber
\en
where $\wht{R}_0$ $=$ $\wht{R}_0(\ph,\tht,0)$ rotates the $z$-axis (like in \reffig{fig:angles}a) in the direction
of the momentum $\pvec$ ($\vec{e}_{\pvec}=\wht{R}_0(\ph,tht,0)\vec{e}_z$.
\bfg[hbt]\bc
(a)\vps\psfig{figure=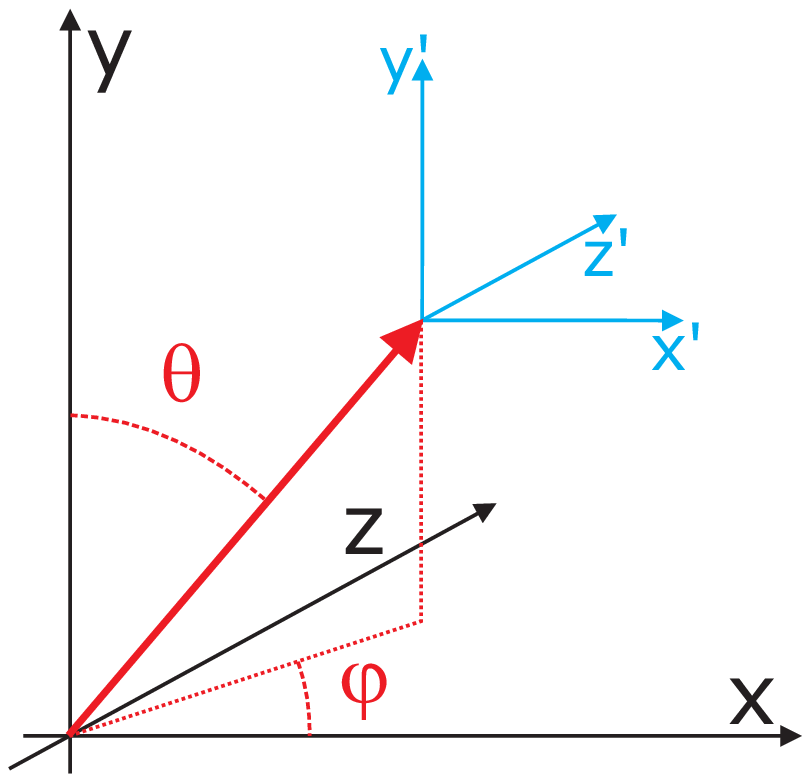,scale=0.5}
\vps(b)\vps\psfig{figure=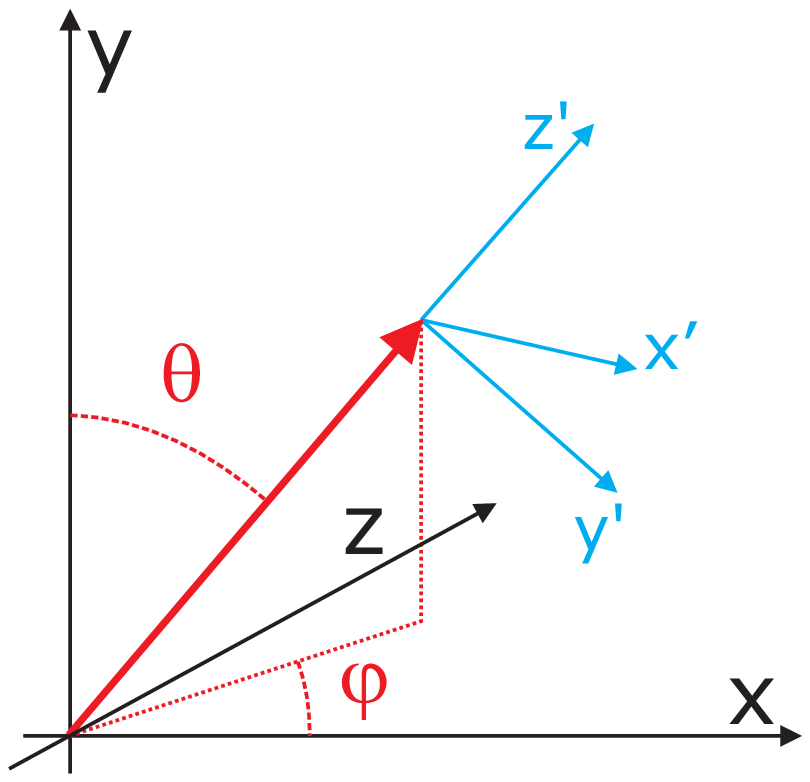,scale=0.5}\ec
\caption[Definition of angles in canonical and helicity formalism]
{Definition of angles in canonical (a) and helicity formalism (b).}
\label{fig:angles}
\efg
In the first step the momentum vector is rotated via $\wht{R}_0\inv$ in the $z$-direction. Secondly
the absolute value of the momentum is Lorentz transformed along $z$ and finally the $z$-axis is
rotated to the $\pvec$ direction via $\wht{R}_0$.
\par
If rotation of the single particle state is derived from the properties of the rotation group. One obtains
\be
\wht{R}\ket{\pvec,m} =\sum_{m\prm} D_{m\prm m}^j\ket{\wht{R}\pvec,m}\label{eq:cano_rotdef}
\ee
where $D_{m\prm m}^j(R)$ are the Wigner $D$-functions for the rotation $R$. From \refeq{eq:cano_rotdef} we see that
the canonical states transform like at-rest states $\ket{jm}$.
\subparagraph{Two-Particle States}
Two-particle systems 1+2 in a rest-system $J$ with respective spins can be constructed directly
with the help of single particle states. One particle $s$ has the momentum $\pvec_s$ and the other
partilce $t$ has the opposite momentum $-\pvec_s$. Apart from that the at-rest system of the two
particles depends only on the stereo angle $\Omega_s$ and the direction of $\pvec_s$. Using \refeq{eq:cano_lordef}
one gets
\be
\ket{\Om_s^0,sm_s tm_t}\stackrel{def}{=}\kappa\left[
L\ub{\pfr_s}_{\dsty(E_s,\pfr_s)}\ket{sm_s}L\ub{\pfr_t}_{\dsty(E_t,-\pfr_s)}\ket{tm_t}
\right].
\ee
The normalization $\kappa$ can be derived from the single particle states:
\be
\kappa=\frac{1}{4\pi}\sqrt{\frac{p_s}{m_J}}=\frac{1}{4\pi}\sqrt{\rho_s}
\ee
where $m_J$ is the invariant mass of the state $J$ and $\rho_s$ the invariant phase-space factor.
\par
To get a practical recipe it is necessary to couple the angular momentum $L$ and the total spin $S=s+t$.
Thus the first step is the coupling of the single particle spins to $S$
\be
\ket{\Om,Sm_S}=\sum_{m_s,m_t} \CG{sm_s}{tm_t}{Sm_S}\ket{\Om,sm_stm_t}
\ee
where $\CG{sm_s}{tm_t}{Sm_S}$ is a Clebsch-Gordan coefficient. A state with a particular total angular momentum
$L$ is then
\be
\ket{Lm_LSm_S}=\int d\Om\,Y_{m_L}^L(\Om)\ket{\Om,Sm_S}
\ee
Finally we couple $L$ and $S$ to $J$:
\bn
\ket{JMLS} &=& \sum_{m_L,m_S} \CG{Lm_L}{Sm_S}{JM} \ket{Lm_LSm_S}\nonumber\\
&=& \sum_{m_L,m_S,m_s,m_t} \CG{Lm_L}{Sm_S}{JM} \CG{sm_s}{tm_t}{Sm_S}\int\Om\,Y_{m_L}^L(\Om)\ket{\Om_s^0,sm_s tm_t}
\en
with $Y_{m_L}^L(\Om)$ being the spherical harmonics. With these definitions we get the following
completeness relation
\be
1=\sum_{J,M,L,S}\ket{JMLS}\bra{JMLS}.
\ee
The normalization of the canonical states is then given by
\bn
\braket{\Om_{s\prm},s\prm m_{s\prm} t\prm m_{t\prm}}{\Om_s,sm_stm_t}
 &=&\delta(\Om_{s\prm}-\Om_s)\delta_{ss\prm}\delta_{tt\prm}\delta_{m_sm_{s\prm}}
\delta_{m_tm_{t\prm}}\\
\braket{J\prm M\prm L\prm S\prm}{JMLS}&=&
\delta_{JJ\prm}\delta_{MM\prm}\delta_{LL\prm}\delta_{SS\prm}.
\en
\paragraph{Helicity Formalism}
\subparagraph{Single Particle States}
The construction of states is similar to the procedure for canonical states.
May $\ket{j\lambda}$ be an at-rest state with spin $j$ and spin projection $\lambda$ on to the $z$-axis
in an euclidean system with coordinates $(x,y,z)$. The single particle state with a momentum
$\pvec$ is constructed through a pure Lorentz transformation $L\pfr$ plus a rotation.
This is due to the fact, that the $z$-axis is firstly rotated to the direction of $\pvec$.
Secondly the result is Lorentz transformed. Therefore the new $z$-axis, $z\prm$, is parallel
to $\pvec$ (see~\reffig{fig:angles}b).
\bn
\ket{\pfr,j\lambda}&\stackrel{def}{=}& L\pfr \ket{j\lambda}\label{eq:heli_lordef}\\
&=& \wht{R}_0 L_zp \wht{R}_0\inv\ket{j\lambda}.
\en
The second line of \refeq{eq:heli_lordef} follows from \refeq{eq:cano_lordef} and the unitarity of rotations.
If the helicity state is rotated, then the momentum $\pvec$ is rotated, but the helicity
is invariant, since the quantization axis is rotated as well. Therefore
\be
\wht{R}\ket{\pvec,\lambda} =\ket{\wht{R}\pvec,\lambda}.\label{eq:heli_rotdef}
\ee
Since \refeq{eq:cano_rotdef} and \ref{eq:heli_rotdef} are complete, it is possible to represent the helicity state
in the canonical basis and vice versa:
\be
\ket{\pvec,\lambda}=\sum_m D_{m\lambda}^j(R_0) \ket{\pvec,m}
\ee
\subparagraph{Two-Particle States}
Analogue to the construction of the canonical two-particle state \refeq{eq:cano_lordef}, the helicity equivalent
is constructed using the states \refeq{eq:heli_lordef}
\bn
\ket{\Om_s,s\lambda_s t\lambda_t}&\stackrel{def}{=}&\kappa\wht{R}_0\left[
L_z p_s\ket{s\lambda_s}L_z p_t\ket{t\lambda_t}\right]\nonumber\\
&=& \wht{R}_0(\Om_s) \ket{\Om=(0,0),s\lambda_s t\lambda_t}\label{eq:cano_twoparticle}
\en
Using a similar procedure to couple all spins one obtains
\be
\ket{JM\lambda_s\lambda_t}=N_J \int d\Om\quad D_{M,\lambda_s-\lambda_t}^{J\ast}\ket{\Om,s\lambda_s t\lambda_t}
\ee
with the normalization factor
\be
N_J=\sqrt{\frac{2J+1}{4\pi}}.
\ee
The choice of $N_J$ was made, so that we get a simple completeness relation
\be
1=\sum_{J,M,\lambda_s,\lambda_t}\ket{JM\lambda_s\lambda_t}\bra{JM\lambda_s\lambda_t}
\ee
The normalization of the helicity states is then given by
\bn
\braket{\Om_{s\prm},s\prm\lambda_{s\prm} t\prm\lambda_{t\prm}}{\Om_s,s\lambda_st\lambda_t}
&=&\delta(\Om_{s\prm}-\Om_s)\delta_{ss\prm}\delta_{tt\prm}\delta_{\lambda_s\lambda_{s\prm}}
\delta_{\lambda_t\lambda_{t\prm}}\\
\braket{J\prm M\prm \lambda_{s\prm} \lambda_{t\prm}}{JM\lambda_s\lambda_t}&=&
\delta_{JJ\prm}\delta_{MM\prm}\delta_{\lambda_s\lambda_{s\prm}}
\delta_{\lambda_t\lambda_{t\prm}}
\en
\subsubsection{Decay Amplitudes}
%
%
The two-particle states can now be used to derive two-body decay formulae in the two 
formalisms.
\paragraph{Canonical (Orbital) Formalism}
May $J$ be the at-rest system of the decaying state $\ket{JM}$ with arbitrary, but well defined 
spin $J$ and projection $M$. The decay amplitudes is derived by using the two-particle states and
summing over all unobservable spin-projections
\be
A=\sum_{m_s,m_t}\bra{\pvec_s,sm_s}\bra{-\pvec_s,tm_t}\MM\ket{JM}\label{eq:cano_decay}
\ee
with $\cal M$ being the unknown decay operator. To get to the final formula we need the following relation
\bn
\braket{\Om_s,sm_stm_t}{JMLS}&=&
\sum_{m_L,m_S,m_{s\prm},m_{t\prm}}\CG{Lm_L}{Sm_S}{JM}\CG{s\prm m_{s\prm}}{t\prm m_{t\prm}}{Sm_S}\nonumber\\
& &\int d\Om_{s\prm} Y_{m_L}^L(\Om_{s\prm})
\braket{\Om_s,sm_s tm_t}{\Om_{s\prm},s\prm m_{s\prm} t\prm m_{t\prm}}\\
&=&\sum_{m_L,m_S,m_{s\prm},m_{t\prm}}\CG{Lm_L}{Sm_S}{JM}\CG{s\prm m_{s\prm}}{t\prm m_{t\prm}}{Sm_S}\nonumber\\
& & \int d\Om_{s\prm} \delta(\Om_s-\Om_{s\prm})
\delta_{ss\prm}\delta_{tt\prm}\delta_{m_sm_{s\prm}}
\delta_{m_tm_{t\prm}}\nonumber\\
&=& \sum_{m_L,m_L}\CG{Lm_L}{Sm_S}{JM}\CG{sm_s}{tm_t}{Sm_S} Y_{m_L}^L(\Om_s)
\en
Now we include the two-particle states and obtain
\bn
A_{m_sm_t}^{JM}&=& \frac{4\pi}{\sqrt{\rho_s}}\bra{\Om_s,s m_s t m_t}\MM\ket{JM}\\
&=&\sum_{L,S} \braket{\Om_{s},s m_s t m_t}{JMLS}\frac{4\pi}{\sqrt{\rho_s}}\bra{JMLS}\MM\ket{JM}\nonumber\\
&\stackrel{def}{=}&\sum_{L,S} \sqrt{4\pi}a_{LS}^J \braket{\Om_{s},s m_s t m_t}{JMLS}\nonumber\\
&\stackrel{def}{=}&\sum_{L,S,m_L,m_S}\sqrt{4\pi}a_{LS}^J 
\CG{Lm_L}{Sm_S}{JM}\CG{sm_s}{tm_t}{Sm_S}Y_{m_L}^L(\Om_s)\label{eq:cano_formalism}
\en
with the canonical partial decay amplitudes $a_{LS}^J$
\be
a_{LS}^J\stackrel{def}{=}\sqrt{\frac{4\pi}{\rho_s}}\bra{JMLS}\MM\ket{JM}
\ee
which contain the physical matrix element of operator $\cal M$.
\paragraph{Helicity Formalism}
The decay helicity amplitude is derived in the same way as \refeq{eq:cano_decay}
\be
A=\sum_{\lambda_s,\lambda_t}\bra{\pvec_s,s\lambda_s}\bra{-\pvec_s,t\lambda_t}\MM\ket{JM}.\label{eq:heli_decay}
\ee
Using
\bn
\braket{\Om_s,s\lambda_s t \lambda_t}{JM\lambda_{s\prm}\lambda_{t\prm}}&=&
N_J\int d\Om\prm D_{M,\lambda_{s\prm}-\lambda_{t\prm}}^{J\ast}(\Om\prm)
\braket{\Om_s,s\lambda_s t \lambda_t}{\Om_{s\prm},s\prm\lambda_{s\prm} t\prm \lambda_{t\prm}}\nonumber\\
&=& N_J\int d\Om_{s\prm} D_{M,\lambda_{s\prm}-\lambda_{t\prm}}^{J\ast}(\Om_{s\prm})
\delta(\Om_{s\prm}-\Om_s)\delta_{ss\prm}\delta_{tt\prm}\delta_{\lambda_s\lambda_{s\prm}}\nonumber\\
&=& N_J D_{M,\lambda_{s\prm}-\lambda_{t\prm}}^{J\ast}(\Om_s)
\en
and inserting the two-particle states we get
\bn
A_{\lambda_s\lambda_t}^{JM}&=&\frac{4\pi}{\rho_s}\bra{\Om_s,s\lambda_s t\lambda_t}\MM\ket{JM}\\
&=& \sum_{\lambda_{s\prm},\lambda_{t\prm}}
\braket{\Om_s,s\lambda_s t\lambda_t}{JM\lambda_{s\prm} m_{t\prm}}\frac{4\pi}{\rho_s}
\bra{JM\lambda_{s\prm} m_{t\prm}}\MM\ket{JM}\nonumber\\
&=&\sqrt{\frac{4\pi}{\rho_s}(2J+1)}\bra{JM\lambda_s m_t}
\MM\ket{JM}D_{M,\lambda_{s\prm}-\lambda_{t\prm}}^{J\ast}(\Om_s)\nonumber\\
&=& N_J f_{\lambda_s\lambda_t} D_{M,\lambda_{s\prm}-\lambda_{t\prm}}^{J\ast}(\Om_s)\label{eq:heli_formalism}
\en
with the helicity amplitudes
\be
N_J f_{\lambda_s\lambda_t}=\sqrt{\frac{4\pi}{\rho_s}(2J+1)}\bra{JM\lambda_s m_t}\MM\ket{JM}
\ee
also containing the physical matrix element of operator $\cal M$.
The two formulae \refeq{eq:cano_formalism} and  \refeq{eq:heli_formalism} have a lot of similarities.
They are both an expansion in terms of spherical functions of different type ($Y_{m_L}^L$ for the canonical
and $D_{m\prm m}^J$ for the helicity formalism) with partial amplitudes which cover the decay
matrix element.
\par
To obtain an intensity it is necessary to include the population of spin states of the initial state. For a particle with spin $J$ there
ae $2J+1$ projections to the quantization axis, thus leading to a spin-density matrix $\rho$ of rank $2J+1$ which is usually diagonal and has the form
\be
\rho_{MM\prm}=\ba{ccc} 1 & & 0\\ & \ddots & \\ 0& & 1\ea
\ee
With this definition the observed number of events is given by
\be
I(\tht)_{\lambda\lambda\prm} = \sum_{M,M\prm,\lambda_s\lambda_{s\prm},\lambda_t\lambda_{t\prm}}
A_{\lambda_s\lambda_t}^{JM}(\ph,\tht)\rho_{MM\prm}A_{\lambda_{s\prm}\lambda_{t\prm}}^{JM\prm\ast}(\ph,\tht)
\ee
where the summation over $\lambda_s$, $\lambda_{s\prm}$, $\lambda_t$, $\lambda_{t\prm}$ is implicitly constrained
to $\lambda=\lambda_s-\lambda_t$ and $\lambda\prm=\lambda_{s\prm}-\lambda_{t\prm}$
\subsubsection{Relations between Canonical and Helicity Base}
%
%
Since both approaches deliver a complete description of the state, it is fairly easy
to go from one representation to another. This is important
if $L$ and $S$ are good quantum numbers for the system. Then the recoupling coefficients
lead to symmetry relations among the helicity amplitudes.
\par
To move from one representation to the other we need the recoupling coefficients. They are
derived using the properties of the $D$-functions:
\be
\braket{JMLS}{JM\lambda_s\lambda_t}=
\sqrt{\frac{2L+1}{2J+1}}\CG{L0}{S(\lambda_s-\lambda_t)}{J(\lambda_s-\lambda_t)}
\CG{s\lambda_s}{t(-\lambda_t)}{S(\lambda_s-\lambda_t)}\label{eq:heli_cano_recoupling}
\ee
With \refeq{eq:heli_cano_recoupling} we obtain a relation between $f_{\lambda_s\lambda_t}$ and
$a_{LS}^J$. The recoupling from the canonical to the helicity representation is
\bn
N_J f_{\lambda_s\lambda_t}^J&=&
\sum_{L,S}\braket{JM\lambda_s\lambda_t}{JMLS}\sqrt{\frac{4\pi}{\rho_s}(2J+1)}
\bra{JMLS}\MM\ket{JM}\nonumber\\
&=& \sum_{L,S}\sqrt{2L+1}\CG{L0}{S(\lambda_s-\lambda_t)}{J(\lambda_s-\lambda_t)}
\CG{s\lambda_s}{t(-\lambda_t)}{S(\lambda_s-\lambda_t)}a_{LS}^J
\en
while from the helicity to the canonical representation is
\bn
a_{LS}^J&=& \sum_{\lambda_s,\lambda_t}\braket{JMLS}{JM\lambda_s\lambda_t}\sqrt{\frac{4\pi}{\rho_s}}
\bra{JM\lambda_s\lambda_t}\MM\ket{JM}\nonumber\\
&=& \sum_{\lambda_s,\lambda_t}\braket{JMLS}{JM\lambda_s\lambda_t}
\frac{N_J}{\sqrt{2J+1}}f_{\lambda_s,\lambda_t}^J\nonumber\\
&=& N_J  \sum_{\lambda_s,\lambda_t}\frac{\sqrt{2L+1}}{2J+1}
\CG{L0}{S(\lambda_s-\lambda_t)}{J(\lambda_s-\lambda_t)}\CG{s\lambda_s}{t(-\lambda_t)}{S(\lambda_s-\lambda_t)}
f_{\lambda_s\lambda_t}^J.
\en
\subsubsection{Symmetry Relations}
%
%
Reactions involving strong or electroweak interactions conserve parity. Parity reverses the direction of $\rvec$ and $\pvec$
while the angular momentum remains unchanged.
Applying the parity transformation on a single particle state one obtains
\be
P\ket{\pvec,j m}=\eta \ket{\ph+\pi,\pi-\tht,p,jm}
\ee
in canonical basis and
\be
P\ket{\pvec,j \lambda}=\eta \Exp{-\imath\pi j}\ket{\ph+\pi,\pi-\tht,|\pvec|,j-\lambda}
\ee
in helicity basis with $\eta$ being the intrinsic parity ($\pm$1) of the particle.
For the two-particle states the parity transformation leads to the following relations
\be
P\ket{JMls}=\eta_1\eta_2(-1)^{l}\ket{JMls}
\ee
for the canonical and
\bn
\ket{JM\lone\ltwo}&=&\sum_{l,s}\sqrt{\frac{2l+1}{2J+1}}\CG{l0}{s\lambda}{JM}\CG{s_1\lone}{s_2(-\ltwo)}{s\lambda}\ket{JMls}\\
P\ket{JM\lone\ltwo}&=&\eta_1\eta_2(-1)^{J+s_1+s_2}\ket{JMls}
\en
for the helicity basis with $\eta_1$, $\eta_2$ and $\eta$ being the intrinsic parities of the two daughters and the decaying particle
respectively. This can be used to derive relations for the helicity amplitudes. If parity conservation holds then
\be
F_{\lone\ltwo}^J=\eta\eta_1\eta_2(-1)^{J+s_1+s_2}F_{(-\lone)(-\ltwo)}^J
\ee
In the special case, that both particles are identical one obtains
\be
F_{\lone\ltwo}^J\stackrel{1\equiv2}{=}\eta(-1)^J F_{\ltwo\lone}.
\ee
\subsubsection{Examples}
%
%
%
In this section several examples are given to illustrate the actual handling of
the helicity formalism in the day-to-day business.
\paragraph{Example: $f_2(1270)\to\pi\pi$}
The $f_2(1270)$ is an isoscalar tensor particle which means that the initial state has $\ig(\jpc)=0^+(\jpctwopp)$.
The two final state pions have $\ig(\jpc)=1^-(0^{-+})$. The intrinsic parity of the $f_2(1270)$ is even, since
$\eta_f=\eta_\pi^2(-1)^l$ and $l=2$. The total spin $s=2s_\pi$ is zero.
Starting with the definition of the formalism
\refeq{eq:heli_formalism}
\[
\AllJM = N_J \FllJ D_{M\lambda}^{J\ast}(\ph,\tht)
\]
and using $\lambda=\lone-\ltwo=0$ and $J=2$ we get
\bn
A_{00}^{2M}(\ph,\tht) &=& N_2 F_{00}^2 D_{M0}^{2\ast}(\ph,\tht)\\
N_2 F_{00}^2&=&\sqrt{5}\ub{\CG{20}{00}{00}}_{1}\ub{\CG{00}{00}{00}}_{1}a_{20}=\sqrt{5}a_{20}\nonumber\\
A_{00}^{2M}(\ph,\tht) &=& \sqrt{5}a_{20}D_{M0}^{2\ast}(\ph,\tht).
\en
Remember that this is a short-hand writing for a whole matrix of amplitudes
\be
A_{00}^{2}(\ph,\tht) = \sqrt{5}a_{20}\ba{c}
d_{(-2)0}^2(\tht)\Exp{-2\imath\ph}\\
d_{(-1)0}^2(\tht)\Exp{-\imath\ph}\\
d_{00}^2(\tht)\\
d_{10}^2(\tht)\Exp{\imath\ph}\\
d_{20}^2(\tht)\Exp{2\imath\ph}\ea.
\ee
The intensity is then derived by the bilinear sum of amplitude and conjugated amplitude, weighted by the spin density matrix
\bn
I(\tht) &=& \sum_{M,M\prm}A_{00}^{2M}(\ph,\tht)\rho_{MM\prm}A_{00}^{2M\prm\ast}(\ph,\tht)\nonumber\\
\rho &=&\frac{1}{5}\ba{ccc} 1 & & \\ & \ldots & \\ & &  1 \ea.
\en
Due to cancellation of $\cth$ and $\sth$ terms, the final result is very simple, it is just a constant:
\bn
I(\tht) &=& |a_{20}|^2\left(\frac{15}{4}\sthn{4}+15\sths\cths+5\left(\threehalf\cths-\frac{1}{2}\right)\right)\nonumber\\
&=& |a_{20}|^2= \cnst
\en
\paragraph{Example: $\omega\to\gamma\pi$}
The $\omega$ is an isoscalar vector particle which means that the initial state has $\ig(\jpc)=0^-(\jpconemm)$.
The final state pion has $\ig(\jpc)=1^-(0^{-+})$ and the photon has $\ig(\jpc)=0(\jpconemm)$. One should keep in mind that
the real photon has no longitudinal component. The intrinsic parity of the $\omega$ is $\eta_\omega=\eta_\pi\eta_\gamma(-1)^l$
and the total spin $s=s_\pi+s_\gamma$ is 1.
Starting with the definition of the formalism
\refeq{eq:heli_formalism}
\[
\AllJM = N_J \FllJ D_{M\lambda}^{J\ast}(\ph,\tht)
\]
and using $\lambda=\lone-\ltwo=\lambda_\gamma=\lone$ and $J=1$ we get
\bn
A_{\lambda0}^{1M}(\ph,\tht) &=& N_1 F_{\lambda0}^1 D_{M\lambda}^{1\ast}(\ph,\tht)\\
N_1 F_{\lambda0}^1&=&\sqrt{3}\ub{\CG{10}{1\lambda}{J\lambda}}_{-\frac{\lambda}{\sqrt{2}}}
\ub{\CG{1\lambda}{00}{1\lambda}}_{1}a_{11}=-\lambda\sqrt{\threehalf}a_{11}\nonumber\\
A_{\lambda0}^{1M}(\ph,\tht) &=& -\lambda\sqrt{\threehalf}a_{11}D_{M\lambda}^{1\ast}(\ph,\tht).\label{eq:heli_ex_omega}
\en
or in matrix representation
\be
A_{\lambda0}^{1}(\ph,\tht)=-\sqrt{\threehalf}\ba{lll}
-d_{(-1)(-1)}^1(\tht)\Exp{-\imath\ph} & 0 & -d_{(-1)1}^1(\tht)\Exp{-\imath\ph}\\
-d_{0(-1)}^1(\tht) & 0 & -d_{01}^1(\tht)\\
-d_{1(-1)}^1(\tht)\Exp{\imath\ph} & 0 & -d_{11}^1(\tht)\Exp{\imath\ph}\ea
\ee
leading to the intensity
\bn
I(\tht) &=& \sum_{M,M\prm,\lambda,\lambda\prm}A_{\lambda0}^{1M}(\ph,\tht)\rho_{MM\prm}A_{\lambda^\prime0}^{1M\prm\ast}(\ph,\tht)
\delta_{\lambda\lambda\prm}\nonumber\\
\rho &=&\frac{1}{3}\ba{ccc} 1 & & \\ & 1 & \\ & &  1 \ea
\en
which finally collapses again to a constant:
\bn
I(\tht) &=& \half|a_{11}|^2\left(
2\left(\frac{1-\cth}{2}\right)^2+2\left(\frac{1+\cth}{2}\right)^2+2\frac{\sths}{2}\right)\nonumber\\
&=& \half |a_{11}|^2(1+\cths+\sths)= |a_{11}|^2= \cnst
\en
\paragraph{Example: $f_{0,2}(any)\to\gamma\gamma$}
The $f_{0,2}$ is an isoscalar scalar or tensor particle which means that the initial state has $\ig(\jpc)=0^+(\jpczeropp$ or $\jpctwopp)$.
The two final state gammas have $\ig(\jpc)=0(\jpconemm)$. The intrinsic parity of the $f_2(1270)$ is even, since
$\eta_f=\eta_\gamma^2(-1)^l$ and $l=2$. The total spin $s=2s_\gamma$ is two and thus $l=0,2$ are allowed for the $f_0$ and $l=0,2,4$ 
are allowed for the $f_2$.
In the first case we investigate $J=0$. Following the definition of the formalism \refeq{eq:heli_formalism}
\[
\AllJM = N_J \FllJ D_{M\lambda}^{J\ast}(\ph,\tht)
\]
and using $\lambda=\lone-\ltwo$ we get with $J=0$
\bn
A_{\lonetwo}^{00}(\ph,\tht) &=& N_0 F_{\lonetwo}^0 D_{0\lambda}^{0\ast}(\ph,\tht)\\
N_0 F_{00}^0&=&\sum_{l,s}(l0 s\lambda|J\lambda)(s_1\lone s_2(-\ltwo)|s\lambda)a_{ls}\nonumber\\
&=&\sqrt{1}(00 00|00)(1\lone 1(-\ltwo)|0\lambda)a_{00}\nonumber\\
& &+\sqrt{5}(20 20|00)(1\lone 1(-\ltwo)|2\lambda)a_{22}\nonumber\\
&=&\sqrt{\frac{1}{3}}a_{00}+\sqrt{\frac{1}{6}}a_{22}\nonumber\\
A_{\lonetwo}^{00}(\ph,\tht) &=& \left(\sqrt{\frac{1}{3}}a_{00}+\sqrt{\frac{1}{6}}a_{22}\right)\ub{D_{00}^{0\ast}(\ph,\tht)}_{1}=\cnst
\en
In the more complicated case for $J=$ we get
\bn
A_{\lonetwo}^{2M}(\ph,\tht) &=& N_2 F_{\lonetwo}^0 D_{M\lambda}^{2\ast}(\ph,\tht)\\
N_2 F_{\lonetwo}^2&=&\sum_{l,s}(l0 s\lambda|J\lambda)(s_1\lone s_2(-\ltwo)|s\lambda)a_{ls}\\
&=&\sqrt{5}(20 00|20)(1\lone 1(-\ltwo)|00)a_{20}\nonumber\\
& &+\sqrt{5}(20 2\lambda|2\lambda)(1\lone 1(-\ltwo)|2\lambda)a_{22}\nonumber\\
& &+\sqrt{9}(40 2\lambda|2\lambda)(1\lone 1(-\ltwo)|2\lambda)a_{42}\nonumber\\
&=&\sqrt{\frac{1}{3}}a_{00}+\sqrt{\frac{1}{6}}a_{22}\\
A_{\lonetwo}^{00}(\ph,\tht) &=& \left(\sqrt{\frac{1}{3}}a_{00}+\sqrt{\frac{1}{6}}a_{22}\right)\ub{D_{00}^{0\ast}(\ph,\tht)}_{1}\cnst.
\en
Because of the complexity of this result we limit the discussion now to the case where $J=\lambda=2$. Then a lot of
terms disappear and the amplitude contracts to
\bn
N_2 F_{1(-1)}^2&=&\sum_{l,s}(l0 s2|22)(s_1 1 s_2 1|s2)a_{ls}\nonumber\\
&=& \sqrt{5}\ub{\CG{20}{22}{22}}_{\sqrt{\frac{2}{7}}}\ub{\CG{11}{11}{22}}_{1}a_{22}
+\sqrt{9}\ub{\CG{40}{22}{22}}_{\frac{1}{3\sqrt{14}}}\ub{\CG{11}{11}{22}}_{1}a_{42}
\en
Using the symmetrization because of the two identical particles in the final state one gets a fairly simple
angular distribution
\be
A^{2M}=N_2\left(F_{1(-1)}+F_{(-1)1}\right)D_{M2}^{2\ast}(\ph,\tht)\propto D_{M2}^{2\ast}(\ph,\tht).
\ee
\paragraph{Example: $\pbarp\to\pi\pi$}
The $\pbarp (\jpczeromp)$ has $\ig(\jpc) =1^+(\jpc)$ with $M=0,\pm1$.
The two final state pions have $\ig(\jpc) =1^-(\jpczeromp)$. 
Following the definition of the formalism \refeq{eq:heli_formalism}
\[
\AllJM(\ph,\tht) = N_J \FllJ D_{M\lambda}^{J*}(\ph,\tht)
\]
and using $\lambda=\underbrace{\lone}_{0}-\ub{\ltwo}_{0}=0$ and $J=l$ we get
\bn
A_{00}^{JM}(\ph,\tht) &=& N_J F_{00}^J D_{M0}^{J*}(\ph,\tht)\\
N_J F_{00}^J&=&\sum_l\sqrt{2l+1}\ub{\CG{l0}{00}{J0}}_{1}\ub{\CG{00}{00}{00}}_{1}a_{l0}=\sqrt{2J+1}a_{J0}\nonumber\\
A_{00}^{JM}(\ph,\tht) &=& \sqrt{2J+1}a_{J0}D_{M0}^{J\ast}(\ph,\tht)=\sqrt{2J+1}a_{J0}d_{M0}^{J}(\tht)\Exp{-\imath M\ph}
\en
The $d$-functions are not orthogonal, if $\ph$ is not observed ambiguities remain in the amplitude and polarization
is needed. to disentangle the waves.
\paragraph{Example: $\pbarp\to \omega\pi^0\to\gamma\piz\piz$}
This is a two step process $\pbarp\to \omega\pi^0$ and $\omega\to\gamma\piz$ and illustrates how subsequent
decays are handled in the helicity formalism. The $\pbarp$ has $\ig(\jpc) =1^+(\jpc)$. The intermediate
$\omega$ has $\ig(\jpc) =0^-(\jpconemm)$, the two pions have $\ig(\jpc) =1^-(\jpczeromp)$ and
the photon has $\ig(\jpc)=0(\jpconemm)$. The angular momentum between the recoil
pion and the $f_2(1270)$ is $L=2$.
Using the result from \refeq{eq:heli_ex_omega} we obtain
\bn
\AllJM(\Om_1,\Om_2)&=&A_{\lambda_\omega\lambda_\gamma}^{JM}(\Om_{\omega\piz},\Om_{\piz\gamma})=
A_{\lambda_\gamma0}^{1\lambda_\omega}(\Om_{\piz\gamma})A_{\lambda_\omega0}^{JM}(\Om_{\omega\piz})\\
&=&N_{\omega,1}F_{\lambda_\gamma0}^1 D_{\lambda_\omega\lambda_\gamma}^{1\ast}(\Om_{\pi0\gamma})
N_{\pbarp J}F_{\lambda_\omega0}^J D_{M\lambda_\omega}^{J\ast}(\Om_{\omega\pi^0})\nonumber\\
&=& -\lambda_\gamma\sqrt{\threehalf}D_{\lambda_\omega\lambda_\gamma}^{1\ast}(\Om_{\pi0\gamma})\nonumber\\
& & \times\left(\sum_l \sqrt{2l+1}(l0 1\lambda_\omega|J\lambda_\omega)a_{\pbarp,l1}D_{M\lambda_\omega}^{J\ast}(\Om_{\omega\piz})\right)\nonumber\\
&=& -\lambda_\gamma\sqrt{\threehalf}D_{\lambda_\omega\lambda_\gamma}^{1\ast}(\Om_{\pi0\gamma})
D_{M\lambda_\omega}^{J\ast}(\Om_{\omega\piz})\nonumber\\
& & \times\left(\sum_l \sqrt{2l+1}(l0 1\lambda_\omega|J\lambda_\omega)a_{\pbarp,l1}\right)
\en
The interesting feature of this amplitude is, that the helicity constant $a_{\omega\pi0,11}$ factorizes and all
helicity amplitudes of the $\pbarp$ system can be modified to
\be
a_{\pbarp,l1}\prm=a_{\pbarp,l1}a_{\omega\pi0,11}
\ee
\paragraph{Example: $\pbarp\to f_2(1270)\pi^0\to\pip\pim\piz$}
This is an two step process $\pbarp\to f_2(1270)\pi^0$ and $f_2(1270)\to\pip\pim$. 
The $\pbarp (\jpczeromp)$ has $\ig(\jpc) =1^-(\jpczeromp)$. The intermediate
$f_2(1270)$ has $\ig(\jpc) =0^+(\jpctwopp)$ and the two pions have $\ig(\jpc) =1^-(\jpczeromp)$. The angular momentum between the recoil
pion and the $f_2(1270)$ is $L=2$.
Starting with the definition of the formalism\refeq{eq:heli_formalism}
\[
\AllJM = N_J \FllJ D_{M\lambda}^{J\ast}(\ph,\tht)
\]
and $\lambda =\lone-\ltwo=0$, $J_{\pbarp}=0$ and $J_{f_2\piz}=2$ we get
The full amplitude is then given as (amplitude tree) 
\bn
\AllJM(\Om_1,\Om_2)&=&\AllJM(\Om_{f_2\piz},\Om_{\pip\pim})=A_{00}^{00}(\Om_{f_2\piz})A_{20}^{00}(\Om_{\pip\pim})\\
&=&
N_{\pbarp,0} F_{00}^0 D_{00}^{0\ast}(\Om_{f_2\piz})N_{f_2,2} F_{00}^2 D_{00}^{2\ast}(\Om_{\pip\pim}p)\nonumber\\
N_{\pbarp,0} F_{00}^0&=&\sqrt{1}\ub{\CG{20}{20}{00}}_{\sqrt{\frac{1}{5}}}\ub{\CG{20}{00}{20}}_{1}a_{\pbarp,22}\nonumber\\
N_{f_2,2} F_{00}^2 &=&\sqrt{1}\ub{\CG{20}{00}{20}}_{1}\ub{\CG{00}{00}{00}}_{1}a_{f_2,20}\nonumber\\
A_{00}^{00}(\Om_{f_2\piz})A_{20}^{00}(\Om_{\pip\pim})&=&\sqrt{5}a_{\pbarp,22}a_{f_2,20}D_{00}^{2\ast}(\ub{\Om_{\pip}}_{=(\ph,\tht)})
\ub{D_{00}^{0\ast}(\Om_{f_2\piz})}_{1}\nonumber\\
&=& \sqrt{5}a_{\pbarp,22}a_{f_2,20} \left(\threehalf\cths-\half\right).
\en
with $\Om_{xy}=(\ph,\tht)$ being the direction of daughter $x$ in the mother system of $xy$.
The final intensity turn out to be simple again:
\be
I(\tht)=5\left|a_{\pbarp,22}a_{f_2,20}\right|^2\left(\threehalf\cths-\half\right)
\ee
\subsubsection{The $\pbarp$ system}
%
%
\paragraph{Proton antiproton at rest}
The proton antiproton system at rest is formed by an incident antiproton beam
on a hydrogen target. Due to Coulomb scattering the antiproton is decelerated
according to Bethe-Bloch's formula until it's energy is small enough to be caught
by a hydrogen atom. The protonium atom (proton antiproton atomic bound state)
is formed at very high $n$ and $l$ of about 30 where it replaces the electron.
The system deexcites through slow radiative transitions and through collisions
(Auger effect). These processes compete with the Stark mixing of the $l$ levels.
Due to the fact, that the protonium atom is much smaller than a usual hydrogen
atom it traverses those and feels a very strong electric field which make the levels
mix (Day, Snow and Sucher, 1960). The average electric field which is seen by the
protonium atom varies with the density of hydrogen atoms and thus depends on the
target density. Since in S-wave orbits a annihilation is very probable high density
targets (like in liquid hydrogen) have a dominant S-wave yield. In low density targets
(like in gaseous hydrogen) more P-wave is present \cite{Day60}.
The advantages of this initial system are manifold
\begin{itemize}
\item $\jpc$ varies with the target density,
\item isospin varies with n (deuterium) or p (hydrogen) targets,
\item atomic (incoherent) initial states enable unambiguous partial wave decomposition.
\end{itemize}
But there are also some disadvantages like
\begin{itemize}
\item limited phase space and a very small kaon yield.
\end{itemize}
The $\pbarp$ system has an invariant mass of $m_{\pbarp}\approx$1877$\,\mevcc$ . Since a recoil
particle is needed to ensure energy and momentum balance investigations are only possible
up to a mass of $m_{Meson}\approx$1700$\,\mevcc$. If due to quantum number requirements another
(heavier) recoil particle is required, the phase space limits are even more severe.
\par
In principle the same kind of physics could be investigated using an $\nbar$ beam. This
has been done by the Obelix collaboration. Since the $\nbar$ as the n is neutral, there
is now deceleration in material and no neutronium atom. But one can do annihilations in flight
which is discussed in the next subsection. This is not true for annihilation of an
antiproton on the neutron of a deuteron, since the deuteron and the antiproton also form
a bound atomic system.
\par
The annihilation - being a strong process - conserves all quantum numbers from the
atomic state. The most important ones are
\begin{itemize}
\item $G=(-1)^{I+L+S}$,
\item $P=(-1)^{L+1}$,
\item $C=(-1)^{L+S}$ and $CP=(-1)^{S+1}$.
\end{itemize}
The isospin of a $\pbarp$ system can be either 0 or 1. Therefore in the isospin space
the protonium system mixes with the neutronium and the wave functions for the initial
state are
\begin{itemize}
\item $\frac{1}{\sqrt{2}}(\ket{\pbarp}+\ket{\nbarn})$ has $I$=0,
\item $\frac{1}{\sqrt{2}}(\ket{\pbarp}-\ket{\nbarn})$ has $I$=1, $I_3$=0 and
\item $\ket{\pbarn}$ has $I$=1, $I_3$=+1.
\end{itemize}
Usually only $S$- and $P$-waves are taken into account. In liquid hydrogen the $S$-wave
is dominant with a yield of more than 90\%. In gaseous hydrogen at NTP the
ratio between $S$- and $P$-wave is around 1. In low pressure targets the $P$-wave
yield may be as much as 90\%~\cite{Batty89}.
Since the atomic system lives quite long compared to the strong interaction 
it rotates many times until it finally annihilates. Therefore the helicity of the initial
antiproton is of no use. The population density is just equally distributed for
all helicities or in other terms for all values of the magnetic quantum number.
\paragraph{Proton antiproton in flight}
With higher antiproton momenta a higher center of mass energy can be reached and thus more
massive mesons can be produced. In addition there is the possibility to create resonances
without recoil particles which makes the analysis much easier and less ambiguous, although not all
quantum numbers are accessible (only fermion anti-fermion quantum numbers remain to be producible).
The annihilation of $\pbarp$ at higher incident momenta is completely different to the situation at rest.
There are no more atomic levels. As soon as the antiproton is to fast to be captured one deals
with a scattering process. The number of partial waves rise with increasing momenta and detailed
investigations show that the maximum needed angular momentum is  $l\approx p_{cms}/$200$\,\mevcc$
which is almost compatible with statistical models. This leads to a huger amount of waves which where
most of them are allowed to interfere. But there are some constraints which limit the amplitudes
to be considered.
\par
Since we have a fermion anti-fermion system helicity conservation reduces the initial helicities
to 0 or 1 with the symmetry relation $H^J_{\nu_1\nu_2}=\eta_J(-1)^JH^J_{-\nu_2(\nu_1)}$
with helicities $\nu_{1,2}$ for the proton and the antiproton respectively and total angular
momentum $J$. From C and CP  invariance on gets in addition $H_{11}=0$ if  $L+S+J$ is odd
and $H_{-1(1)}=0$ if $S=0$ and/or $J=0$. Since C and CP are conserved interference
is only possible in between waves where C and CP are equal. This decouples singlet and 
triplet states as well as it decouples even from odd $L$. In total four incoherent sets of amplitudes remain. 
The final state poses additional constraints so that many amplitudes vanish due to other conservation laws.
Nevertheless the number of parameters for a partial wave analysis is still very high and can easily exceed
100. Therefore it might be possible that in many cases the amplitudes have to be simplified with
the cost of giving informations and constraints away.
\par
If there is a formation process with two particles in the final state a full featured amplitude is usually
applicable and produces unambiguous results \cite{Peters01}.
But if there are three and more particles involved and the number of resonances is large
it is necessary to reduce the complexity of the fit by integrating out the production process and
concentrating on the final state. Many constraints from the initial state are lost that way but the number
of parameters can be handled easier. The main caveat is the handling of coherence. For two fully coherent
amplitudes the intensity looks like $I=|A+e{i\phi}B|^2$ = $|A|^2+|B|^2+2(\RE{AB^\ast}\sin\varphi+\IM{AB^\ast}\cos\varphi)$.
If there is no coherence it is $I=|A|^2+|B|^2$. In the case of a mixture of coherent and incoherent terms an
effective coherence is needed which can be defined like $I=|A|^2+|B|^2+c(\RE{AB^\ast}\sin\varphi+\IM{AB^\ast}\cos\varphi)$
with $c$=-2,$\ldots$,2. This technique leaves in the possibility to check whether or not coherence between two
amplitudes is needed at all \cite{Abele99}.
\paragraph{Other anti-nucleon nucleon systems}
There are actually two other techniques which have been used apart from $\pbarp$ annihilation.
One is the $\nbarn$ annihilation by Obelix. This reaction cannot occur from an atomic state, but
as an in flight reaction. This is done by using a antiproton  beam to produce antineutrons in the reaction
$\pbarp\rightarrow\nbarn$. This produces antineutrons in a broad energy range. They annihilate
on a deuterium target and the neutral events are selected with zero net charge in the final event.
The quantum numbers can be applied as in the $\pbarp$ case in flight. The usual energy range does
not exceed 300$\,\mevc$ so angular momenta up to $D$-wave are sufficient to fit the data.
The other technique is to use a antiproton beam on a deuterium target and by selecting the net
charge of the event a selection of $\pbarp$ and $\pbarn$ reactions is possible. In addition information
comes from the spectating nucleon, either a proton or a neutron. The antiproton is in an atomic
orbit of the deuteron before annihilation, but the levels are somewhat different (due to the different
reduces mass of the system) and the base system for the waves is the center of mass of the deuteron,
and not the individual nucleon which takes part of the annihilation. In addition, there is Fermi motion
of the nucleons inside the deuteron. To select the initial angular momentum of the
whole system the breakup momentum of the spectator is used. If the momentum is small ($p_{spec}<$100$\,\mevc$)
the process is dominated by S-wave. Due to the momentum range all phase space boundaries (like in Dalitz plots)
remain fuzzy.

\subsection{Moments Analysis}
%
%
The basic idea of this approach is to do a Fourier decomposition of
of he final state. This is done for each bin of invariant masses of the
exclusive final state system. This method is explained best by using an example.
For an unpolarized target and without measuring the recoil polarization of the neutron
in a charge exchange reaction of the type
\be
\pim+\mbox{p}\to M^0 +\mbox{n}\qquad\mbox{and}\qquad M^0\to a+b
\ee
the differential cross section can be written~\cite{Martin76} as
\be
I(t,M,\tht,\ph)= \frac{\partial^4\sigma}{\partial t\partial M\partial\cth\partial\ph}= \half\sum_{\lamP,\lamN}
|H_{\lamP\lamN}(t,M,\tht,\ph)|^2\label{eq:moments_def1}
\ee
where $t$ is the momentum transfer between the incident $\pim$ and the $M^0$ system, M is the $M^0$ invariant mass,
$\lamP$, $\lamN$ are the proton and neutron helicities and $\tht,\ph$ are the polar and azimuthal angles of either
particle $a$, or $b$ in the Gottfried-Jackson frame~\footnote{The Gottfried-Jackson frame is the $M^0$ frame rotated
so that its $z$-axis is parallel to the incident $\pim$ direction.}.
The full helicity amplitude $H_{\lamP\lamN}(t,M,\tht,\ph)$ which is the sum of the terms corresponding to all possible
intermediate spin $j$ and helicity $m$ states of $M^0$ is given by
\be
\sqrt{4\pi}H_{\lamP\lamN}(t,M,\tht,\ph)=\sum_{j=0}^\infty\sum_{m=-j}^j\sqrt{2j+1}H_{\lamP\lamN,m}^j d_{m0}^j(\tht)\Exp{\imath m\ph}.\label{eq:moments_def2}
\ee
Inserting \refeq{eq:moments_def2} in \refeq{eq:moments_def1} one gets
\bn
4\pi I(\tht,\ph)&=& \half\sum_{\lamP,\lamN}\sum_{j_2,m_2}\sum_{j_1,m_1}\sqrt{2j_1+1}\sqrt{2j_2+1}\Exp{\imath(m_1-m_2)\ph}\nonumber \\
&=& \times H_{\lamP,\lamN,m_1}^{j_1\ast} H_{\lamP,\lamN,m_2}^{j_2} d_{m_10}^{j_1}(\tht)d_{m_20}^{j_2}(\tht)\label{eq:moments_def3}
\en
The helicity products of relation \refeq{eq:moments_def3} can be expressed in terms of the density matrix
\be
\rho_{m_1m_2}^{j_1j_2}=\frac{1}{2N}\sum_{\lamP,\lamN}H_{\lamP,\lamN,m_1}^{j_1\ast} H_{\lamP,\lamN,m_2}^{j_2}\label{eq:moments_def4}
\ee
where $\dsty N=\frac{\partial^2\sigma}{\partial t \partial M}$. Therefore \refeq{eq:moments_def3} can be rewritten as
\be
I(\tht,\ph)= \frac{N}{4\pi}\sum_{j_2,m_2,j_1,m_1}\sqrt{2j_1+1}\sqrt{2j_2+1}\rho_{m_1m_2}{j_1j_2}
d_{m_10}^{j_1}(\tht)d_{m_20}^{j_2}(\tht)\Exp{\imath(m_1-,_2)\ph}
\ee
and its integral gives the trace equality
\be
\sum_{j,m}\rho_{mm}^{jj}=1.
\ee
As the strong interaction conserves parity, for the process $1+2\to 3+4$, the helicity amplitudes satisfy the relation
\be
H_{\lambda_3,\lambda_4,\lone,\ltwo}=\eta(-1)^k H_{-\lambda_3,-\lambda_4,-\lone,-\ltwo}\label{eq:moments_def7}
\ee
whre $\eta$ is the intrinsic parity product of the particles 1, 2, 3 and 4 and $k=\sum_i(s_i+\lambda_i)$ is the sum of their
spins and helicities. Combining \refeq{eq:moments_def7} and \refeq{eq:moments_def4} one obtains
\be
\rho_{m_1m_2}^{j_1j_2}=(-1)^{m_1+m_2}\rho_{(-m_1)(-m_2)}^{j_1j_2}.
\ee
\par
In any reference frame where the $z$-axis is in the production plane, the most general angular distribution of the $M^0$
system can be written, according to parity conservation, as a sum of the real parts $\RE{Y_l^m}$ of the spherical harmonic moments.
For simplicity we will write $Y_l^m$ instead of $\RE{Y_l^m}$
\be
I(t,M,\tht,\ph)=N\sum_{l=0}^\infty\sum_{m=-l}^l \langle Y_l^m \rangle Y_l^m(\tht,\ph)
\ee
where $\langle Y_l^m \rangle$ are the normalized moments. In particular $\langle Y_0^0 \rangle=\frac{1}{4\pi}$ and $Y_0^0=1$.
\par
Using standard relations amongst rotation matrix products, the $\langle Y_l^m \rangle$ coefficients can be expressed as a sum of the real parts
of the elements $\rho_{m_1m_2}^{j_1j_2}$
\bn
\langle Y_l^m \rangle&=&\frac{1}{4\pi}\sum_{j_2,m_2,j_1,m_1}
(-1)^{m-m_2}\sqrt{(2l+1)\frac{2j_1+1}{2j_2+1}}\nonumber\\
& &\CG{lm}{j_1(-m_1)}{j_2(-m_2)}\CG{l0}{j_10}{j_20}\RE{\rho_{m_1m_2}^{j_1j_2}}.\label{eq:moments_def10}
\en
Together with \refeq{eq:moments_def4} it is easy to see, that the $\langle Y_l^m \rangle$ can be expressed as bilinear forms of helicity amplitudes.
\par
Including the factor $N$ in the coefficients $\langle Y_l^m \rangle$,one obtains a new set of coefficients $t_l^m$ which depend on the momentum
transfer $t$ and the mass $M$ of the $M^0$ system. The angular distribution of the produced events can then be written as
\be
I_{prod}(t,M,\tht,\ph)=\sum_{l=0}^\infty\sum_{m=0}^l t_l^m Y_l^m(\tht,\ph)\label{eq:moments_def11}
\ee
This set of coefficients is obtained by fitting \refeq{eq:moments_def11}. To obtain the different waves, which contribute to a specific moment
the relation \refeq{eq:moments_def10} has to inverted. Since this is a non-linear system of equation there are several solutions and it has
to be ensured in the analysis process, that the different solutions show the same physical behavior. Otherwise the data will not be
conclusive.
\par
An example with only $S$- and $P$-wave may serve to illustrate the details. For $l\le1$ there are only two components
\be
f(\tht)= \langle S\rangle P_0(\cth)+3\langle P\rangle P_1(\cth)
= \langle S\rangle+3\langle P\rangle \cth
\ee
%

%

%
\section{Dynamical Amplitudes}
\subsection{S-Matrix}
%
%
\bfg[hbt]\bc
\psfig{figure=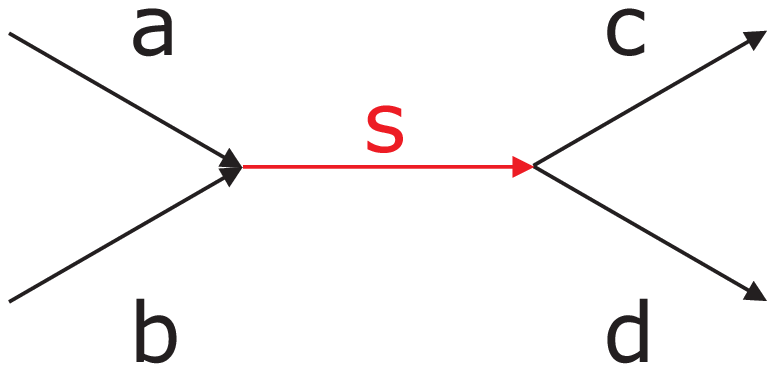,scale=0.5}\ec
\caption[s-Channel scattering]{s-Channel scattering.}
\label{fig:reaction_s_channel}
\efg
Consider a two-body scattering of the type $ab\to cd$ (like in \reffig{fig:reaction_s_channel}).
The differential cross section is given in terms of the
invariant amplitude $\MM$ and the `scattering amplitude'  $f$ through
\be
  {d\sigma_{fi}\over d\Omega}
={1\over (8\pi)^2\,s}\left(q_f\over q_i\right)|\MM_{fi}|^2
=|f_{fi}(\Omega)|^2\label{eq:sm_crosssection}
\ee
where `$i$' and `$f$' stand for the initial and final states;
$\Omega=(\theta,\phi)$ denotes the usual spherical coordinate system;
and $s=m^2$ is the square of the center-of-mass (CM) energy.
The $q_i$($q_f$) is the breakup momentum in the initial(final) system.
[The observed cross section is in reality the average of the
initial spin states and the sum over all final spin states---
this is suppressed here for simplicity.]
The scattering amplitude can be expanded in terms of the partial-wave
amplitudes
\be
 f_{fi}(\Omega)={1\over q_i}\sum_J (2J+1)T^J_{fi}(s)
 D^{J\,*}_{\lambda\mu}(\phi,\theta,0)\label{eq:scattering_amplitude}
\ee
where
$\lambda=\lambda_a-\lambda_b$
and $\mu=\lambda_c-\lambda_d$ in terms of the helicities
of the particles involved in the scattering $ab\to cd$.
Note that this `scattering amplitude' is a factor of two
bigger than that with a more common
definition (for example, see Section 5.1, Chung \cite{Chung71}).
One may in addition note
that the argument of the $D$-function is frequently given
as $(\phi,\theta,-\phi)$ (see Jacob and Wick \cite{Jacob59}
and Martin and Spearman \cite{Martin70}).
Integrating the differential cross section over the angles,
one finds, for the cross section in the partial wave $J$,
\be
   \sigma^J_{fi}=\left(4\pi\over q^2_i\right)
   (2J+1)|T^J_{fi}(s)|^2\label{eq:sm_tm_matrix}
\ee
Note that $T^J$ has no unit; the unit for the cross section is
being carried by $q^2_i$.  It is necessary to define more precisely
the initial and the final states
\bn
  \keti &=& \ket{ab,JM\lambda_a\lambda_b}\\
  \ketf &=& \ket{cd,JM\lambda_c\lambda_d} \label{eq:if_def}
\en
where $M$ is the z-component of total spin $J$ in a coordinate system
fixed in the overall CM frame and the notations $\{ab\}$ and
$\{cd\}$ designate additional informations needed to fully specify
the initial and the final states.
Because of conservation of angular
momentum, an initial state in $\ket{JM}$ remains the same
in the scattering process. Note the normalization
(see Section 4.2, Chung \cite{Chung71})
\be
   \braket{f}{i} =\delta_{ij}\label{eq:if_deltaif}
\ee
In the remainder of this section and in subsequent sections,
it is be understood that the ket states mentioned always refer to those of \refeq{eq:if_def}.  In particular,
explicit references to the total angular momentum $J$ will be
suppressed.  Note that, with this convention, one has eliminated
the necessity of specifying continuum variables such as angles and
momenta.
\par
   In general, the amplitude that an initial state
$\keti$ will be found in the final state $\ketf$ is
\be
   S_{fi}=\braf S \keti   \label{eq:sm_def}
\ee
where $S$ is called the scattering operator.
One may remove
the probability that the initial and final states do not interact
at all, by defining the transition operator $T$ through
\be
   S=I+2\imath\ T \label{eq:sm_tmdef}
\ee
where $I$ is the identity operator.
The factors 2 and $\imath$ have been introduced for convenience.
From conservation of probability, one deduces that the
scattering operator $S$ is unitary, i.e.
\be
   S\ S^\dagger = S^\dagger S=I 
\ee
\subsection{T-Matrix}
\subsubsection{Harmonic Oscillator and the Lorentz-Function}
%
%
\paragraph{Harmonic Oscillator}
As an academic example toward the discussion about the properties of the $T$-matrix and
how to obtain a reasonable parametrization we go back to the classical harmonic oscillator.
from the Lagrange function
\be
L=\frac{m}{2}\xdot^2-\frac{kx^2}{2}
\ee
and the Lagrange equations
\be
\ddt\frac{\dL}{\dxidot}=\frac{\dL}{\dxi}
\ee
one can derive the equation of motion, which is in this case the well known formula
\be
\xddot+\oZ^2=0
\ee
If the oscillator is damped and periodic forces are applied to the damped harmonic system like
\be
\xddot+2\lambda\xdot+\oZ^2x=\frac{f}{m}\cos \oR t=\frac{f}{m}\RE{\Exp{ \imath\oR t}}
\ee
leads to the time averaged intensity (over a half period) in
the proximity of the resonance 
\be
I(\oR)=\frac{f^2}{4m}\frac{\lambda}{(\oR-\oZ)^2+\lambda^2}.
\ee
This function is called Lorentz function and has the phase $\delta$ with
\be
\tan\delta=\frac{2\lambda\oR}{\oZ^2-\oR^2}\label{eq:osc_phase}.
\ee
and has a maximum for $\oR=\oZ$ with a value of
$\dsty I_0=I(\oZ)=\frac{f^2}{4m}$.
\paragraph{Primer in Scattering Theory}
In scattering theory a line shape of a resonance is obtained by
solving the Schr\"odinger equation. For two spinless particles with masses
$m_1$ and $m_2$ the interaction can be described by a spherical potential with
$V(\rvec)=V(|\rvec|)=V(r)$. The Schr\"odinger equation is defined like
\be
-\frac{\hbar}{2\mu^2}\Nabla^2\Psi+V(r)\Psi = E\Psi
\ee
with the reduced mass $\dsty\mu = \frac{m_1 m_2}{m_1+m_2}$. The incoming particle is
assumed to be a free particle, represented by an incoming wave. If the particle
moves in $z$ we get
\be
\Psi_i=\Exp{\imath kz} \mbox{\ with \ } \kvec=\frac{\pvec}{h}.
\ee
The incoming wave can be expanded in partial waves
\be
\Psi_i =\suml U_l(r)P_l(\cos\theta)
\ee
where $U_l$ are the solutions of the free ($V=0$) radial equation.
\be
\frac{d^2}{dr^2}U_l(r)+\left(k^2-\frac{l(l+1)}{r}\right)U_l(r)=0.
\ee
The asymptotic behaviors of the wave for $kr\to\infty$ is
\be
\Psi_{i,kr\to\infty} = \Exp{\imath kz}
= \suml\frac{\imath^l(2l+1)}{2\imath kr}
\left[\Exp{\imath\left(kr-\frac{l\pi}{2}\right)}
-\Exp{\imath\left(kr+\frac{l\pi}{2}\right)}\right].
\ee
Without interaction we get $\Psi_f=\Psi_i$.
Even with interaction the incoming wave remains unchanged, while the phase
of the outgoing wave is moved by $2\delta$ and the amplitude is reduced
by a factor $\eta$. The asymptotic behavior is
\be
\Psi_{i,kr\to\infty} = \suml\frac{\imath^l(2l+1)}{2\imath kr}
\left[\etl\Exp{ 2\imath\dll}\Exp{\imath\left(kr-\frac{l\pi}{2}\right)}-\Exp{\imath\left(kr+\frac{l\pi}{2}\right)}\right].
\ee
The difference between the incoming and outgoing wave defines
the scattering wave.
\bn
\Psi_s &=& \Psi_f-\Psi_i\nonumber\\
&=& \frac{1}{k}\suml(2l+1)\frac{\etl\Exp{ 2\imath\dll}-1}{2\imath}
P_l(\cos\theta)\frac{\Exp{\imath kr}}{r}\\
&=& \frac{1}{k}\suml(2l+1)\Tl P_l(\cos\theta)\frac{\Exp{\imath kr}}{r}\\
&=& f(\theta)\frac{\Exp{\imath kr}}{r}.\label{eq:psi_scattering}
\en
Eq. \ref{eq:psi_scattering} contains two important definitions:
\bn
\Tl &=& \frac{\etl\Exp{ 2\imath\dll}-1}{2\imath}\\
f(\theta) &=& \frac{1}{k}\suml(2l+1)\Tl P_l(\cos\theta)
\en
Another important result, is that the total cross
section is proportional to the imaginary part of the forward scattering amplitude.
This is well known as the optical theorem:
\be
\sigma_T=\frac{4\pi}{k}\IM{f(0)}\label{eq:tm_optical}
\ee
with the scattering amplitude $f(\tht)$ defined as
\bn
\Psi_S&=&f(\tht)\frac{\Exp{\imath kr}}{r}\nonumber\\
f(\tht)&=&\frac{1}{k}\sum_{l=0}^\infty(2l+1)\frac{\etl\Exp{2\imath\dll}-1}{2\imath}P_l(\cth)\\
f(\tht)&=&\frac{1}{k}\sum_{l=0}^\infty(2l+1)T^l P_l(\cth).\nonumber
\en
Using  $\theta=0\Rightarrow \cos\theta=1 \Rightarrow P_l(1)\equiv1$ we get
\be
\sigma_T=\sigma_E+\sigma_I=\frac{4\pi}{k}\suml(2l+1)\IM{\Tl}.\label{eq:elast_inelast_crosssection}
\ee
To derive explicit forms for $\sigma_E$ and $\sigma_I$ we now 
investigate the imaginary part of the $T$-matrix:
\bn
\IM{\Tl} &=& \IM{\frac{\etl\Exp{ 2\imath\dll}-1}{2\imath}}\\
&=& \IM{\frac{-\etl\Exp{ 2\imath\dll}+\imath}{2}}\nonumber\\
&=& \frac{1}{2}\IM{\etl\sin(2\dll)+\imath(-\etl\cos2\dll+1)}\nonumber\\
&=& \frac{1}{4}\left(-\etl\cos2\dll+2\right)\nonumber\\
&=& \frac{1}{4}\left(-\etl\cos2\dll+1+\etl^2+1-\etl^2\right)\nonumber\\
&=& \frac{1}{4}\left(\left|\frac{\etl\Exp{ 2\imath\dll}-1}{\imath}\right|^2+(1-\etl^2)\right).
\en
From this we get for the elastic cross section (using \refeq{eq:elast_inelast_crosssection})
\bn
\sigma_E &=& \fpk\suml(2l+1)\left|\frac{\etl\Exp{ 2\imath\dll}-1}{2\imath}\right|^2\nonumber\\
&=& \fpk\suml(2l+1)|\Tl|^2\label{eq:elast_crosssection}
\en
and correspondingly for the inelastic cross section
\be
\sigma_I=\frac{\pi}{k^2}\suml(2l+1)(1-\etl^2).\label{eq:inelast_crosssection}
\ee
For the fully elastic case the $T$-matrix reduces to the simple form
\bn
\Tl &=& \frac{\Exp{2\imath\dll}-1}{2\imath} = \Exp{\imath\dll}\sin(\dll) \\
&=& \frac{\sin(\dll)}{\Exp{\imath\dll}} = \frac{\sin\dll}{\cos\dll-\imath\sin\dll} \nonumber\\
&=& \frac{1}{\cot\dll-\imath}.
\en
\subsubsection{Simple and Relativistic Breit-Wigner Function}
%
%
\paragraph{Decay of Unstable States}
We assume a state which can be either a resonance in scattering theory or 
a metastable state in an atom.
The wave function of the non-stationary state of frequency
$\oR=\frac{E_R}{\hbar}$ and lifetime $=\frac{\hbar}{\Gamma}$ can be written as
\bn
\psit &=& \Psi_0\Exp{-\oR t}\Exp{-\frac{t}{2\tau}} \nonumber\\
&=& \Psi_0\Exp{-\oR t}\Exp{-\gamhalf t}.
\en
$\Psi$ as a function of the frequency is derived from a Fourier transformation
\bn
\Psi(\omega)&=&\frac{1}{\sqrt{2\pi}}\int_{-\infty}^{\infty}\psit\Exp{\imath\omega t}dt\\
&=& \frac{\Psi_0}{\sqrt{2\pi}}\int_{-\infty}^{\infty}\Exp{\dsty\imath\left(\omega-\oR+\imath\gamhalf\right)t}dt\nonumber\\
&=& \frac{\Psi_0}{\omega-\oR-\imath\gamhalf}
\left[\frac{1}{\sqrt{2\pi}}\Exp{\dsty\imath\left(\omega-\oR+\imath\gamhalf\right)t}\right]_{-\infty}^{\infty}\nonumber\\
&=& \frac{\kappa}{(E_R-E)-\imath\gamhalf}
\en
where the constant $\kappa$ is determined by the normalization of the wave function. To get the connection to
scattering theory, we determine $\kappa$ with the formulae of the previous paragraphs.
In the case of elastic scattering (the resonance decays only in one elastic channel $m_1+m_2\to R\to m_1+m_2$)
the maximum cross section for a given $l$ is (see \refeq{eq:elast_crosssection})
\be
(\sigma_E^l)_{\mbox{\tiny max}} =\fpk(2l+1).
\ee
On the other hand the elastic cross section is proportional to the absolute square of the wave function
$\sigma_E\propto \Psi\Psi^\ast$. Since $\dsty(\Psi\Psi^\ast)_{\mbox{\tiny max}}=\frac{4\kappa^2}{\Gamma^2}$,
we get the Breit-Wigner formula with $\kappa=\gamhalf$
\be
\Psi(E)= \frac{\gamhalf}{(E_R-E)-\imath\gamhalf}.\label{eq:bw1}
\ee
The elastic cross section is then
\be
\sigma_E^l=\fpk(2l+1)\frac{\frac{\Gamma^2}{4}}{(E_R-E)^2+\frac{\Gamma^2}{4}}.
\ee
\paragraph{Breit-Wigner from Phase Movement}
In the case of elastic scattering of spinless particles via a resonance of spin $J=l$
we start with $T=\frac{1}{\cot\delta-\imath}$. In the proximity of the resonance
it is well known that $\dsty\delta\approx\frac{\pi}{2}\Rightarrow \cot\delta\approx 0$
and $E_R=E(\delta=\frac{\pi}{2})$. We expand $\cot\delta$ and get
\bn
\cot\delta(E) &=& \cot\delta(E_R)+(E-E_R)\left[\frac{d}{dE}\cot\delta(E)\right]_{E=E_R}\\
&=& 0+(E-E_R)(-\gamhalf)
\en
where $\gamhalf$ was defined as the first derivative of the $\cot\delta$. From this we obtain
\be
T=\frac{1}{\cot\delta-\imath}=\frac{\gamhalf}{(E_R-E)-\imath\gamhalf}\label{eq:bw2}
\ee
which is only valid for $|E-R_R|\approx\Gamma\ll E_R$.
\paragraph{Breit-Wigner from Field Theory}
A different way is to start from field theory and to to derive a dynamic function, e.g. a Breit-Wigner,
from a propagator approach.
\bfg[hbt]
\bc\psfig{figure=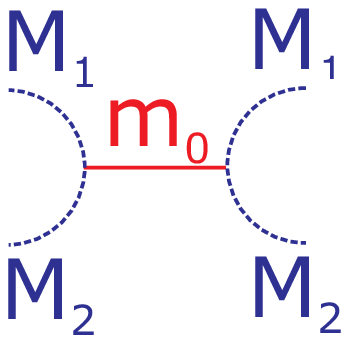,scale=0.5}\ec
\caption[Resonant Scattering]
{Resonant scattering. A pair of mesons undergoes a resonant scattering and appears again in the final state.}
\label{fig:reso_scatt}
\efg
\bfg[hbt]\bc
\psfig{figure=fig/loop_r_loop.eps,height=1.5cm} {\large +}
\psfig{figure=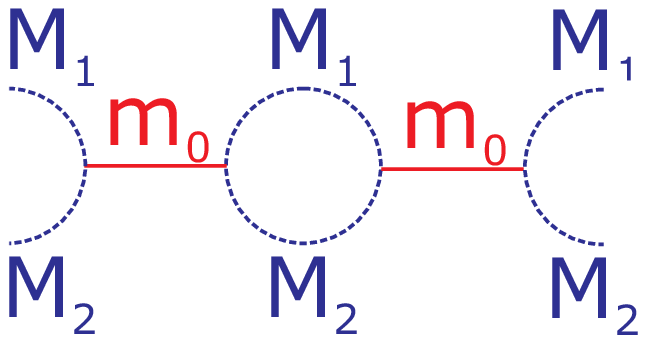,height=1.5cm} {\large +} \\
\psfig{figure=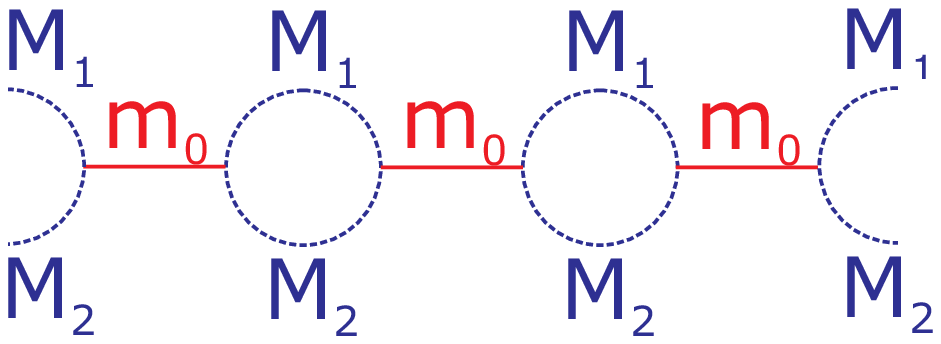,height=1.5cm} {\large + $\ldots$=} 
\psfig{figure=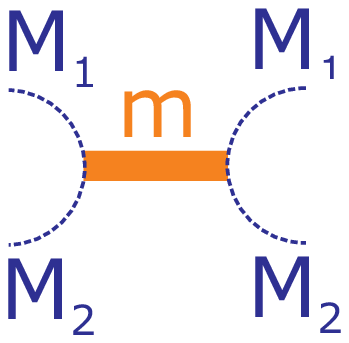,height=1.5cm}\ec
\caption[$T$-Matrix from perturbation.]{$T$-Matrix from perturbation.}
\label{fig:tm_perturb}
\efg
Let's suppose we start with a process like the one in \reffig{fig:reso_scatt}. The transition amplitude, respective
the $T$-matrix can be written like
\be
T=V_{12}\frac{1}{E_0-E}V_{12}
\ee
where the transition is constructed from a propagator and the two couplings at the vertices.
But that is not the end of the story since more loops can open up like
\be
T=V_{12}\frac{1}{E_0-E}b\frac{1}{E_0-E}V_{12}=\frac{V_{12}bV_{12}}{(E_0-E)^2}
\ee
and finally the transition amplitude is constructed from an infinite series of propagations (see \reffig{fig:tm_perturb})
where the strength gets smaller by the factor $b$ for each additional loop:
\be
T=V_{12}\frac{1}{E_0-E}V_{12}+\frac{V_{12}bV_{12}}{(E_0-E)^2}+\frac{V_{12}b^2V_{12}}{(E_0-E)^3}+\ldots
\ee
This can be written as a geometric series
\bn
T &=& \frac{V_{12}V_{12}}{E_0-E}\left(1+\frac{b}{E_0-E}+\frac{b^2}{(E_0-E)^2}+\ldots\right)\nonumber\\
&=& \frac{V_{12}V_{12}}{E_0-E}\left(\frac{1}{1-\frac{b}{E_0-E}}\right)\nonumber\\
&=& \frac{V_{12}V_{12}}{E_0-E-b}.
\en
Since $b$ is complex we get
\bn
T &=& \frac{V_{12}V_{12}}{E_0-\RE{b}-E-\imath\IM{b}}\nonumber\\
&=& \frac{V_{12}V_{12}}{E_R-E-\imath\IM{b}}.
\en
\paragraph{Relativistic Breit-Wigner}
All discussions so far were based on non-relativistic forms. Since most reactions in meson spectroscopy
involve relativistic particles a relativistic formulae is mandatory for any serious data analysis.
Without prove and omitting some details here is sketch toward a covariant description.
\par
Starting with the optical theorem in a slightly different form
\be
\IM{T(s,0)}=\frac{q}{4\pi}\frac{\sqrt{s}}{2}\sigma_T
\ee
we derive
\be
\Tl(s)=\frac{\sqrt{s}}{2}\frac{1}{\cot\delta-\imath}
\ee
with the inverse $T$-matrix (dropping $l$ for ease of writing)
\be
T\inv(s)=\frac{q}{\sqrt{s}}(\cot\delta-\imath)=k(s)-\imath\frac{2q}{\sqrt{s}}.
\ee
The matrix $k(s)$ will later be introduced as $K$-matrix (see \refsec{sec:kmatrix}).
$k(s)$ can be any real valued function. $T(s)$ has a `peak', e.g. a resonance, if
$k(s)=0$. The simplest covariant function with this property is $\dsty k(s)=\frac{\dsty s_r-s}{\dsty \gamma}$
from which we obtain the relativistic Breit-Wigner function
\be
T(s)=\frac{\gamma}{s_r-s-\imath\frac{2q\gamma}{\sqrt{s}}}.
\ee
\subsubsection{Centrifugal Barrier}
%
%
\label{sec:barrier}
From the covariant description of the decay amplitudes, one can show
that the mass dependent width of the resonances are proportional to
$q^{2l+1}$. This is known as the centrifugal barrier.
Taking out the phase space factor $\rho=\sqrt{2q/m}$ the remaining
width is still proportional to $q^{2l}$. This is only valid for low energies (where they are very important)
and a more general formalism is needed to accommodate for the properties of a scattered wave
far away from thresholds where the $q^{2l+1}$ rule is invalid.
\par
For fixed-channel orbital angular momentum $L$ these effects are correlated in a simple
way with the semi-classical impact parameter
\be
b=\left[L(L+1)\right]^{\half }/q
\ee
where $q$ is the channel wave number. When $b$ is large ($q$ small), cross sections and partial decay widths
are relatively suppressed.
In any two-body resonance decay channel $n$ with orbital momentum $L\neq0$, the form of the radial wave function will
be determined dominantly by the repulsive centrifugal barrier and the channel wave number $q_n$ for a particle
separation $r$ greater than a certain `interaction radius' ($R$). The approximate wave equation which holds outside $R$ is
\be
\frac{\partial^2}{\partial \rho^2}U_l^n{\rho}\simeq \left(\frac{b_n^2}{r^2}-1\right)U_l^n{\rho}\label{eq:barrier_diffeq}
\ee
where $\rho=q_n r$ and $b_n=\left[L(L+1)\right]^{\half}/q_n$ is again the impact parameter. The solution for
\refeq{eq:barrier_diffeq} is proportional to the spherical Hankel function
\be
U_l^n{\rho}\stackrel{r>R}{\simeq} i C_n\rho h_l^{(1)}(\rho)~\sim C_n \Exp{\imath\left(\rho-\frac{1}{2}L\pi\right)}
\ee
where $C_n$ is a constant. Consider the situation if $R$ is very small compared to $b$: for $r<b$, the 
radial probability density grows rapidly with decreasing $r$. The reason is that the outgoing wave is reflected inward
by the centrifugal barrier. At $r=R$, the radial probability density is
\be
\left[\Tl^n(R/b)\right]\inv \equiv\rho^2|h_l^{(1)}(\rho)|^2.
\ee
For a decay, we are interested on the lifetime which is involved with the centrifugal barrier. In general we obtain
a reduced width $\gamma_n$. In addition the factor $q_n/m$ is taken out of the physical partial width $\Gamma_n$ so that
\bn
\Gamma_n &=& \gamma_n\frac{q_n}{m}\Tl^n(R/b_n)\\
\Gamma_n(q_n) &=& \Gamma_n^0 \frac{\frac{q_n}{m}\Tl^n(R/b_n)}{\frac{q_n^0}{m}\Tl^n(R/b_n^0)}
\en
where $m$ is the mass of the decaying particle and $q_n/m$ accounts for the two-body phase space. How do they actually look like.
The spherical Hankel functions are defined in terms of ordinary Hankel functions of half-odd-integer order:
\bn
j_l(x)&\equiv& \frac{\pi}{2x}^\half J_{1+\half }(x)\nonumber\\
n_l(x)&\equiv& \frac{\pi}{2x}^\half N_{1+\half }(x)\nonumber\\
h_l^{(1,2)}(x)&\equiv& \frac{\pi}{2x}^\half \left[J_{1+\half }(x)\pm N_{1+\half }(x)\right]
\en
For real values of $x$, $h_l^{(1)}(x)=[h_l^{(2)}(x)]^\ast$. The explicit forms of the first three spherical Hankel functions are
\bn
h_0^{(1)}(x)&=& \frac{\Exp{\imath x}}{\imath x} \label{eq:barrier_hankel0}\\
h_1^{(1)}(x)&=& \frac{-\Exp{\imath x}\left(1+\frac{\imath}{x}\right)}{x}\\
h_2^{(1)}(x)&=& \imath\frac{\Exp{\imath x}\left(1+\frac{3\imath}{x}-\frac{3}{x^2}\right)}{x} \label{eq:barrier_hankel2}
\en
The asymptotic behavior of the spherical Hankel functions for small 
arguments $x\ll l$ may be inferred from the forms
of the spherical Bessel functions
\bn
j_l(x)&\to&\frac{x^l}{(2l+1)!!}\\
n_l(x)&\to&\frac{-(2l-1)!!}{x^{l+1}}\nonumber
\en
where the double factorial is defined as $x!!=x(x-2)(x-4)\ldots$.
The \refeq{eq:barrier_hankel0}--\ref{eq:barrier_hankel2} then translate into
\bn
F_l(q)&=& F_l(x=\frac{q}{q_{scale}})=\sqrt{\frac{|h_l^{(1)}(x)|^2}{|h_l^{(1)}(x=1)|^2}}\nonumber\\
F_l(q)&\stackrel{q\to q_{scale}}{=} & 1\\ 
F_l(q)&\stackrel{q\to 0}{=} & q^l.\nonumber
\en
And finally we obtain the damping functions for $l=0$ to $3$.
\bn
F_0(x)&=&1\nonumber\\
F_1(x)&=&\sqrt{\frac{x}{x+1}}\nonumber\\
F_2(x)&=&\sqrt{\frac{13 x^2}{(x-3)^2+9x}}\\
F_3(x)&=&\sqrt{\frac{277 x^3}{x(x-15)^2+9(2x-5)^2}}\nonumber
\en
They are applied to the natural width with the factor $B_l^2$ which are defined as
\be
B_l(q,q_R)=\frac{F_l(q)}{F_l(q_R)}
\ee
%

%
\subsubsection{Required Properties for $T$}
%
%
\label{sec:tm_def}
The main requirement is the analyticity of $T$. An empirical fact is
the appearance of singularities, e.g. poles, which are accommodated by allowing
a finite number of poles in $T$, thus $T$ is meromorph (or has maximum analyticity).
In addition $T$ is unitary and fulfills a dispersion relation which is discussed below.
\paragraph{Unitarity}
Starting from the optical theorem \refeq{eq:tm_optical} it follows that
\be
\IM{T^l}=|T^l|^2
\ee
or for multiple channels and dropping $L$
\bn
\IM{T_{jj}}&=&\sum_{j=0}^{n}|T_{ij}|^2\\
\frac{\imath}{2}(T_{jj}^\ast-T_{jj})&=& \sum_{j=0}^{n}T_{ij}^\ast T_{ij}.
\en
Collecting terms on the left-hand side, we have
\bn
1-2\imath T_{jj}^\ast+2\imath T_{jj}+4\sum_{j=0}^{n}T_{ij}^\ast T_{ij}&=&1
\sum_{j=0}^{n}(\delta_{ij}-2\imath T_{ij}^\ast)(\delta_{ij}+2\imath T_{ij}^\ast)=1
\en
This becomes equal to
\be
\sum_{j=0}^{n}S_{ij}^\ast S_{ij}=1
\ee
Since the outgoing channels are different they cannot interfere ($\braket{i}{j}=0$)
when $i\ne j$. Therefore, we can write the unitarity condition on the $S$-matrix
\be
\sum_{j=0}^{n}S_{kj}^\ast S_{ij}=\delta_{ik}
\ee
and obtain
\be
\IM{T_{ij}}=\sum_{n}T_{nj}^\ast T_{ni}.
\ee
\paragraph{Dispersion Relations}
The derivation of the dispersion relation requires a lot of complex variable analysis, so we present
only a sketch how to obtain the relations. A detailed description can be found in~\cite{Burkhardt69}.
\par
From the assumption of analyticity we deduce the integral around a closed contour ($C$) for a point inside
the contour. This is done with a Cauchy integral
\be
\Tl(s)=\frac{1}{2\imath\pi}\int_{C}\frac{\Tl(s\prm)ds\prm}{s\prm-s}
\ee
which can be expressed in terms of the individual parts of the contour as
\bn
\Tl(s)&=&\frac{1}{2\imath\pi}\int_{-\infty}^{s_L}ds\prm\frac{\Tl(s\prm+\imath\eta)-\Tl(s\prm-\imath\eta)}{s\prm-s}\nonumber\\
& +&\frac{1}{2\imath\pi}\int_{(m_1+m_2)^2}^{\infty}ds\prm\frac{\Tl(s\prm+\imath\eta)-\Tl(s\prm-\imath\eta)}{s\prm-s}
+\frac{1}{2\imath\pi}\left[\int_{C_0}+\int_{C_1}+\int_{C_2}\right]
\en
where $C_0$, $C_1$, $C_2$ are the infinite circle and the small circles surrounding $s_L$ and $(m_1+m_2)^2$. 
$s_L$ is where the left-hand cuts start and are process dependent. 
$(m_1+m_2)^2$ defines the right-hand cuts (see \refsec{sec:resodef} for details)
We have assumed $\eta$ is small so
that it can be neglected in the denominators. One can show that if
\bn
|\Tl(s)|\to & 0 & |s|\to\infty\nonumber\\
&< O((s-s_L)\inv) & s\to s_L\\
&< O((s-(m_1+m_2)^2)\inv) & s\to (m_1+m_2)^2\nonumber
\en
independently of the direction in the complex plane from which the limits are taken, then the integrals over 
$C_0$, $C_1$, and $C_2$ make a vanishing contribution in the limit as the circles get very large or very small respectively;
We shall assume this behavior to hold, as it does in analytic potential theory.
Using the hermitian analyticity (e.g. $\Tl(s+\imath\varepsilon)-\Tl(s-\imath\varepsilon)=2\imath\IM{\Tl(s+\imath\varepsilon)}$)
we have
\be
\Tl(s)=\frac{1}{\pi}\int_{-\infty}^{s_L}\frac{\IM{\Tl(s\prm)}}{s\prm-s}ds\prm
+\frac{1}{\pi}\int_{(m_1+m_2)^2}^{\infty}\frac{\IM{\Tl(s\prm)}}{s\prm-s}ds\prm
\ee
where the imaginary parts are evaluated at $s\prm+\imath\eta$, $\eta\to0$ (the physical amplitude is $\Tl(s+\imath\varepsilon)$, $\varepsilon\to0$.
$s$ is greater than $(m_1+m_2)^2$) which may be written
\be
\Tl(s)=\frac{1}{\pi}\int_{-\infty}{\infty}\frac{\IM{\Tl(s\prm)}}{s\prm-s-\imath\varepsilon}ds\prm
\ee
since $\IM{\Tl(s)}=0$ between two cuts, $s_L<s<(m_1+m_2)^2$.
\subsubsection{Factorization}
%
%
Eq. \ref{eq:bw1} and \refeq{eq:bw2} show that resonances are related to singularities of the $T$-matrix.
The task of this section is to show that the $T$-matrix factorizes in the case of many open channels.
From \refeq{eq:bw2} follows that the $T$-matrix has a singularity at
\be
\cot\delta=\frac{E_R-E}{\gamhalf}\Rightarrow E=E_R-\imath\gamhalf.
\ee
The determinant of the inverse matrix $T\inv$ has a zero at the singularity of $T$:
\be
\Det{T\inv\left(E=E_R-\imath\gamhalf\right)}=0
\ee
After diagonalization of the inverse $T$-matrix with a unitary matrix $U$ we get
\be
T_D\inv=\ba{ccc} \lone & 0 & 0 \\ 0 & \ltwo & 0 \\ 0 & 0 & \ddots \ea
\Rightarrow \Det{T_D\inv}=\lone\ltwo\ldots
\ee
For a single singularity there only one Eigen-value  vanishes. For a singularity of higher order
several Eigen-values can be zero.
For $\lone=C(E-E_R)$ we get
\be
T_D\inv=\ba{ccc} C(E-E_R) & 0 & 0 \\ 0 & \ltwo & 0 \\ 0 & 0 & \ddots \ea
\Rightarrow
T_D=\ba{ccc} \frac{1}{C(E-E_R)} & 0 & 0 \\ 0 & \frac{1}{\ltwo} & 0 \\ 0 & 0 & \ddots \ea.
\ee
For $R_D$ being a matrix with
\bn
R_D&\equiv& \lim_{E\to E_R} (E-E_R)T_D\nonumber\\
&=& \lim_{E\to E_R}\ba{ccc} \frac{1}{C} & 0 & 0 \\ 0 & \frac{E-E_R}{\ltwo} & 0 \\ 0 & 0 & \ddots \ea
= \ba{ccc} \frac{1}{C} & 0 & 0 \\ 0 & 0 & 0 \\ 0 & 0 & \ddots \ea
\en
the transformation
\be
R=U^T\frac{1}{C}U \qquad\mbox{hence}\qquad R_{ij}=\sum_k U_{ik}\frac{1}{C}U_{kj}
\ee
proves that $R$ can be factorized:
\be
R_{ij}=\sum_k\frac{1}{\sqrt{C}}U_{ik}\frac{1}{\sqrt{C}}U_{kj}
\ee
Therefore also $T$ can be factorized like
\be
T_{ij}=\frac{\sqrt{\frac{\dsty\Gamma_i}{\dsty2}}\sqrt{\frac{\dsty\Gamma_j}{\dsty2}}}%
{E_R-E-\imath\sum_{r=1}^n\frac{\dsty\Gamma_R}{\dsty2}}=
\frac{\sqrt{\frac{\Gamma_i\Gamma_j}{4}}}{E_R-E-\imath\sum_{r=1}^n\frac{\dsty\Gamma_R}{\dsty2}}.
\ee
\subsection{K-Matrix Formalism}
\subsubsection{Definitions and Properties}
%
%
\label{sec:kmatrix}
\bfg[hbt]
\bc\psfig{figure=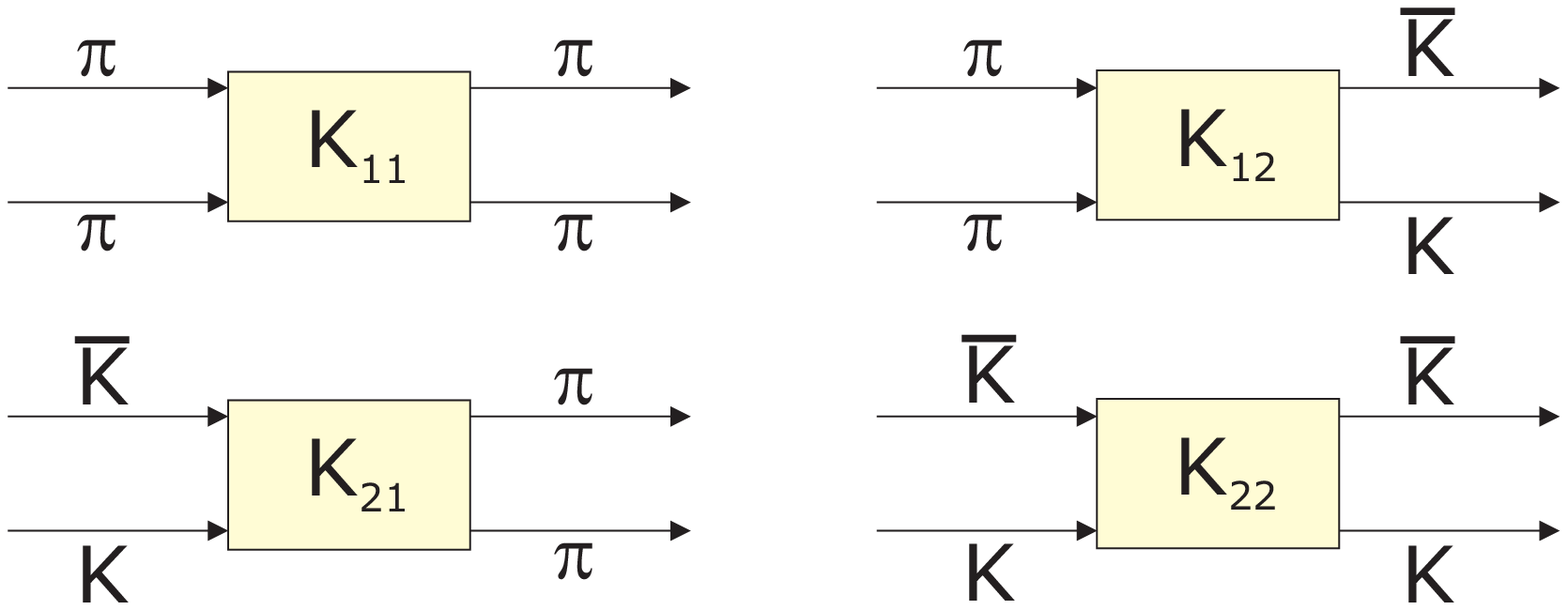,scale=0.5}\ec
\caption[Scattering process using a $K$-matrix propagator]{Scattering process using a $K$-matrix propagator. The boxes hide the actual process.}
\label{fig:km_propagator_box}
\efg
The $K$-matrix formalism provides an elegant way of
expressing the unitarity of the $S$-matrix for the processes of
the type $ab\to cd$. It has been originally introduced
by Wigner~\cite{Wigner46} and Wigner and Eisenbud~\cite{Wigner47} for
study of resonances in nuclear reactions.  The first use in particle
physics goes back to an analysis of resonance production in $Kp$ scattering
by Dalitz and Tuan~\cite{Dalitz60}. A comprehensive review is found in \cite{Badalyan80}.
In this paper we give a concise description of the $K$-matrix formalism
for ease of reference. Its generalizations to arbitrary
production processes are covered in some detail.
\par
The reader is referred to the text book by
Martin and Spearman \cite{Martin70} for some of the material
covered in this note.  However, one must note that the definitions
given in this paper are different from those used by Martin
and Spearman.
From the unitarity of the
$S$, one gets
\be
   T-T^\dagger=2\imath\ T^\dagger T=2\imath\ TT^\dagger\label{eq:tm_unitarity1}
\ee
In terms of the inverse operators, \refeq{eq:tm_unitarity1} can be rewritten
\be
  (T^\dagger)\inv -T\inv =2\imath I\  \label{eq:tm_unitarity2}
\ee
which may further be transformed into
\be
  (T\inv +\imath I)^\dagger=T\inv +\imath I.\label{eq:tm_unitarity3}
\ee
   One is now ready to introduce the $K$ operator via
\be
   K\inv =T\inv +\imath I \label{eq:km_def}
\ee
From \refeq{eq:tm_unitarity3} one finds that the $K$ operator is Hermitian, i.e.
\be
       K=K^\dagger. \label{eq:km_hermitian}
\ee
From time reversal invariance of $S$ and $T$ it follows, that the $K$~operator 
must be symmetric, i.e. the $K$-matrix may be chosen to be real and 
symmetric. 
One can eliminate the inverse operators in \refeq{eq:km_def} by
multiplying by $K$ and $T$ from left and right and vice versa,
to obtain
\be
  T=K+\imath TK=K+\imath KT\label{eq:ktm_commutation1}
\ee
which shows that $K$ and $T$ operators commute, i.e.
\be
  [K,T]=0\label{eq:ktm_commutation2}
\ee
and that, solving for $T$, one gets
\be
  T=K(I-\imath K)\inv =(I-\imath K)\inv K\label{eq:ktm_commutation3}
\ee
and
\be
  S=(I+\imath K)(I-\imath K)\inv =(I-\imath K)\inv (I+\imath K).\label{eq:ksm_relation}
\ee
Note that the $T$-matrix is complex only through the $\imath$ that
appears in this formula, i.e. $T\inv$ has been explicitly broken up
into its real and imaginary parts, see equation \refeq{eq:km_def}.
\par
   It is also useful to split $T$ into its real and imaginary part.
From \refeq{eq:ktm_commutation1} one finds, noting that $K$ is a real matrix,
\bn
   \RE{T}&=&(I+K^2)\inv K=K(I+K^2)\inv \label{eq:ktm_commutation4}\\
   \IM{T}&=&(I+K^2)\inv K^2=K^2(I+K^2)\inv 
\en
Combining this result with \refeq{eq:tm_unitarity1},
one finds that the unitarity takes on the simple form
\be
   \IM{T}=T^\ast T=TT^\ast\label{eq:tm_unitarity_im}
\ee
or, from \refeq{eq:km_def}, one gets
\be
   \IM{T\inv} =-I.\label{eq:tm_unitarity_im1}
\ee
\par
   Consider now an isoscalar $\pi\pi$ scattering
in S-wave below $\sqrt{s}=1\,$GeV.
This is a single-channel problem and unitarity is rigorously maintained.
From \refeq{eq:sm_unitarity}, one may set
\be
   S=e^{2\imath\delta}\label{eq:sm_expidelta1}
\ee
where $\delta$ is the familiar phase shift.
The transition amplitude $T$ is given, from \refeq{eq:sm_tmdef},
\be
   T=e^{\imath\delta}\sin{\delta}\label{eq:tm_expidelta1}
\ee
Note that the factors 2 and $\imath$ in \refeq{eq:sm_tmdef} make the $T$ attain
the simple, familiar form.  This formula shows that the trajectory of
$T$ in the complex plane (Argand diagram)
is a circle of a unit diameter with its
center at $(0,\imath/2)$.  This is the so-called unitarity circle and
the physically allowed $T$ should remain at or within this circle.
The $S$-wave cross section is, from \refeq{eq:sm_tm_matrix},
\be
  \sigma=\left(4\pi\over q^2_i\right)\sin^2{\delta}.\label{eq:swave_crosssection}
\ee
The $K$-matrix for this case is simply
\be
   K=\tan{\delta}\label{eq:km_expidelta}
\ee
and a pole in $K$ is therefore associated with $\delta=\pi/2$.
\par
   Consider next a two-channel problem in which
the $S$-matrix may be expressed as $2\times 2$ matrices.
Let $S_{ij}$ be symmetric; then it would take six parameters to
represent three complex variables: $S_{11}$, $S_{12}$, and $S_{22}$.
The unitarity relationship
\be
   S_{ik}S^{^\ast}_{jk}=\delta_{ij}\label{eq:sm_unitarity}
\ee
imposes three independent equations.  This shows that the $S$-matrix
depends on just three parameters.  It can be shown readily that
the matrix elements are
\bn
    S_{11}&=&\eta\ e^{2i\delta_1}\nonumber\\
    S_{22}&=&\eta\ e^{2i\delta_2}\label{eq:sm_expidelta2}\\
    S_{12}&=&\imath\sqrt{1-\eta^2}\ e^{\imath\Phi_{12}}, \qquad
      \Phi_{12}\ =\ \delta_1+\delta_2\nonumber 
\en
where $\delta_i$ is the phase shift for channel $i$ and $\eta$
is the inelasticity ($\eta \leq 1$).
Note that there exists only one inelasticity in the two-channel case.
\par
   Turning to the $K$-matrix representation of the $T$-matrix, let
\bn
   K=\ba{cc} K_{11}&K_{12}\\
              K_{21}&K_{22}\ea\label{eq:km_2channel}
\en
where $K_{12}=K_{21}$ and $K_{ij}=\mbox{real}$.
Then, from \refeq{eq:km_def}, one finds
\bn
   T={1\over 1-D-\imath(K_{11}+K_{22})}
\ba{cc} K_{11}-iD&K_{12}\\
              K_{21}&K_{22}-iD\ea\label{eq:tm_2channel}
\en
where
\be
   D=K_{11}K_{22}-K_{12}^2.\label{eq:km_2channel_det}
\ee
\subsubsection{Lorentz-invariant Description}
%
%
The transition amplitudes $T$ as defined in \refeq{eq:sm_tmdef} is
not Lorentz invariant.  The invariant amplitude is defined through
two-body wave functions for the initial and the final state, and
the process of the derivation involves proper normalizations
for the two-particle states (see Section 5.1, Chung \cite{Chung71}).
The resulting invariant amplitude
contains the inverse square-root of the two-body phase space
elements in the initial and the final states.
The Lorentz-invariant amplitude, denoted
$\wht T$, is thus given by
\be
   T_{ij}=\{\rho_i\}^\half\ \wht T_{ij}\
\{\rho_j\}^\half.\label{eq:ltm_def}
\ee
In matrix notation, one may write
\be
    T = \{\rho\}^\half\ \wht T \
\{\rho\}^\half\label{eq:ltm_matrixdef}
\ee
and
\be
   S=I+2\imath\ \{\rho\}^\half\ \wht T\
\{\rho\}^\half\label{eq:ltm_sdef}
\ee
where the phase-space `matrix' is diagonal by definition, i.e.
\bn
   \rho =\ba{cc} \rho_1&0\\
             0&\rho_2\ea\label{eq:rho_def}
\en
and
\be
      \rho_1={2q_1\over m}\quad\mbox{and}\quad
     \rho_2={2q_2\over m}.\label{eq:rho_2channel}
\ee
The $q_i$ is the breakup momentum in channel $i$.
(Here one considers a two-channel problem for simplicity without loss
of generality.)
The unitarity demands that, from \refeq{eq:tm_unitarity_im} and \refeq{eq:tm_unitarity_im1},
\be
   \IM{\wht T}=\wht T^\ast\rho\wht T=\wht T\rho\wht T^\ast\label{eq:ltm_unitarity2}
\ee
and
\be
   \IM{\wht T\inv} =-\rho\label{eq:ltm_unitarity3}
\ee
It is in this form one encounters most frequently the unitary conditions
of the transition matrix in the literature.
\par
   The cross section in the $J$th partial wave is
given by, from \refeq{eq:sm_tm_matrix},
\be
   \sigma^J_{fi}=\left(16\pi\over s\right)\left(\rho_f\over\rho_i\right)
(2J+1)|\wht T^J_{fi}(s)|^2\label{sJdf1}
\ee
Note that this formula embodies the familiar presence of
the flux factor of the initial system and the phase-space
factor of the final system in the process $ab\to cd$.
In the $K$-matrix formalism, one allows for $\rho$ to become
imaginary below a given threshold; however, the cross section above
has no meaning below a threshold, and one could then modify the expression
above by multiplying it with two step functions: $\theta(\rho^2_i)$
and $\theta(\rho^2_f)$.
\par
   One may recapitulate the expressions for the differential cross section
and its partial-wave expansion in terms of the invariant amplitudes
$\wht T^J_{fi}(s)$.
For the purpose, one defines the `invariant scattering amplitude'
\be
 \wht T_{fi}(\Omega)=\sum_J (2J+1)\wht T^J_{fi}(s)
 D^{J\,*}_{\lambda\mu}(\phi,\theta,0)\label{eq:ltm_partialwaves}
\ee
and the differential cross section is given by
\be
  {d\sigma_{fi}\over d\Omega}
=\left(4\over s\right)\left(\rho_f\over \rho_i\right)|\wht T_{fi}(\Omega)|^2.
\label{eq:ltm_crosssection}
\ee
The initial and final density of states are, with $s=m^2$,
\bn
  \rho_i&=&\sqrt{\left[1-\left(m_a+m_b\over m\right)^2\right]
  \left[1-\left(m_a-m_b\over m\right)^2\right]}\\
  \rho_f&=&\sqrt{\left[1-\left(m_c+m_d\over m\right)^2\right]
  \left[1-\left(m_c-m_d\over m\right)^2\right]}\label{eq:rho_if}
\en
in terms of the particle masses involved in the scattering
$ab\to cd$.
Note that these phase-space factors are normalized such that
\be
   \rho_i\to 1\quad\mbox{as}\quad m^2\to\infty.\label{eq:rho_norm}
\ee
The invariant amplitude $\wht T_{fi}(\Omega)$ is dimensionless, and
has a partial-wave expansion \refeq{eq:ltm_partialwaves}.  The partial-wave
amplitude $\wht T^J_{fi}(s)$ is related to the $\wht K$-matrix
via \refeq{eq:lkm_tdef}, and unitarity is preserved if the $\wht K$-matrix
is taken to be real and symmetric.  It should be noted that
the formula for the differential cross section \refeq{eq:ltm_crosssection}
has no `arbitrary' numerical factors.  The `conventional'
invariant amplitude, introduced in \refeq{eq:sm_crosssection}, is given by
\be
  \MM_{fi}=16\pi\,\wht T_{fi}(\Omega)\label{eq:matrix_element}
\ee
\par
   One may consider again the isoscalar $\pi\pi$ scattering
in S-wave below 1.0 GeV.  In terms of the phase-shift $\delta$,
the invariant amplitude is given by, from \refeq{eq:tm_expidelta1},
\be
   \wht T={1\over \rho}e^{\imath\delta}\sin{\delta}\label{eq:ltm_pipi}
\ee
and when substituted into \refeq{sJdf1} the cross section
\refeq{eq:swave_crosssection} results.  These expressions are very familiar, and
they demonstrate clearly the interplay between the phase shifts,
the invariant amplitudes and the cross sections.

\par
   One can similarly define the invariant analogue of the $K$-matrix
through
\be
    K = \{\rho\}^\half\ \wht K \ \{\rho\}^\half\label{eq:lkm_def}
\ee
translating the inverse $K$-matrix to (\refeq{eq:km_def})
\be
   \wht K\inv =\wht T\inv +\imath\rho \label{eq:lkm_tdef}
\ee
which leads to
\be
  \wht T=\wht K+\imath\wht K\rho\wht T
        =\wht K+\imath\wht T\rho\wht K\label{eq:ltm_lkdef}
\ee
and
\be
    \wht T\rho\wht K=\wht K\rho\wht T\label{eq:ltm_lkdef2}
\ee
Solving for $\wht T$, one obtains
\be
  \wht T=\wht K(I-\imath\rho\wht K)\inv 
  =(I-\imath\wht K\rho)\inv \wht K\label{eq:ltm_lkdef3}
\ee
and, from \refeq{eq:ksm_relation},
\bn
  S &=&(I+\imath\{\rho\}^\half\ \wht K\{\rho\}^\half)
       (I-\imath\{\rho\}^\half\ \wht K\{\rho\}^\half)\inv \nonumber\\
    &=&(I-\imath\{\rho\}^\half\ \wht K\{\rho\}^\half)\inv 
       (I+\imath\{\rho\}^\half\ \wht K\{\rho\}^\half).\label{eq:sm_km_2channel}
\en
Note that $\wht K$ and $\rho$ do not commute.
The Lorentz-invariant $T$-matrix is then given by
\bn
   \wht T={1\over 1-\rho_1\rho_2\wht D-\imath(\rho_1\wht K_{11}+\rho_2\wht K_{22})}
\ba{cc} \wht K_{11}-\imath\rho_2\wht D&\wht K_{12}\\
              \wht K_{21}&\wht K_{22}-\imath\rho_1\wht D\ea\label{eq:ltm_lkm_2channel}
\en
where
\be
   \wht D=\wht K_{11}\wht K_{22}-\wht K_{12}^2.\label{eq:lkm_det_2channel}
\ee
\subsubsection{Implementation of Dynamics (and/or Resonances)}
%
%
   There are two possibilities for parameterizing resonances in $K$-matrix 
formalism: Resonances can arise from constant $K$-matrix elements with the 
energy variation supplied by phase space or from strongly varying 
$K$-matrix elements (pole terms) supplying  a phase motion~\cite{Au87}. 
They differ in their dynamical character. The former are assumed to arise 
from exchange forces in the corresponding hadronic channels (molecular 
resonances), so that great effects are expected near corresponding 
thresholds, whereas the latter (normal resonances) correspond to dynamical 
sources at the constituent level, coupling to the observed hadrons through
decay~\cite{Au87}. The dynamical origin of resonances has to be determined by
experiment. In the approximation of resonance domination for the 
amplitudes, one has therefore
\be
  K_{ij}=\sum_R {g_{R i}(m)g_{R j}(m)
   \over m^2_R - m^2} + c_{ij}\label{eq:km_poledef}
\ee
and
\be
 \wht K_{ij}=\sum_R {g_{R i}(m)g_{R j}(m)
   \over (m^2_R - m^2)\sqrt{\rho_i\rho_j}} + \hat c_{ij}\label{eq:lkm_poledef}
\ee
where the sum on $R$ goes over the number of poles
with masses $m_R$, and the residue functions
(expressed in units of energy) are given by
\be
     g^2_{R i}(m)=m_R\Gamma_{R i}(m)\label{eq:km_coupling}
\ee
where $g_{R i}(m)$ is real (but it could be negative) above
the threshold for channel $i$. The constant $K$-matrix elements have to be
dimensionless and real to preserve unitarity. The corresponding width
$\Gamma_R(m)$ is
\be
  \Gamma_R(m)=\sum_i \Gamma_{R i}(m)\label{eq:km_gamma}
\ee
for each pole $R$.
\par
   Consider now a normal resonance $R$ coupling to $n$ open
two-body channels, i.e. the mass $m_R$ is above the
threshold of all the two-body channels.
The partial widths may be given an expression
\be
  \Gamma_{R i}(m)={g^2_{R i}(m)\over m_R}=\gamma^2_{R i}
  \Gamma^0_R \left[B^l_{R i}(q,q_R)\right]^2\rho_i\label{GBdf}
\ee
and the residue function by
\be
    g_{R i}(m)=\gamma_{R i}\sqrt{m_R\Gamma^0_R}
          \ B^l_{R i}(q,q_R)\sqrt{\rho_i}.\label{rsd1}
\ee
The $B(m)$'s are ratios of usual centrifugal barrier factors in terms of the 
breakup momentum in channel $i$ and the resonance breakup momentum $q_R$ 
for the orbital angular momentum $l$:
\be
    B^l_{R i}(q,q_R)={F_l(q)\over F_l(q_R)}\label{Bra}
\ee
Widely used are Blatt-Weisskopf barrier factors~\cite{Hippel72} (see \ref{sec:barrier}) for more details
The $\gamma$'s are real constants (but they can be negative)
and may be given the normalization
\be
   \sum_i\gamma^2_{R i}=1.\label{eq:km_gammanorm}
\ee
In practice, it is probably better to avoid this normalization condition
by using the parameters
\be
   g^0_{R i}=\gamma_{R i}\sqrt{m_R\Gamma^0_R}\label{eq:km_gamma2}
\ee
as variables in the fit.
The residue function is then given by
\be
    g_{R i}(m)=g^0_{R i}
          B^l_{R i}(q,q_R)\sqrt{\rho_i}.\label{eq:km_gamma3}
\ee
\par
We define a $K$-matrix total width $\tilde\Gamma_R$ and
the $K$-matrix partial widths $\tilde\Gamma_{R i}$ by
\be
  \tilde\Gamma_R=\sum_i\tilde\Gamma_{R i}
=\Gamma_R (m_R)
=\Gamma^0_R\sum_i\gamma^2_{R i}\ \rho_i(m_R)\label{eq:km_partialgamma}
\ee
From these one finds
\bn
  \Gamma^0_R&=&
  \sum_i{\tilde\Gamma_{R i}\over\rho_i(m_R)},\\
  \gamma^2_{R i}&=&
  {\tilde\Gamma_{R i}\over\Gamma^0_R\rho_i(m_R)},\label{eq:km_partialgammai}\\
  g^0_{R i}&=&
  \sqrt{m_R\tilde\Gamma_{R i}
         \over\rho_i(m_R)}.
\en
Note that the  $K$-matrix width do not need to be identical with the width
which is observed in an experimental mass distribution nor with the
width of the $\wht T$-matrix pole in the complex energy plane. We 
will refer the former as $\Gamma_{obs}$, the latter as $\Gamma_{pole}$.
In the limit in which the masses of the decay particles can be
neglected compared to $m_R$,
one has $\Gamma_R(m_R)\simeq\Gamma^0_R$.
\par
   In terms of the $\gamma$'s and $g^0$'s, the invariant $K$-matrix now has 
a simpler form
\bn
 \wht K_{ij}&=&\sum_R {\gamma_{R i}\gamma_{R j}
m_R\Gamma^0_R
B^l_{R i}(q,q_R)B^l_{R j}(q,q_R)\over m^2_R-m^2}
\label{eq:km_barrier}\\
          &=&\sum_R {g^0_{R i}g^0_{R j}
B^l_{R i}(q,q_R)B^l_{R j}(q,q_R)\over m^2_R-m^2}\nonumber
\en
allowing for the possibility that $\gamma$'s
and $g^0$'s can be negative.
\par
   Consider now an isovector $P$-wave $\pi\pi$ scattering at or near
the $\varrho$ mass.  Then the $\pi\pi$ elastic scattering amplitude is given by
\be
   K={m_0\Gamma(m)\over m_0^2-m^2}=\tan{\delta}\label{eq:km_pipidelta}
\ee
where $m_0$ is the mass of the $\varrho$ and $\delta$ is the usual
phase shift.  The  mass-dependent width is given~by
\be
   \Gamma(m)=\tilde\Gamma_0\left(\rho\over\rho_0\right)
 \left[B^1(q,q_0)\right]^2 \label{eq:km_barrier1}
\ee
where $\tilde\Gamma_0$ is the $K$-matrix width and $q$ ($q_0$) is the 
$\pi\pi$ break-up momentum for the $\pi\pi$ mass $m$ ($m_0$).
Neglecting the angular dependence of the amplitude, one obtains
\be
   T=e^{\imath\delta}\sin\delta
=\left[m_0\tilde\Gamma_0\over m_0^2-m^2-im_0\Gamma(m)\right]
[B^1(q,q_0)]^2 \left(\rho\over \rho_0\right)\label{eq:tm_pipidelta}
\ee
The first bracket in \refeq{eq:tm_pipidelta} contains the usual Breit-Wigner form
and the last bracket expresses the two-body phase-space factor. In this
simple case, the $K$-matrix width and the observed width are identical.
Note that the phase-space factor is absent in the Lorentz-invariant
amplitude $\wht T$ given by \refeq{eq:ltm_def}.
The $q^2$ dependence of the amplitude (in $[B^1(q,q_0)]^2$) reflects
the fact that
both the initial and the final $\pi\pi$ systems are in $P$-wave.
The normalization for the transition amplitude
has been chosen such that
\be
   T=+i \quad\mbox{and}\quad \wht T={+i\over\rho}
\quad \mbox{at}\quad m=m_0. \label{eq:tm_pipinorm}
\ee
It is seen that the invariant amplitude $\wht T$ is
not normalized to 1 but to $\rho\inv $.
It is for this reason that the Argand diagram is usually plotted
with $T$ and not $\wht T$.
\subsubsection{Some Examples}
%
%
\paragraph{Coupled Channel Case}
In this example we investigate the influence of coupling
of a resonance to a second open channel at the $T$-matrix parameters
(phase-shift $\delta$ and inelasticity $\eta$).
Consider an $S$-wave resonance at 1500$\,\mevcc$  
decaying into $\pi\pi$ and $K\bar K$ with a total width
of 100$\,\mevcc$. In the first example we haven chosen the 
$K$-matrix widths to be $\tilde\Gamma_{\pi\pi} = 20$$\,\mevcc$ and 
$\tilde\Gamma_{K\bar K} = 80$$\,\mevcc$,in the second one we 
choose $\tilde\Gamma_{\pi\pi} = 80$$\,\mevcc$ and $\tilde\Gamma_{K\bar K} 
= 20$$\,\mevcc$. The change
of the coupling to the channel~$K\bar K$ has no influence
on the inelasticity $\eta$ and on the line shape of $|T|^2$. 
The only visible difference is the behavior of the 
phase motion $\delta$ (see \reffig{fig:km_2channel}(b)). In the
case of a strong coupling to the $K\bar K$ channel 
($\tilde\Gamma_{K\bar K} = 80$$\,\mevcc$ )
one cannot decide whether there is a resonance or not by 
measuring the first channel only (here $\pi\pi$), if the errors 
of the phase-shift are large. Note the different scales for the 
phase-shift $\delta$~for the two examples.
\bfg[hbt]\bc
(a)\vps\psfig{figure=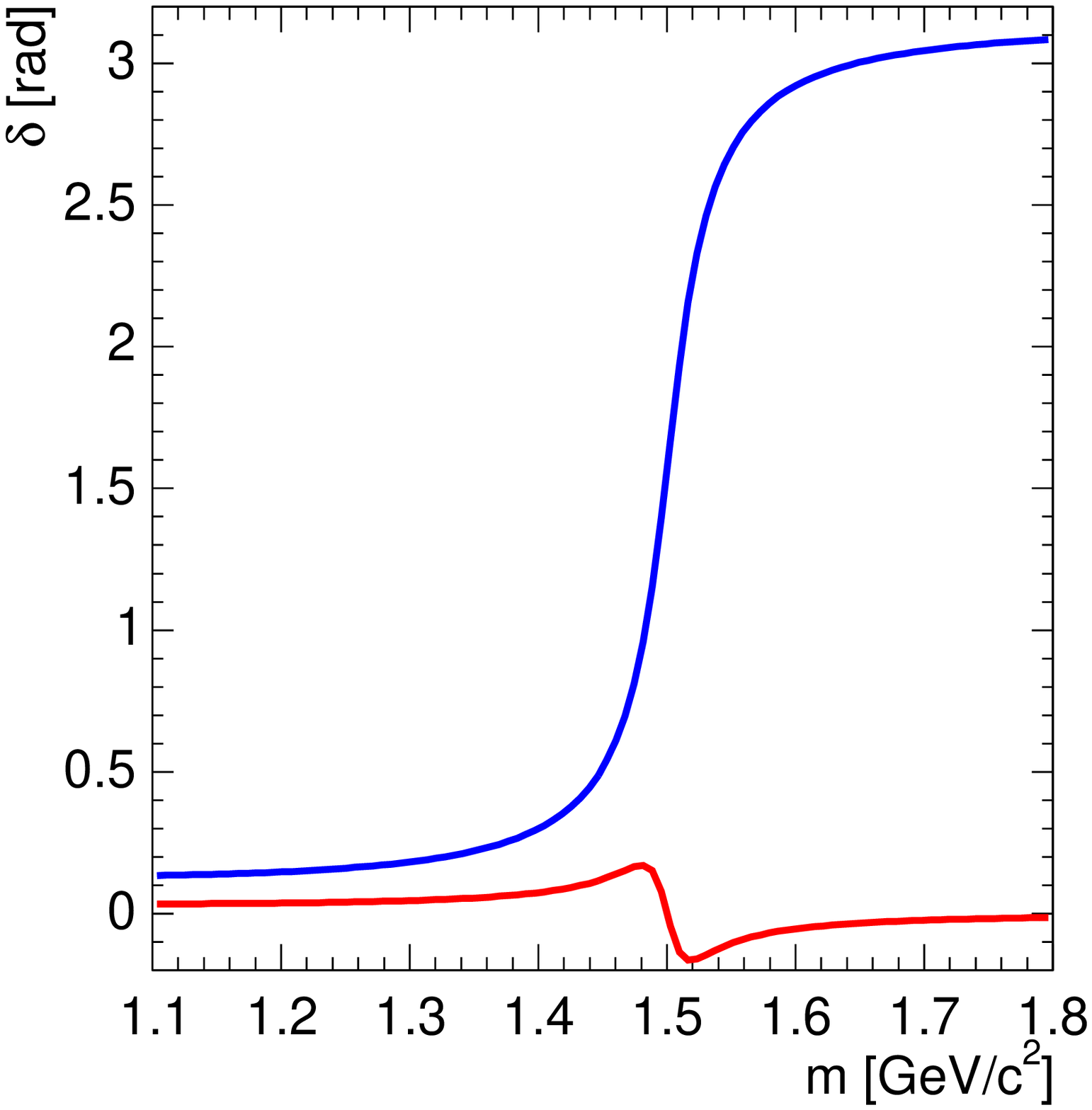,scale=0.2}
\vps(b)\vps\psfig{figure=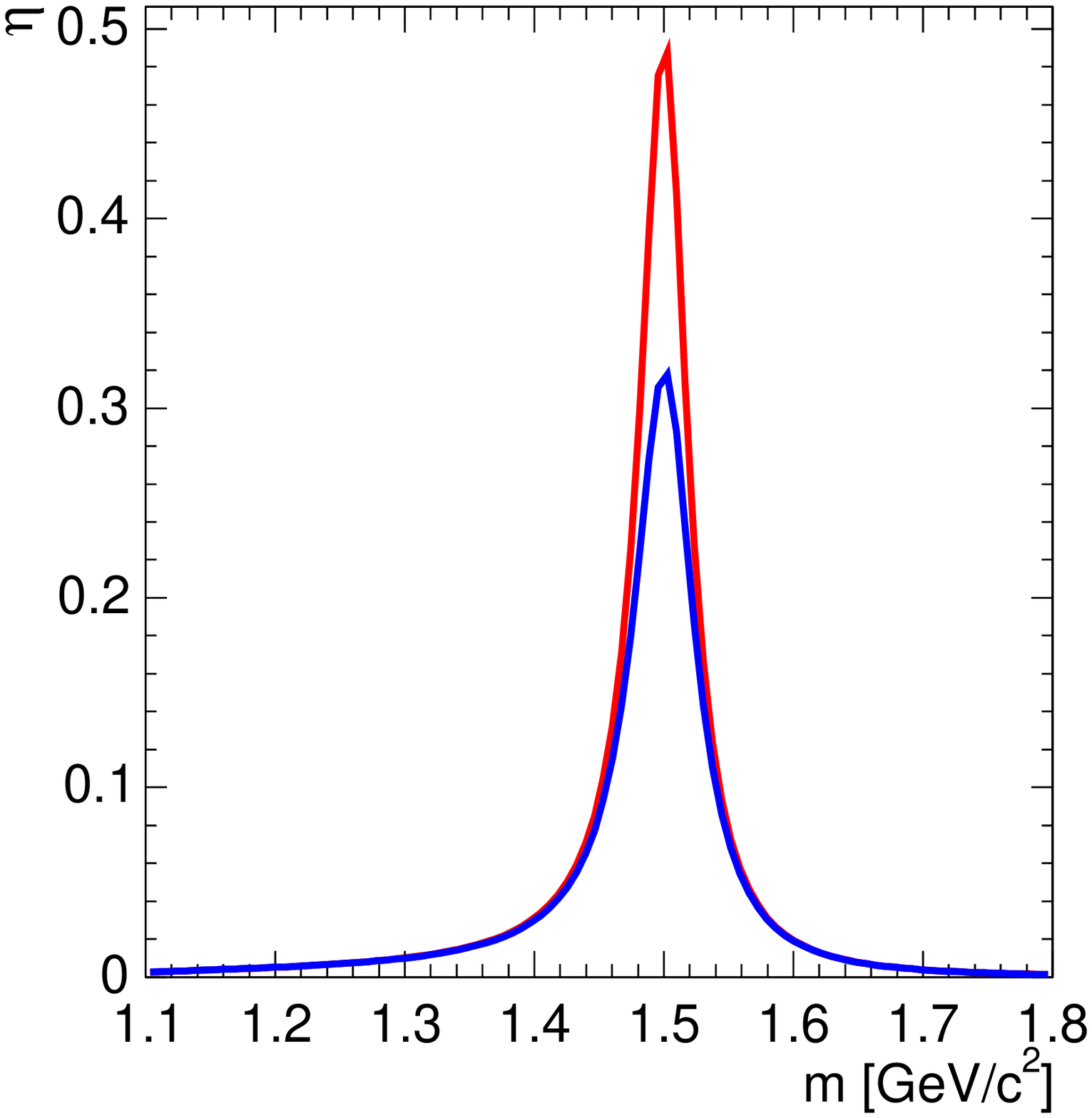,scale=0.2}
\vps(c)\vps\psfig{figure=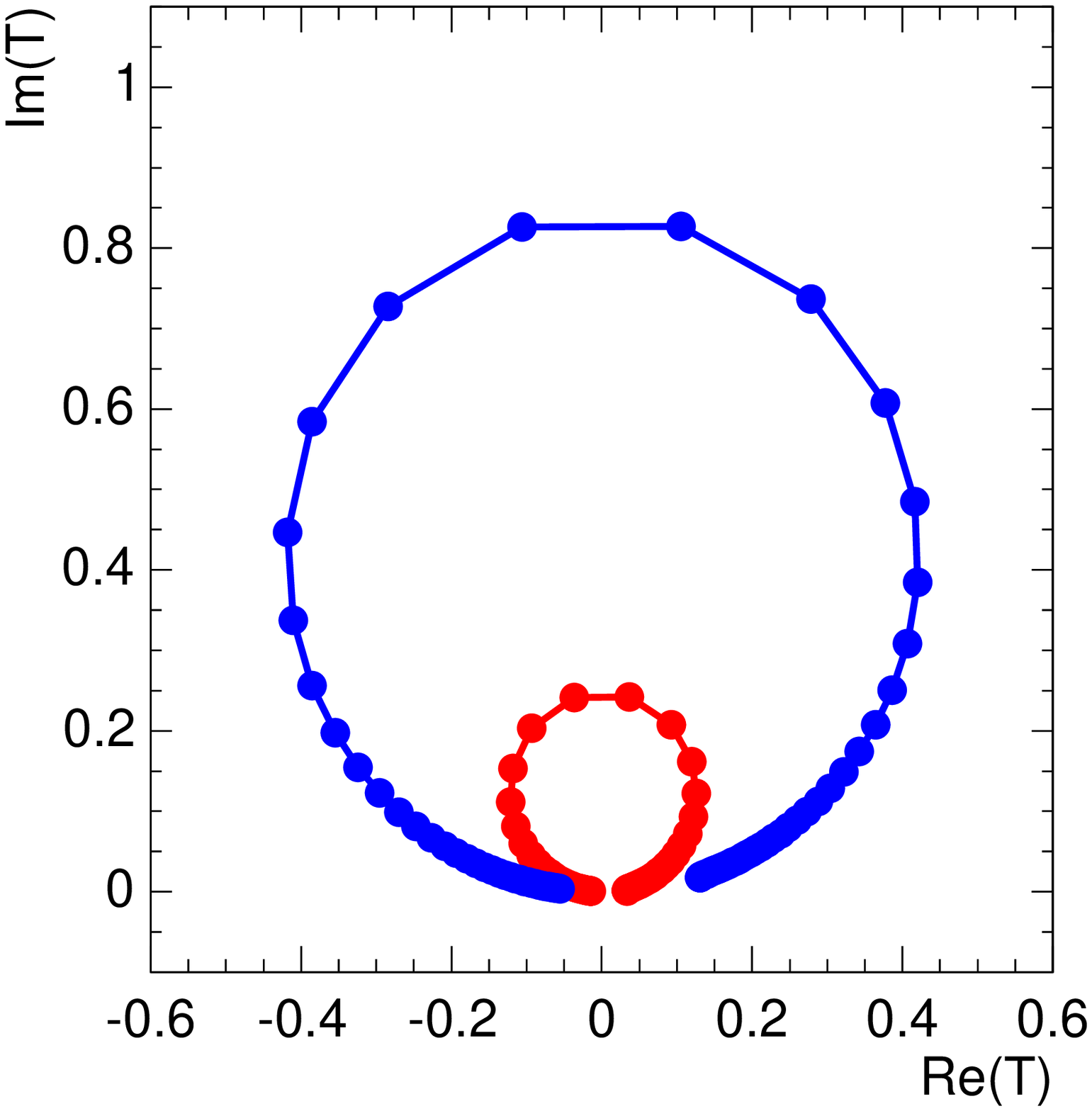,scale=0.2}
\ec
\caption[Coupled channel Breit-Wigner]
{$\pi\pi$ dominated decays of a resonance at 1500$\,\mevcc$ with a total with of 100$\,\mevcc$. (light curve $K\bar{K}$ dominates, black curve $\pi\pi$ dominates). In the case of a dominant $\pi\pi$ width, the phase (a) looks normal in $\pi\pi$, while it changes dramatically and gets tiny if the $K\bar{K}$ channel dominates. (b) shows the line shape and (c) the Argand plot for both channels}
\label{fig:km_2channel}
\efg
\paragraph{Nearby Resonances}
Consider again a $\pi\pi$ scattering at mass $m$.
But suppose there exist two resonances with masses $m_a$ and $m_b$ coupling to
the isoscalar $D$-wave channel. The prescription for the $K$-matrix
in this case is that
\be
   K={m_a\Gamma_a(m)\over m_a^2-m^2}
   +{m_b\Gamma_b(m)\over m_b^2-m^2}\label{eq:km_2poles}
\ee
i.e. the poles are summed in the $K$-matrix. The mass-dependent widths are given~by
\be
   \Gamma_R(m)=\tilde\Gamma^0_R\left(m_a\over m\right)
       \left(q\over q_R\right) [B^2(q,q_R)]^2\label{eq:gamma_spin2}
\ee
where $r=a$ or $r=b$ and $\tilde\Gamma^0_a$ and $\tilde\Gamma^0_b$ are the
two observed widths in the problem.
$q_r$ is the $\pi\pi$ breakup momentum at $m=m_r$.  
If $m_a$ and $m_b$ are far apart relative to the widths, then $K$ is dominated 
either by the first or the second resonance depending on whether $m$
is near $m_a$ or $m_b$.  The transition amplitude is then given merely by 
the sum
\bn
   T&\simeq&\left[m_a\tilde\Gamma^0_a\over m_a^2-m^2-im_a\Gamma_a(m)\right]
   \left[\left(m_a\over m\right)\left(q\over q_a\right)\right]
      [B^2(q,q_a)]^2\\
&+&\left[m_b\tilde\Gamma^0_b\over m_b^2-m^2-im_b\Gamma_b(m)\right]
   \left[\left(m_b\over m\right)\left(q\over q_b\right)\right]
      [B^2(q,q_b)]^2.\label{eq:tm_2poles}
\en
In the limit in which the two states have the same mass, i.e.
$m_0\equiv m_a=m_b$, then the transition amplitude becomes
\be
   T={m_0[\Gamma_a(m)+\Gamma_b(m)]
 \over m_0^2-m^2-im_0[\Gamma_a(m)+\Gamma_b(m)]}.\label{eq:tm_2poles_match}
\ee
This shows that the result is a single Breit-Wigner form but
its total width is now the sum of the two individual widths.
In case of two nearby resonances \refeq{eq:tm_2poles} is not strictly valid.
For a specific example \reffig{fig:km_nearbypoles} 
shows the transition amplitude $T$~from the correct equation \refeq{eq:km_2poles} 
and from the approximate \refeq{eq:tm_2poles}. Note
that \refeq{eq:tm_2poles} exceeds the unitary circle.
\bfg[hbt]\bc
(a)\vps\psfig{figure=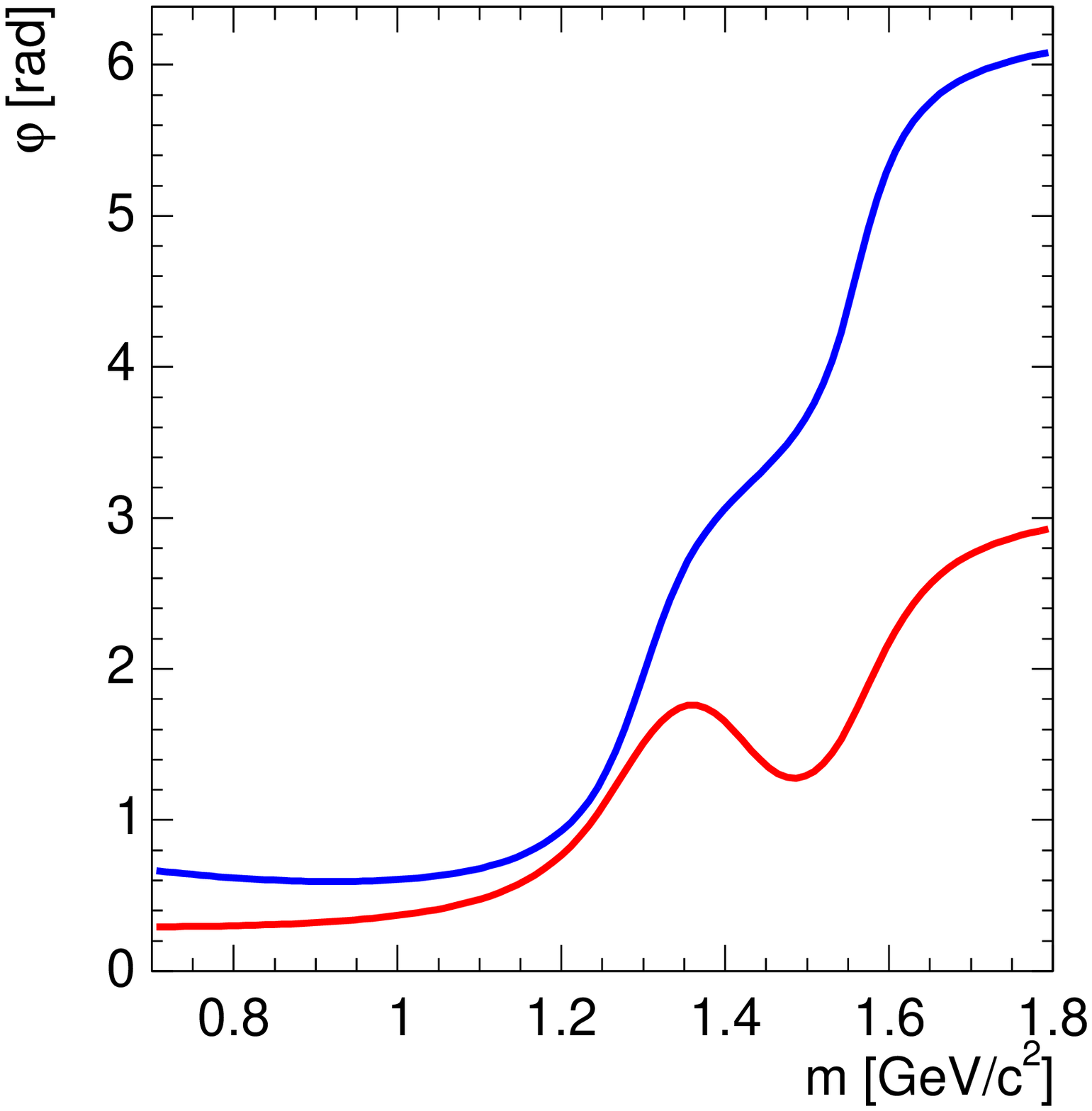,scale=0.2}
\vps(b)\vps\psfig{figure=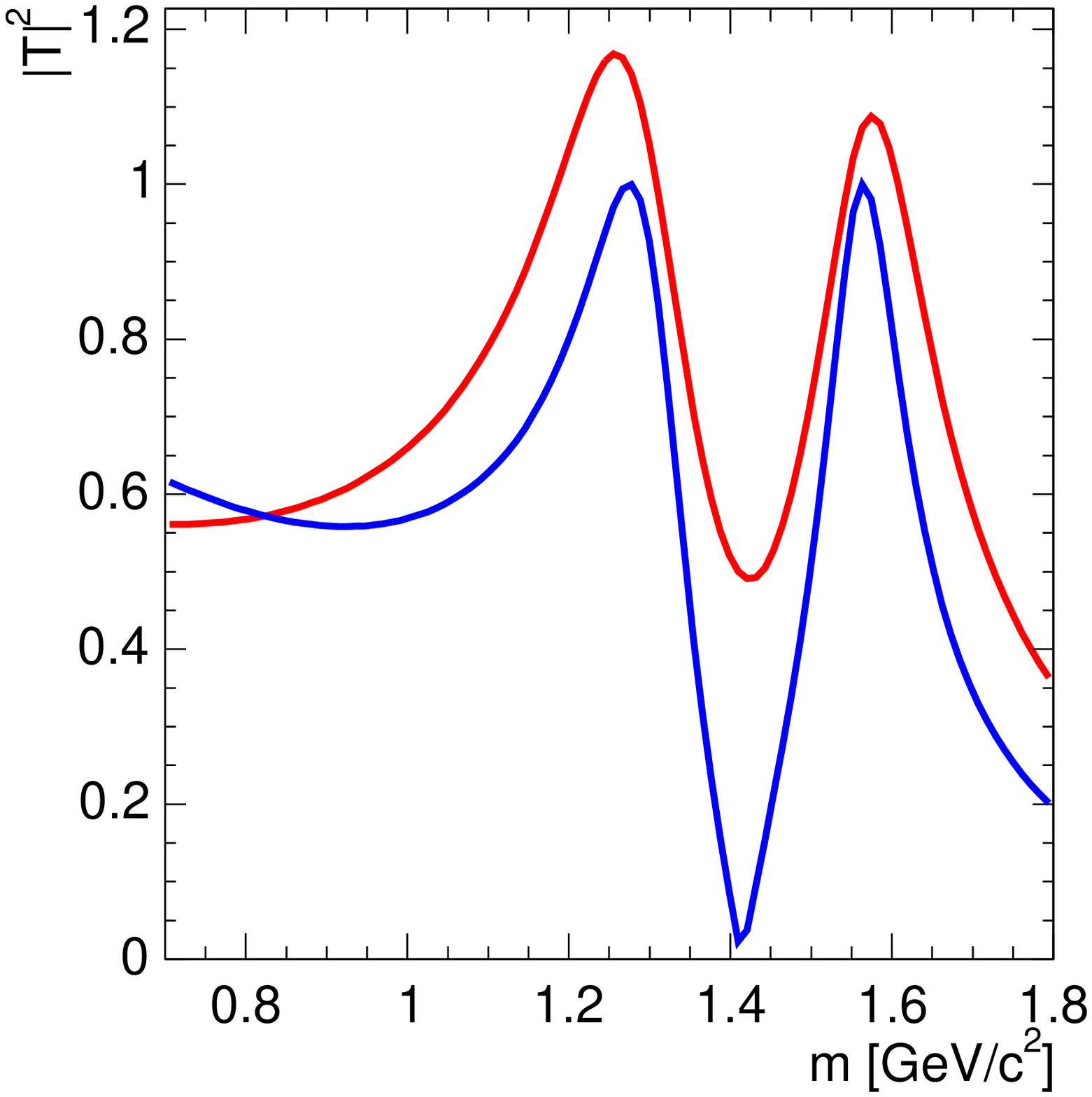,scale=0.2}
\vps(c)\vps\psfig{figure=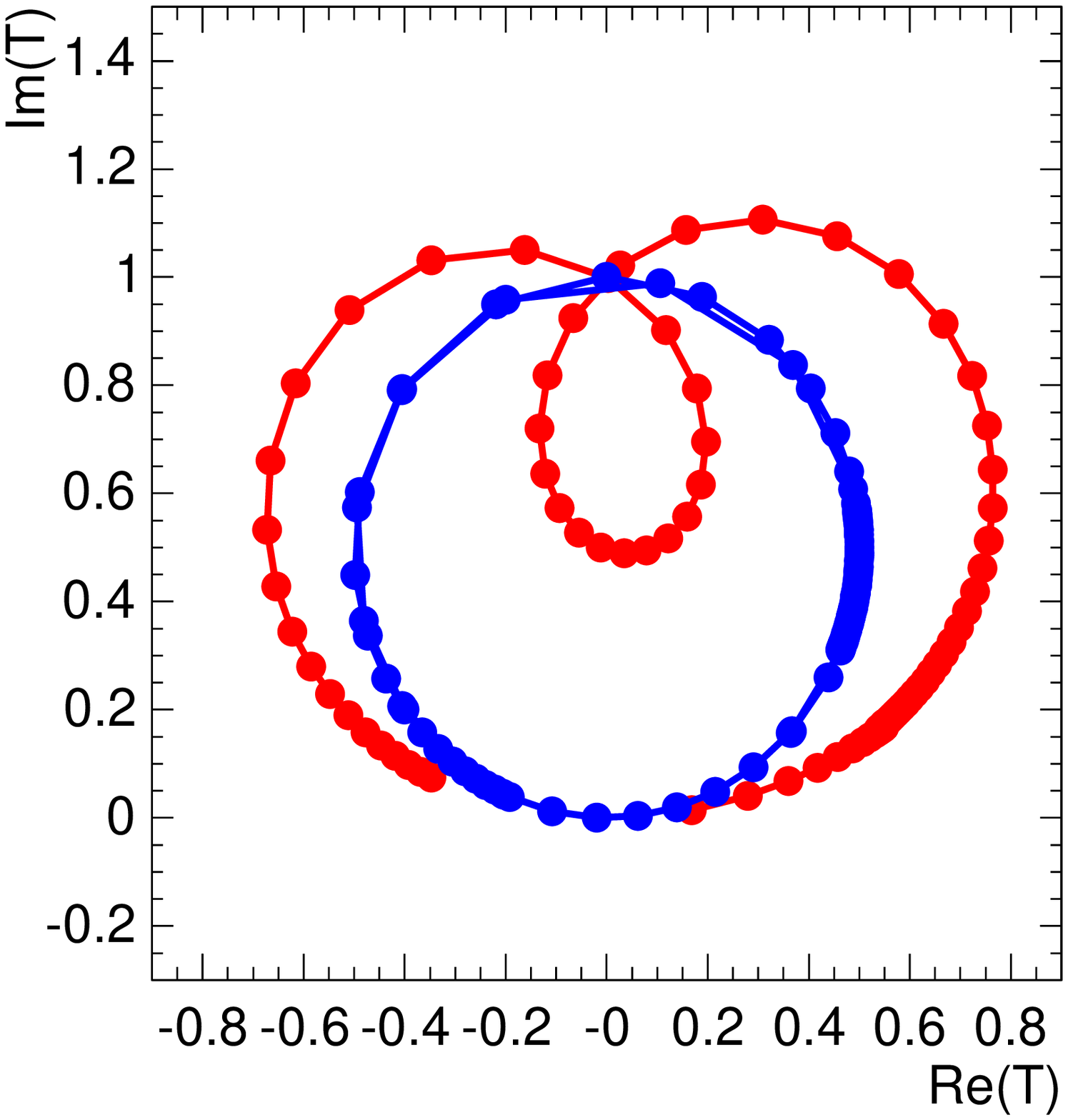,scale=0.2}
\ec
\caption[Nearby resonances in $K$-matrix formalism compared to sum of Breit-Wigner]
{Nearby resonances in $K$-matrix formalism compared to sum of Breit-Wigner. The light curves show the sum of Breit-Wigner functions and the black curves demonstrate the effect of a unitarized K-matrix approach. The masses are
$m_a$=1.275$\,\mevcc$, $\Gamma_a^0$=185$\,\mevcc$ and $m_a$=1.565$\,\mevcc$, $\Gamma_a^0$=150$\,\mevcc$.
Both resonances decay only to $\pi\pi$. It is clearly visible how the sum of Breit-Wigner functions
violate unitarity (c), while the $K$-matrix formalism leads to a nice circle. The effect is also visible in the phases (a)
which moves backward for while and in the intensity (b) where it exceeds unity.}
\label{fig:km_nearbypoles}
\efg
\paragraph{Flatt\'e Formula}

\label{sec:flatte}
\par
   As a next example we take the isovector $S$-wave scattering
with the $a_0(980)$ coupling to the $\pi\eta$ (channel 1) and
$K\bar K$ (channel 2) final states.
Then the elements of the invariant $K$-matrix~\refeq{eq:km_barrier} are
\bn
   \wht K_{11}&=&{\gamma_1^2m_0\Gamma_0\over m_0^2-m^2}\\
   \wht K_{22}&=&{\gamma_2^2m_0\Gamma_0\over m_0^2-m^2}\label{eq:km_flatte}\\
   \wht K_{12}&=&\wht K_{21}\ =\ {\gamma_1\gamma_2m_0\Gamma_0
        \over m_0^2-m^2}.
\en
The normalized couplings
are denoted by $\gamma_1^2$ and $\gamma_2^2$, which are both dimensionless
and satisfy
\be
  \gamma_1^2+\gamma_2^2=1.\label{eq:flatte_norm}
\ee
Then the $\wht T$-matrix \refeq{eq:ltm_lkm_2channel} is given as
\bn
\wht T={m_0\Gamma_0\over m_0^2-m^2-im_0\Gamma_0(\rho_1\gamma_1^2+\rho_2\gamma_2^2)}
\ba{cc} \gamma_1^2&\gamma_1\gamma_2\\
\gamma_1\gamma_2&\gamma_2^2\ea\label{eq:tm_flatte}
\en
If one sets~\refeq{eq:km_gamma2}
\be
    g_i=\gamma_i\sqrt{m_0\Gamma_0}\label{eq:flatte_gamma}
\ee
so that
\be
   g_1^2+g_2^2=m_0\Gamma_0\label{eq:flatte_gamma2}
\ee
then
\bn
   \wht T={\ba{cc} g_1^2&g_1g_2\\
              g_1g_2&g_2^2\ea
\over m_0^2-m^2
-\imath(\rho_1g_1^2+\rho_2g_2^2)}.\label{eq:tm_flatte2}
\en
This is the Flatt\'e formula~\cite{Flatte76}.
\par
   The $a_0(980)$ appears as a `regular' resonance in the
$\pi\eta$ system (channel 1). The comparable Breit-Wigner
denominator, for $m$ near $m_r$, is
$
      m^2_r-m^2-im_r\Gamma_r
$
in the resonance approximation.  One finds, therefore,
\bn
    m^2_0&=&m^2_r+\left(\gamma_2\over \gamma_1\right)^2
\left[|\rho_2(m_r)|\over \rho_1(m_r)\right]m_r\Gamma_r\\
  \Gamma_0&=&{m_r\Gamma_r\over m_0\rho_1(m_r)\gamma^2_1}\label{eq:flatte_massformula}
\en
in terms of the mass $m_r$ and width $\Gamma_r$.
Note that $\rho_i$'s have been evaluated
at $m=m_r$ where $\wht T$ is expected to attain its maximum value.
The above formulas give merely a good starting point; in practice
one must vary $m_0$ and $\Gamma_0$ to fit the $\pi\eta$ spectrum.
The ratio $(\gamma_2/\gamma_1)^2$
is an unknown (commonly fixed at the $SU(3)$ value of 1.5),
but the shape of the square of the amplitudes depends only
weakly on this value.  Once the ratio is fixed, then
$\gamma^2_1$ and $\gamma^2_2$ are fixed through the normalization
condition \refeq{eq:flatte_norm}.
\bfg[hbt]\bc
(a)\vps\psfig{figure=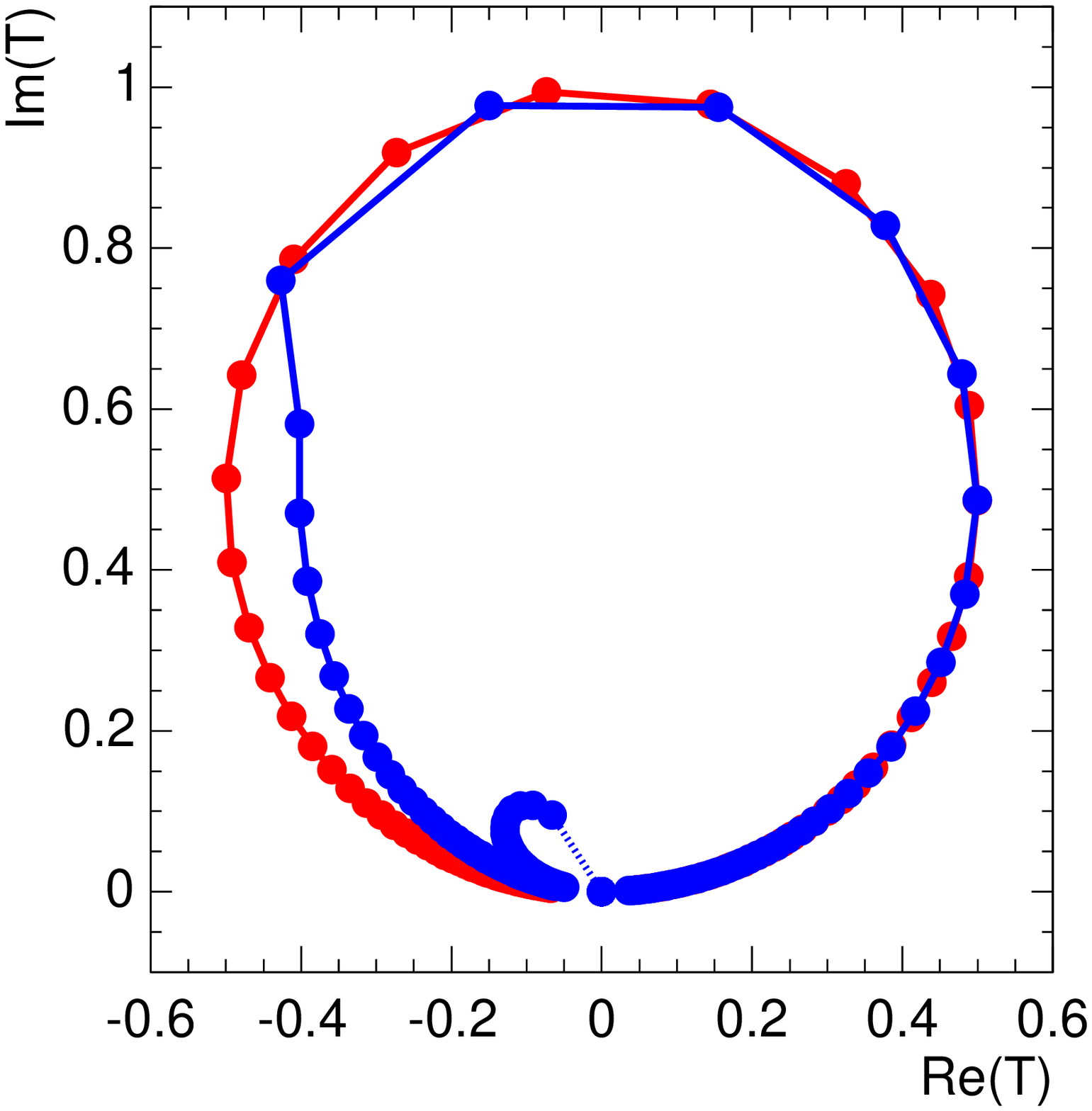,scale=0.2}
\vps(b)\vps\psfig{figure=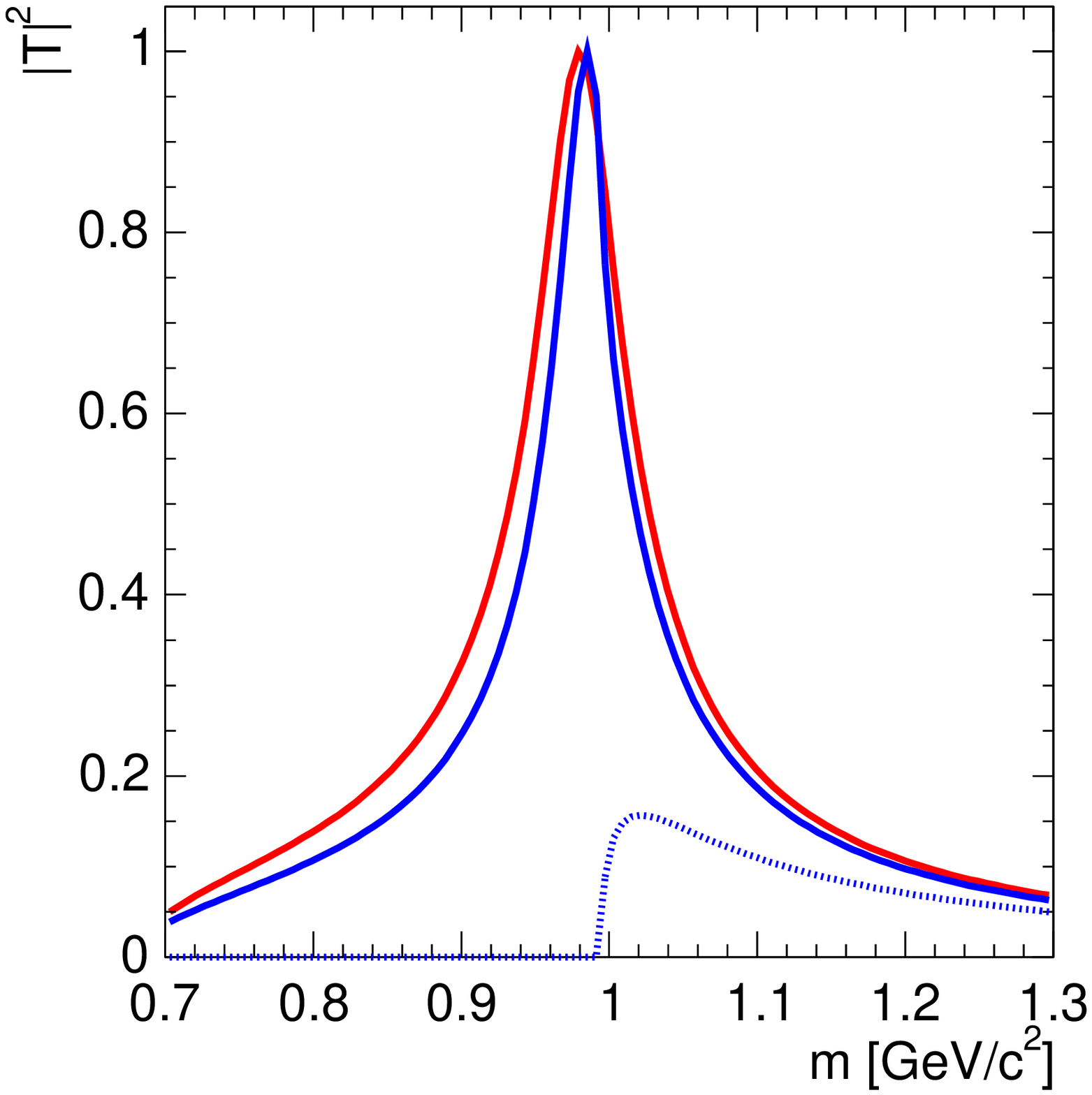,scale=0.2}
\ec
\caption[Difference between Flatt\'e and Breit-Wigner]{Difference between Flatt\'e and Breit-Wigner. (a) The Breit-Wigner function (grey curve) shows a unit circle, while in the Flatt\'e function (black curves)
one sees in channel 1 a drop in elasticity at threshold of channel 2 and a small circle for
channel 2 starting at the respective threshold. (b)
Interestingly the line shape for channel 1 does not change very much, while for channel 2 there is a dramatic
effect at threshold.}
\label{fig:flatte_relbw}
\efg
\bfg[hbt]\bc
\psfig{figure=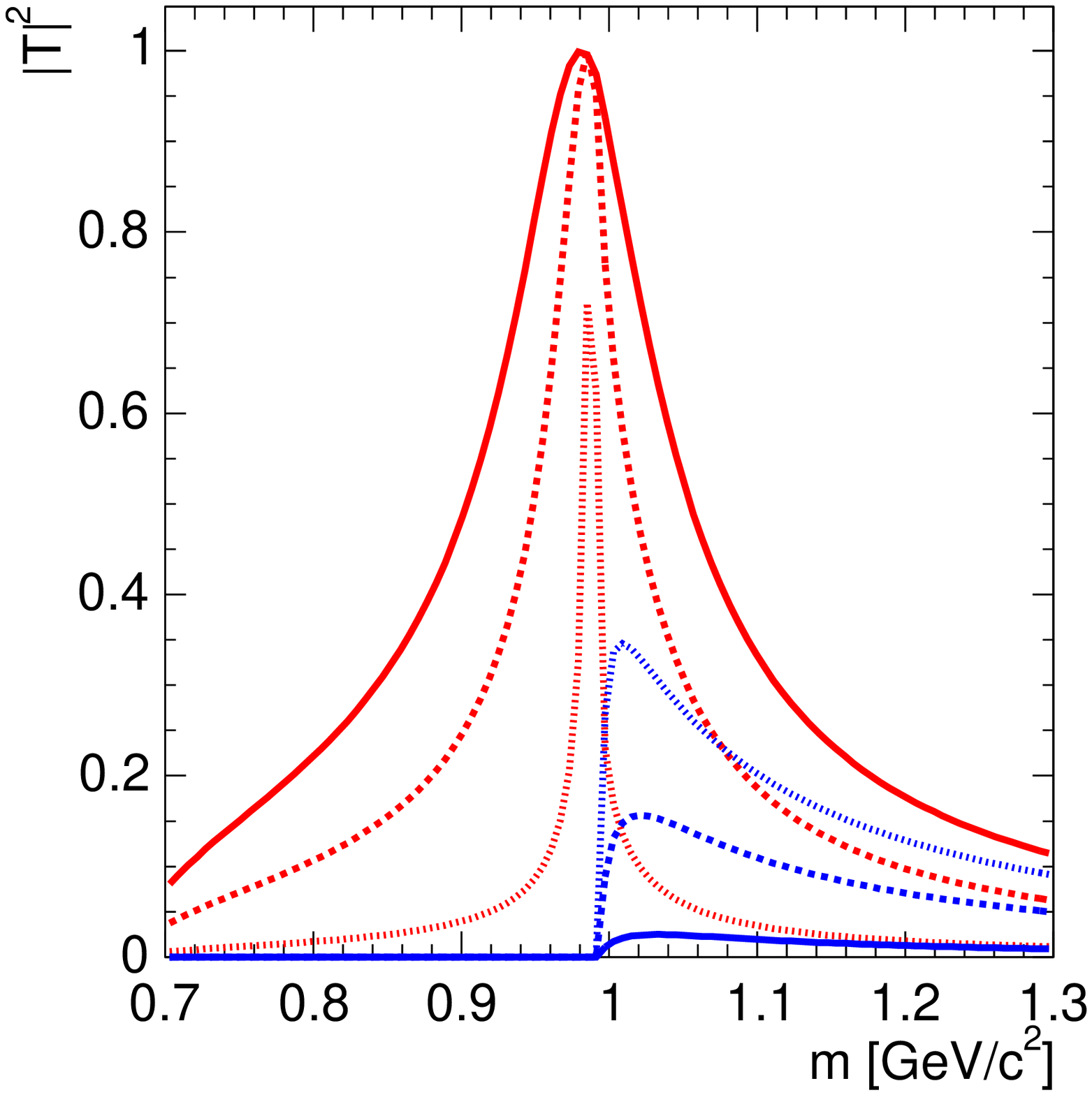,scale=0.2}
\ec
\caption[Effect of coupling variations in the Flatte\'e formula]{Effect of coupling variations in the Flatte\'e formula. The stronger the coupling the more dramatic are the effects on the line shape.
The solid, dashed and dotted curves represent coupling ratios from $\gamma_1^2/\gamma_2^2$=10 to 1 and 0.1. respectively for the two channels.}
\label{fig:flatte_couplings}
\efg
Now we turn to few examples how $K$-matrices have actually been used in various analysis.
\paragraph{Example: $\pi\pi$ S-Wave \'{a} la Au, Morgan and Pennington}
%
%
\cite{Au87}
The work from Au, Morgan and Pennington combines most of the data on $\pi\pi$ and $K\bar{K}$ scattering
at that time. For the $I=$ $\pi\pi$ $S$-wave the following parametrization was chosen
\bn
K_{ij}&=& \frac{s-s_0}{4m_K^2}\sum_r\frac{g_{r,i}g_{r,j}}{(s_r-s)(s_r-s_0)}+\sum_n c_{ij}^n
\left[\frac{s}{4m_K^2}-1\right]\\
&\equiv& (s-s_0) \wht{K}_{ij}
\en
Alternatively they parametrize also an $M$-matrix. The $M$-matrix approach is defined by $T = (M-i\rho)\inv$.
They use
\be
M_{ij}=\frac{a_{ij}}{s-s_0}+\sum_r \frac{g_{r,i}\prm g_{r,j}\prm}{s^\prime_r}
+\sum_n c^{\prime n}_{ij}\left[\frac{s}{4m_K^2}-1\right]
\ee
As we will see in most parameterizations of the $\pi\pi$ $S$-wave a parameter like $s_0$ appears to
satisfy the condition of the Adler zero.
\paragraph{Example: $\pi\pi$ S-Wave \'{a} la Crystal Barrel}
%
%
This parametrization was used to analyze the Crystal Barrel data of the
channels $\pbarp\to 3\pi0$, $2\pi^0\eta$ and $2\eta\pi0$ with an open $K\bar{K}$ channel~\cite{Amsler95}
being constrained by scattering data \cite{Rosselet77,Grayer74}. A $K$-Matrix formalism was
applied using
\be
K_{ij}=\sum_r\frac{g_{r,i}g_{r,j}B_i^l(q_i)B_i^l(q_j)}{m_r^2-m^2}+c_{ij}
\ee
with $K$ being a 3$\times$3 matrix with the channels 1=$\pi\pi$, 2=$K\bar{K}$ and 3=$\eta\eta$.
The centrifugal barrier was chosen according to the work of Hippel and Quigg~\cite{Hippel72}. For scalars
this factor is equal to one. 
\paragraph{Example: $\pi\pi$ S-Wave \'{a} la Anisovich and Sarantsev}
%
%
The work of Anisovich and Sarantsev is the most sophisticated so far and combines all Crystal Barrel data as
well as scattering data and other information. The parametrization is very complex and involves a lot
of parameters to accomodate the different reactions~\cite{Anisovich02}.
\be
K_{ij}(s)=\left(\sum_{r}\frac{g_{r,i}g_{r,j}}{m_r^2-s}
+f_{ij}\frac{s_1+s_0}{s+s_0}\right)\frac{s-s_A}{s+s_{A0}}
\ee
where $K_{ij}$ is a 5$\times$5 matrix ($i$, $j$=1,2,3,4,5), with the following notations for meson states:
1=$\pi\pi$, 2=$K\bar{K}$, 3=$\eta\eta$, 4 =$\eta\eta\prm$ and 5= multi-meson states (like dominant four-pion
state for $s<$1.6$\,\gevcc$). The $g_{r,i}$ is the the coupling constant of the bare state $r$ to a particular channel $i$.
The parameters $f_{ij}$ and $s_0$ describe a smooth part of the $K$-matrix elements. The factor $(s-s_A)/(s-s_{A0})$ is used to
suppress the false kinematic singularity at $s=0$ in the physical region near the $\pi\pi$ threshold. The parameters $s_A$ and 
$s_{A0}$ are kept to be small.
\subsection{P-Vector Approach}
%
%
\bfg[hbt]\bc
\psfig{figure=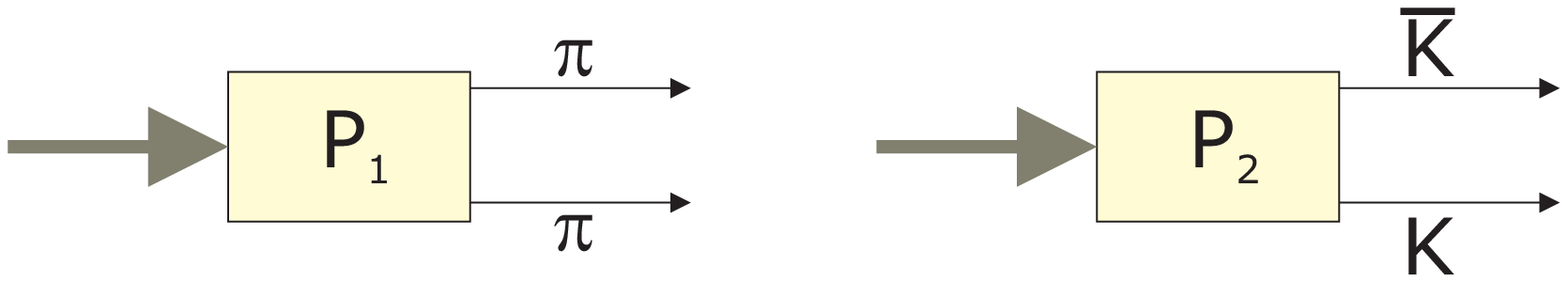,scale=0.5}\ec
\caption[Production process]{Production process. The boxes hide the actual process. The source of production
is unknown and therefore replaced by a strength parameter.}
\label{fig:pv_production_box}
\efg
So far one has considered $s$-channel resonances, or `formation'
of resonances, observed in the
two-body scattering of the type $ab\to cd$.  The $K$-matrix
formalism can be generalized to cover the case of `production'
of resonances in more complex reactions.  The key assumption is
that the two-body system in the final state is an isolated one
and that the two particles do not simultaneously interact with
the rest of the final state in the production process.
\par
According to Aitchison \cite{Aitchison72}, the production amplitude $P$ should be 
transformed into $F$ in the presence of two-body final state interactions, 
as follows:
\be
   F=(I-\imath K)\inv P=TK\inv P\label{eq:pv_def}
\ee
Or, taking the invariant form, it may be written
\be
   \wht F=(I-i\wht K\rho)\inv \wht P
    =\wht T\wht K\inv \wht P\label{eq:lpv_def}
\ee
where $\wht P$ characterizes production of a resonance and
$\wht F$ is the resulting invariant amplitude.
Note the following relationships:
\be
  F=\{\rho\}^{\half}\ \wht F \quad \mbox{and}\quad
  P=\{\rho\}^{\half}\ \wht P\label{eq:lfv_def}
\ee
\par
   Consider first a single-channel problem, e.g. the isoscalar $\pi\pi$ system
in $S$-wave below $K\bar K$ threshold.  Then, the $K$ is simply given 
by~\refeq{eq:km_expidelta} and one finds
\be
   \wht F=e^{\imath\delta}\cos{\delta}\ \wht P.\label{eq:fv_expidelta}
\ee
Unitarity demands that in this case the phase of the total amplitude $\wht F$
and the $\pi\pi$ scattering phase $\delta$ should be identical. Thus the 
final-state interaction brings in a factor
$e^{\imath\delta}$---this is the familiar Watson's theorem \cite{Watson52} and the 
production amplitude $\wht P$ has to be a real function.
It is emphasized that $\wht P$ must have the same poles
as those of the $K$-matrix; otherwise $\wht F$ would vanish at the pole
position ($\delta=\pi/2$).
\par
In general, $\wht P$ and $\wht F$ are both
column vectors, $n$-dimensional for an $n$-channel problem.
If the $K$-matrix is given as a sum of the poles as in \refeq{eq:km_poledef},
then the corresponding $P$-vector is
\be
  P_i=\sum_R {\beta^0_R \ g_{R i}(m)
    \over m^2_R - m^2}\label{eq:pv_poledef}
\ee
and
\be
\wht P_i=\sum_R {\beta^0_R \ g_{R i}(m)
    \over (m^2_R - m^2)\sqrt{\rho_i}}\label{eq:lpv_poledef}
\ee
where $\beta^0_{R }$ (expressed in units of energy), carries
the production information of the resonance $R$.
The constant $\beta^0_R$ is in general complex, but it can be set to be
real under certain assumptions. 
One should note, that Longacre \cite{Longacre82,Longacre86,Lindenbaum92}
used complex $\beta$ throughout his analysis. This ansatz was heavily criticize by Morgan and Pennington~\cite{Morgan93}
\par
   The $P$-vector should contain the same set of poles as those
found in the $K$-matrix. However, according to Aitchison, constant terms
(or a polynomial in energy) can always be added to the poles in the 
$P$-vector (as in the $K$-matrix) elements, without destroying the unitarity of
the $F$-vector (or $T$-matrix),
\be
   P_i\to P_i + d_i\label{eq:pv_polynomial}
\ee
where the constant $d_i$ is in general complex.
\par
In \cite{Foster68} the production process $\ppbar\to\pip\pim\piz$
was described by a
production amplitude for the $\rho$ $\pi\pi$~and $f_2(1270)$ $\pi\pi$~intermediate
states. The data required introduction of an additional constant
amplitude which was interpreted as that for direct three-pion production.
This direct production amplitude can be described in our
formalism with an additional constant in the $F$-vector. 
\par 
It is often more convenient to rescale $\beta^0$'s
\be
  \beta^0_R=\beta_R\sqrt{m_R\Gamma^0_R}\label{eq:pv_betadef}
\ee
so that $\beta$'s are dimensionless.
Then the $\wht P$-vector reads, from \refeq{eq:km_gamma2},~\refeq{eq:km_gamma3},
\be
  \wht P_i=\sum_R {\beta_R \gamma_{R i}\,
  m_R \Gamma^0_R B^l_{R i}(q,q_R)\over m^2_R -m^2}\label{eq:lpv_fullformula}
\ee
where, once again, $\gamma$'s are real but $\beta$'s could be complex.
If the
production process has some known dependence on momentum or angular
momentum, the production strength $\beta$~should be modified
accordingly.
\par
It is instructive to work out the above formula in the case of
a single resonance in a single channel.  Then one has
$\dsty\wht P={\beta m_0\Gamma_0\over m^2_0-m^2} B^l(q,q_0)$
so that, with $\wht K\rho$ given by \refeq{eq:km_pipidelta},
\be
  \wht F(m)=\beta {m_0\Gamma_0\over m^2_0-m^2-im_0\Gamma(m)}B^l(q,q_0).
  \label{eq:lfv_breitwigner}
\ee
This is exactly what one writes down for a Breit-Wigner form, except
that one has multiplied by an arbitrary constant $\beta$ and the
centrifugal damping factor $B^l(q,q_0)$.
This provides a $K$-matrix justification of the traditional
`isobar' model.
Note that the numerator is a constant (independent of $m$).
\par
The scattering process may involve particle exchange in the $t$-channel (see \reffig{fig:reaction_t_channel})
and the left-hand singularities reach into the physical region. Since the formalism does not have any
left-hand-cuts, they lead to artificial poles in a $P$-vector approach. Since these artificial poles
depend on the physical processes involved and differ from reaction to reaction. If such poles exist, they
can be different in the definition of the $K$-matrix and the $P$-vector.
\bfg[hbt]\bc
\psfig{figure=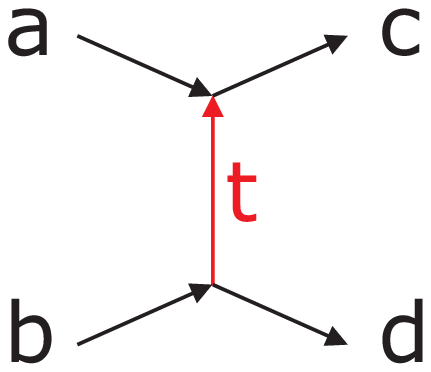,scale=0.5}\ec
\caption[t-Channel scattering]{t-Channel scattering.}
\label{fig:reaction_t_channel}
\efg
\par
Cahn and Landshoff \cite{Cahn86} state that in some approximations
the column vector
\be
   Q=K\inv P \quad{\mbox{and}}\quad
   \{\rho\}^{\half} Q=\wht Q   \quad\mbox{and}\quad
   \wht Q=\wht K\inv \wht P
\ee
may be considered a
constant in a given limited energy range.
Then, one has
\be
     F=TQ \quad{\mbox{and}}\quad
        \wht F=\wht T\wht Q\label{eq:qv_def}
\ee
i.e. the two-body final-state interaction may be expressed as
a product of the $\wht T$-matrix and a constant column vector.
The $\wht Q$-vector is devoid of the threshold singularities
(i.e., no dependence on $\rho$) and does not contain pole terms.
It therefore depends in general on $s=m^2$ only.
In a single-channel problem, e.g. the isoscalar $\pi\pi$ system
in $S$-wave below 1 GeV, we now obtain, instead of \refeq{eq:fv_expidelta} derived for
the $P$ vector approach:
\be
   \wht F={1\over\rho}e^{\imath\delta}\sin{\delta}\ \wht Q\label{eq:qv_expidelta}
\ee
This amplitude contains the familiar scattering amplitude
$e^{\imath\delta}\sin{\delta}$.
\par
   The $P$- and $Q$-vector approaches, even though both being approximations 
for production of multi particle final states, correspond to different
interpretations of the physical processes. For clarity we consider a specific
reaction, e.g. channel 1: $\bar{p}p \rightarrow\ \pi\pi\eta$
and channel 2: $\bar{p}p \rightarrow$ $K\bar{K}\eta$
from which we want to extract information on
$(\pi\pi)$ interactions. In the $Q$-vector approach, the
amplitude is given by $F_1 = T_{11} Q_1 + T_{12} Q_2$: \
the $\pi\pi$  system is produced with an amplitude $Q_1(s)$.
The two--pion interactions are then taken into account by the
scattering amplitude $T_{11}$. Alternatively, a $K$ and a $\bar{K}$ are
produced with amplitudes $Q_2$ which scatter via $T_{12}$ into the outgoing
pions. This picture has to be contrasted to that which one may have in mind 
in applying the $P$-vector approach, $F = (I - \imath K)\inv  P$. A resonance is 
produced with an amplitude P, the term $(I - \imath K)\inv $ may be considered as 
propagator for this particle which then decays.
\subsection{N/D-Method by Example}
%
%
The parametrization in terms of $K$-matrices with or without production vectors
is only one approach to describe two-body scattering. Under the assumption of unitarity
and maximum analyticity it is possible to derive the N/D method from the dispersion relations.
The amplitude $\Tl$ for a given angular momentum fulfills a dispersion relation derived from the
optical theorem. This is an integral equation for which has no explicit solution. But $\Tl$ 
can be expressed in terms of a nominator $N_l$ and a denominator $D_l$
\be
\Tl(s)=\frac{N_l(s)}{D_l(s)}
\ee
where $N_l$ contains only left-side singularities and $D_l$ only right-side singularities. The functions
$N_l$ and $D_l$ are correlated by a system of integral equations. The advantage is, that this system can be
reduced to a differential equation of the the Fredholm type, which is known to be solvable. The solution
is a infinite series, where the actual form depend on the integration kernel. In practice this
method is realized by a finite sum, with parameters being tuned to fit the experimental data.
\par
The work by Anisovich, Bugg, Sarantsev and Zou~\cite{Anisovich94}
may serve as an example for such an analysis. In order to fit the Crystal barrel data on $\pbarp\to3\pi^0$,
and $2\eta\pi^0$ they used the following Ansatz below 1.1$\,\gevcc$
\bn
K_{ij}&=& \left(\frac{s-2m_\pi^2}{s}\right)\left(\frac{\alpha_i\alpha_j}{s_A-s}
\frac{\beta_i\beta_j}{s_B-s}\frac{\gamma_i\gamma_j}{s_C-s}+a_{ij}+b_{ij}s\right)\nonumber\\
N_{\pi\pi}(s)&=& N_{11}(s)=(c_1+c_2s)K_{11}+i\rho_2(c_3+c_4s)(K_{11}K_{22}-K_{12}K_{21})\\
N_{\eta\eta}(s)&=& N_{22}(s)=c_1 K_{22}+i\rho_2 c_3(K_{11}K_{22}-K_{12}K_{21})\nonumber
\en
where $\rho_i$ are the usual phase space factors and $a_{ij}$, $b_{ij}$ and $c_i$ are arbitrary complex parameters
and $s_i$ are pole positions. $\alpha_i$, $\ldots$, $\gamma_i$ are coupling constants.
\subsubsection{What is a Resonance}
%
%
\label{sec:resodef}
So far we collected several approaches to obtain a dynamical function.
But all formalisms are only parameterizations which are more or less
motivated by models. The main physical content is hidden in the
$T$-matrix, however it was obtained.
\par
The instructive Breit-Wigner function is well suited for low statistics,
where complicated methods fail because of the many unconstrained parameters,
or in the case of very well separated resonances. In this case the fitted
mass and width are easily mapped to the physical properties of the state.
In general this is more complicated and lead s to the investigation of the
singularities of the meromorph $T$-matrix.
\par
The $K$-matrix poles as well as other parameterization deliver only in rare
cases singularities which are near to them of the $T$-matrix. This is for example the
case for
\bi
\item resonances which are very far apart,
\item minimal coupling to sub-dominant channels and
\item far away from thresholds.
\ei
If at least one of those is violated interference terms appear which move the unphysical
poles.
\begin{table}[hbt]
\begin{narrowtabular}{5cm}{ccc} \hline
Sheet & $\IM{q_1}$ & $\IM{q_2}$ \\ \hline
I & + & + \\
II & - & +\\
III & - & - \\
IV & + & - \\ \hline
\end{narrowtabular}
\caption[Defintion of Riemann sheets]{Definition of Riemann sheets (2 channels).}
\label{tbl:riemann}
\end{table}
\par
The energy of a particle is function of $p^2$. The inverse function, the root, has no
unique solution. The sign can be positive or negative. If the momentum is a complex number, the two roots lie
$\pi$ apart on the same circle. This is easily illustrated for $w=z^2$. If $z=\Exp{\imath\ph}$ then
$w=\Exp{2\imath\ph}$. If $\ph$ moves only from 0 to $\pi$ then $w$ moves around the whole circle. If $\ph$ moves on
$w$ starts a second turn. The difference between the lower and upper function value is
\be
\sqrt{w}-\sqrt{w^\ast}=\pm\sqrt{|w|}\left(
\Exp{\imath\frac{\ph}{2}}+\Exp{-\imath\frac{\ph}{2}}\right)=\left.\cosh\frac{\ph}{2}\right|_{\ph=0}\neq 0.
\ee
To resolve this ambiguity one introduces for each threshold involved in the process a new pair of
Riemann sheets, which represent each one possible root of the breakup momentum. The convention is
(1) that every crossing of the real energy axis involves a sign change for the open channel and
(2) that for every passed threshold the sign of the threshold channel changes. Please see \reftbl{tbl:riemann} and
\reffig{fig:cplx_sheets} for illustration. This means for a two channel problem
\be
\left(\wht{T}^{III}\right)\inv=\left(\wht{T}^{II}\right)\inv+\imath\rho_2
\ee
\bfg[hbt]\bc
(a)\vps\psfig{figure=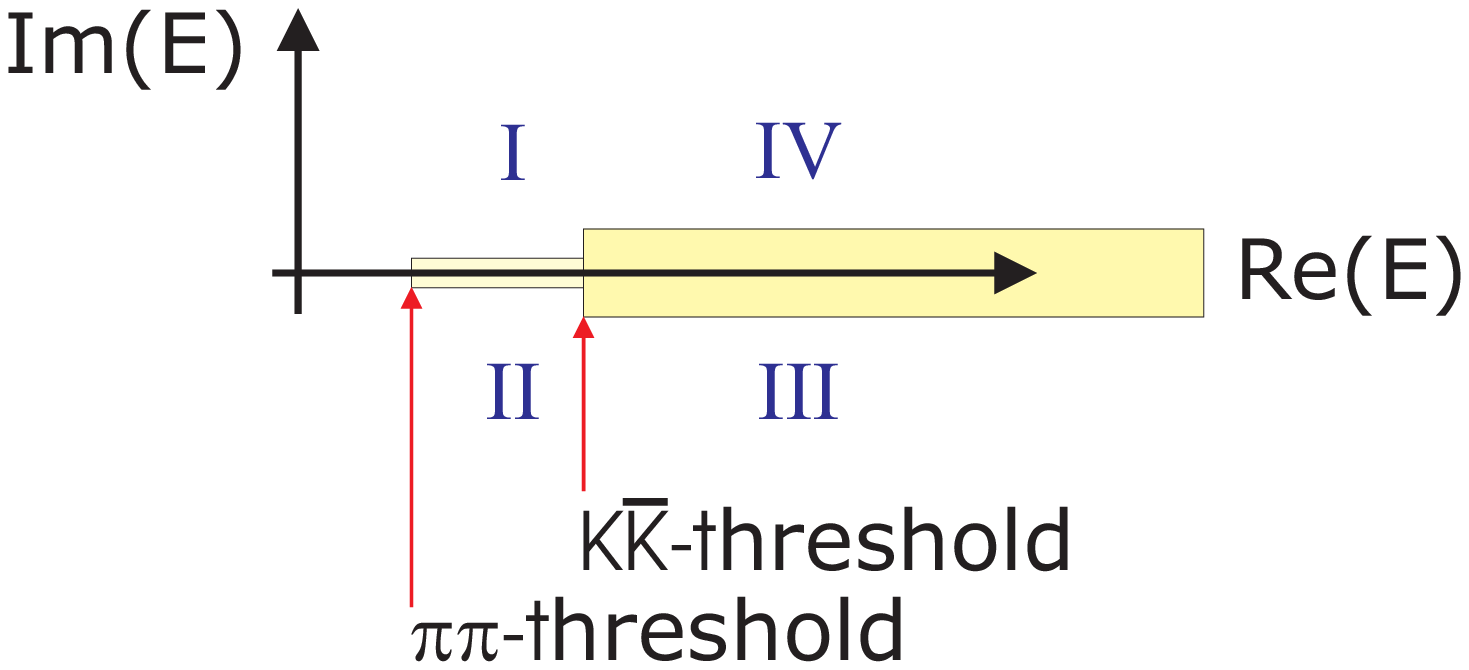,scale=0.5}
\vps(b)\vps\psfig{figure=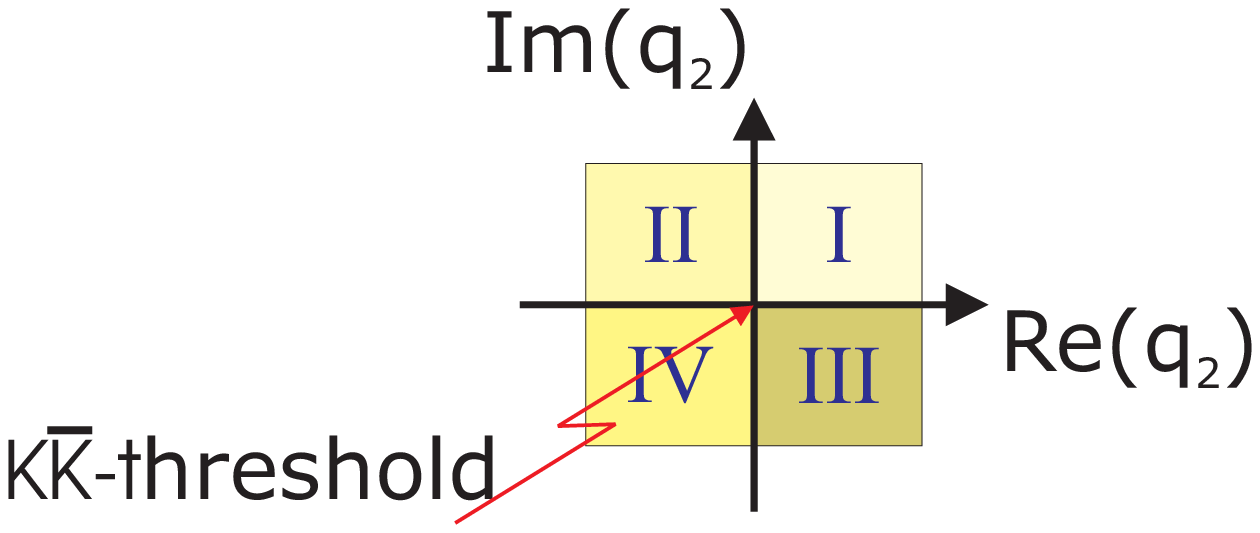,scale=0.5}\ec
\caption[Numbering of Riemann-sheets (2 channels)]{Numbering of Riemann-sheets (2 channels).}
\label{fig:cplx_sheets}
\efg
where (III) and (II) denote Riemann sheets according to the definitions in \reftbl{tbl:riemann}.
The different sheets in the complex energy and momentum plane are shown in \reffig{fig:cplx_sheets}. 
The different
sheets are separated by cuts. The convention is that the cuts lie on the real axis and start at the threshold.
In s-channel it is common to defines the real axis above the threshold to be the right-hand cut (RHC).
It contains all singularities of the s-channel scattering processes. The $t$- and $u$-channel can have resonances
and bound-states too, which depend upon the reaction under investigation and are defined as left-hand cuts (LHC), since
they appear usually below threshold (see \reffig{fig:cplx_cuts} for illustration). 
One should note, that these singularities are not taken into account in $K$-matrix
formalisms. Furthermore, the LHC may reach far out into the physical region, where they might imitate  resonances.
\bfg[hbt]\bc
\psfig{figure=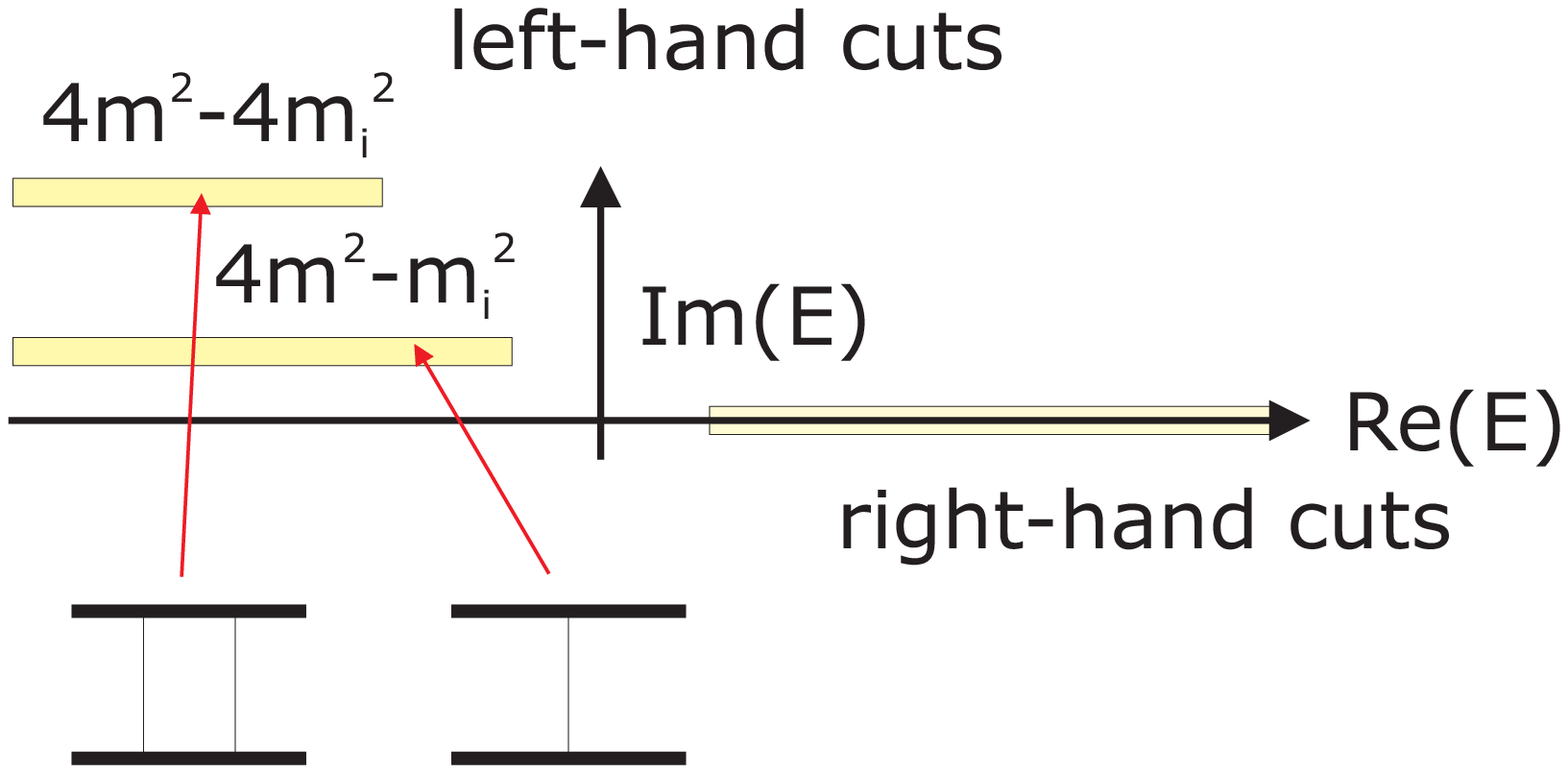,scale=0.5}\ec
\caption[Left-hand and right-hand cuts]{Illustration of left-hand and right-hand cuts.}
\label{fig:cplx_cuts}
\efg
This has dramatic effects for the interpretation of poles in a transition amplitude. To clearly identify resonances
it is mandatory to investigate the convergence radius of the pole. The convergence radius has to reach out to
the real axis. Unfortunately the amplitude on the real axis is dominated by the nearest singularity and effects from
the LHC if they influence the physical region.
In summary this means, that the question of resonances is translated into the question of how many poles does the
$T$-matrix have in the complex energy plane and what are their properties.
\par
The basic idea is simple. The $T$-matrix contains a complex denominator $D$. A singularity appears if $D$ vanishes.
The mass of a resonance is then the real part of this point and the width is twice the imaginary part
\be
T(E+\imath\gamhalf)=0
\ee
which has a similar singularity as a non-relativistic Breit-Wigner. The fundamental statement is fact that
the relation between the position of the pole and the according Riemann sheet characterize the resonance~\cite{Morgan93}.
The cross section (real axis) is mainly influenced by the nearest pole (see \reffig{fig:real_bw}). In principle
all poles influence the real axis, which is easily seen if they are close together. In general the following approximation
holds near the threshold (at the boundary between sheets II and III)
\be
\Gamma_r^{\mbox{\tiny{BW}}}\approx\half\left( \Gamma_r^{II}+\Gamma_r^{III}\right)
\ee
where $\Gamma_r^{III}$ is usually larger than $\Gamma_r^{II}$. The empirical principle of meromorph
functions for $T$ (maximum analyticity or analytic with a finite number of poles) requires that only physical singularities
are allowed in the analytic continuation into the complex energy and momentum plane like
\bi
\item kinematic singularities which are defined via unitarity relations. They appear at every threshold, where the
two-body breakup gets real,
\item dynamic singularities due to the interaction, like in exchange processes, and
\item poles which correspond to resonances or bound states. 
\ei
\bfg[hbt]\bc
\psfig{figure=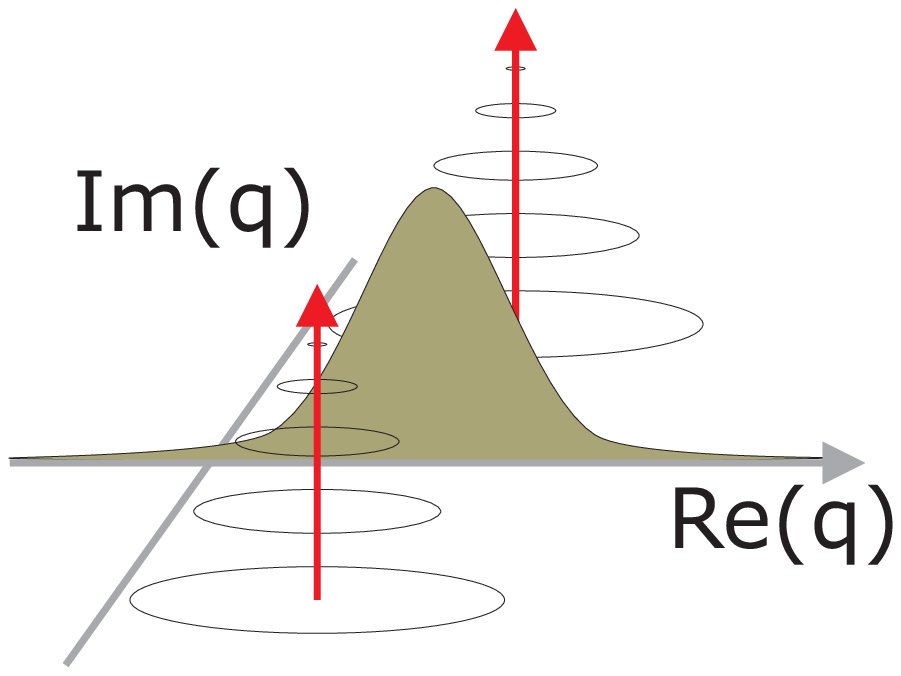,scale=0.5}\ec
\caption[From a pole to a real Breit-Wigner]
{From a pole to a real Breit-Wigner. The nearest pole dominates the properties of a point on the real axis.}
\label{fig:real_bw}
\efg
\par
The time dependence of a decay amplitude is the Fourier transform of an exponential decay ($\dsty\propto \Exp{\imath q_rt}$).
Thus, the imaginary part of $q_r$ is negative, which means, that the pole lies in the unphysical half of the momentum plane
($\IM{q}<0$, see \reffig{fig:cplx_planes}a). In contrast to that, a bound state lies on the positive imaginary axis. Due
to unitarity and hermitian analyticity the analytic continuation leads to
\be
\wht{T}_l(q)=\wht {T}_l^\ast(-q^\ast)\qquad\mbox{and}\qquad \wht{T}_l(s)=\wht {T}_l^\ast(s^\ast).
\ee
Therefore we have two poles for a resonance symmetric around the imaginary axis. In the energy plane they lie symmetric around
the real axis.
\bfg[hbt]\bc
(a)\vps\psfig{figure=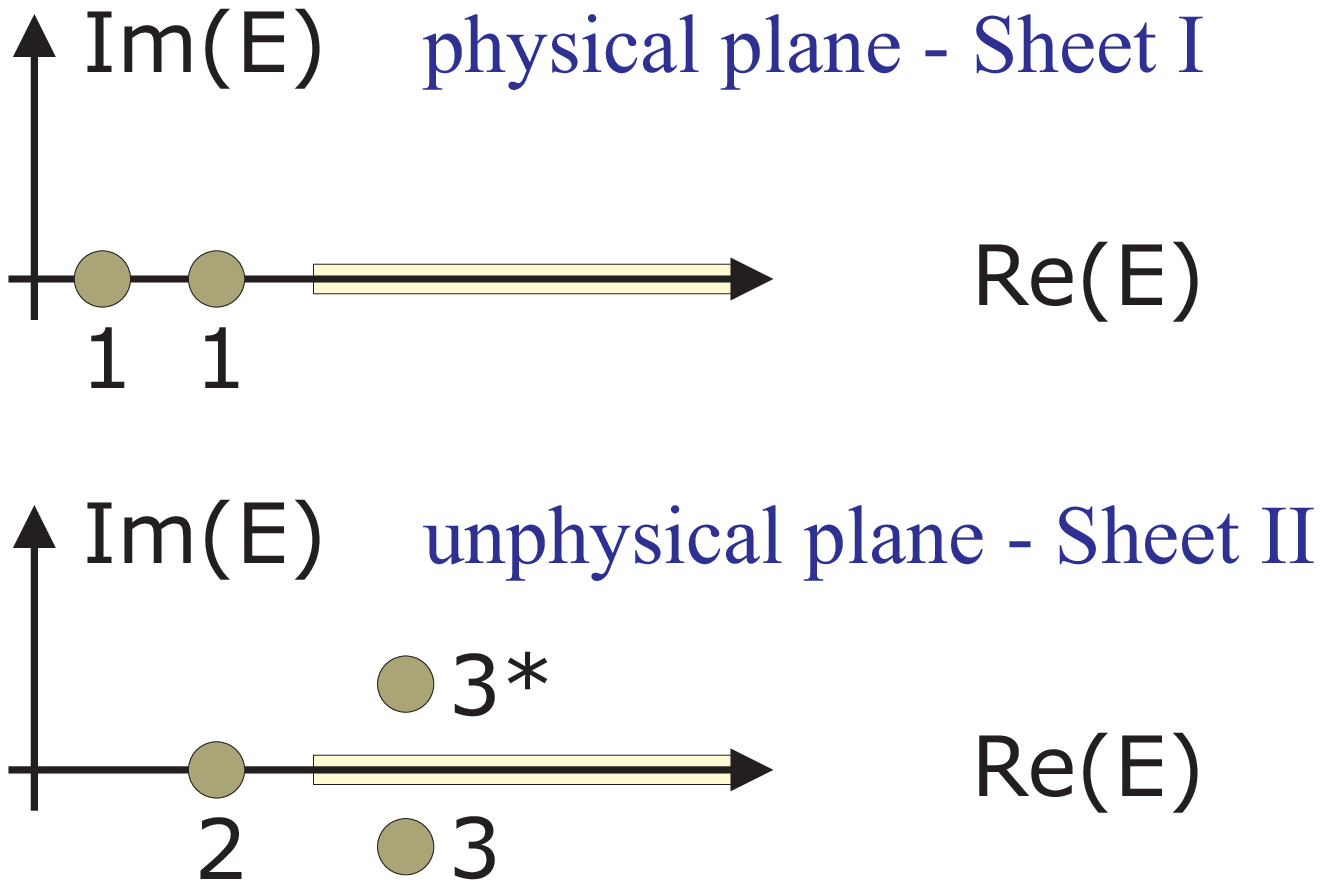,scale=0.5}
\vps(b)\vps\psfig{figure=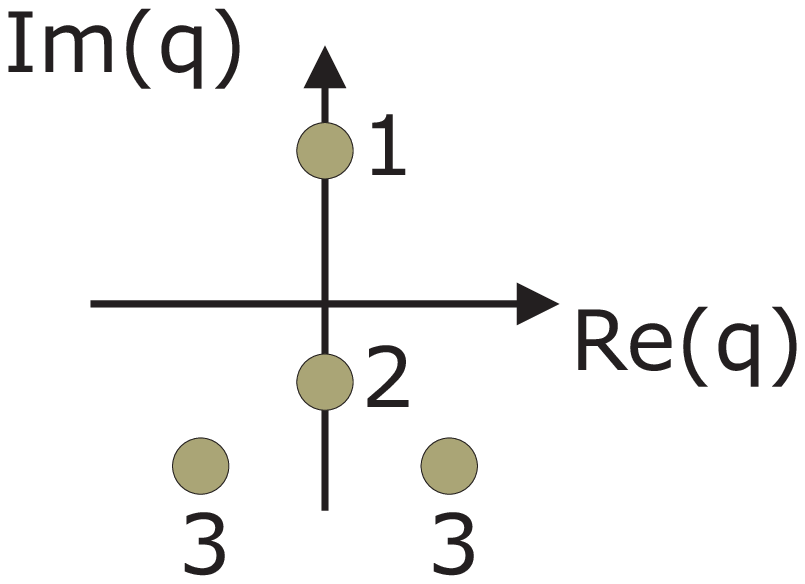,scale=0.5}\ec
\caption[Singularities in the complex energy and momentum plane]
{Singularities in the (a) complex energy and (b) momentum plane.
(1) bound states, (2) anti-bound states and (3) resonances.}
\label{fig:cplx_planes}
\efg
\par
Let's go back to poles at threshold, since they are the most complicated ones and the respective
resonances remain a mystery in meson spectroscopy. If the poles are far away from thresholds, they
are no really affected by them (see \reffig{fig:cplx_pole_thres}a). This is true even for the
$\rho(770)$ which lies exactly between two threshold. For the $a_0$/$f_0$ at the $\K\bar{K}$ threshold, the
situation is more puzzling. Without thresholds, the poles on sheet II and III are usually identical and lie
on top of each other. At threshold, they do not and it is important to verify that the two poles which
one finds near the threshold correspond to each other. A resonance can only be quoted if the poles belong to each other.
Since the two sides of the threshold correspond to a different sign for $\rho$ (sheet II: $(\imath\sqrt{\rho})^2=-\rho$ and
sheet III: $\sqrt{\rho}^2=+\rho$) it is sufficient to show, that while changing $\rho$ smoothly to $-\rho$ the poles
can be connected (see \reffig{fig:cplx_pole_thres}b).
\bfg[hbt]\bc
(a)\vps\psfig{figure=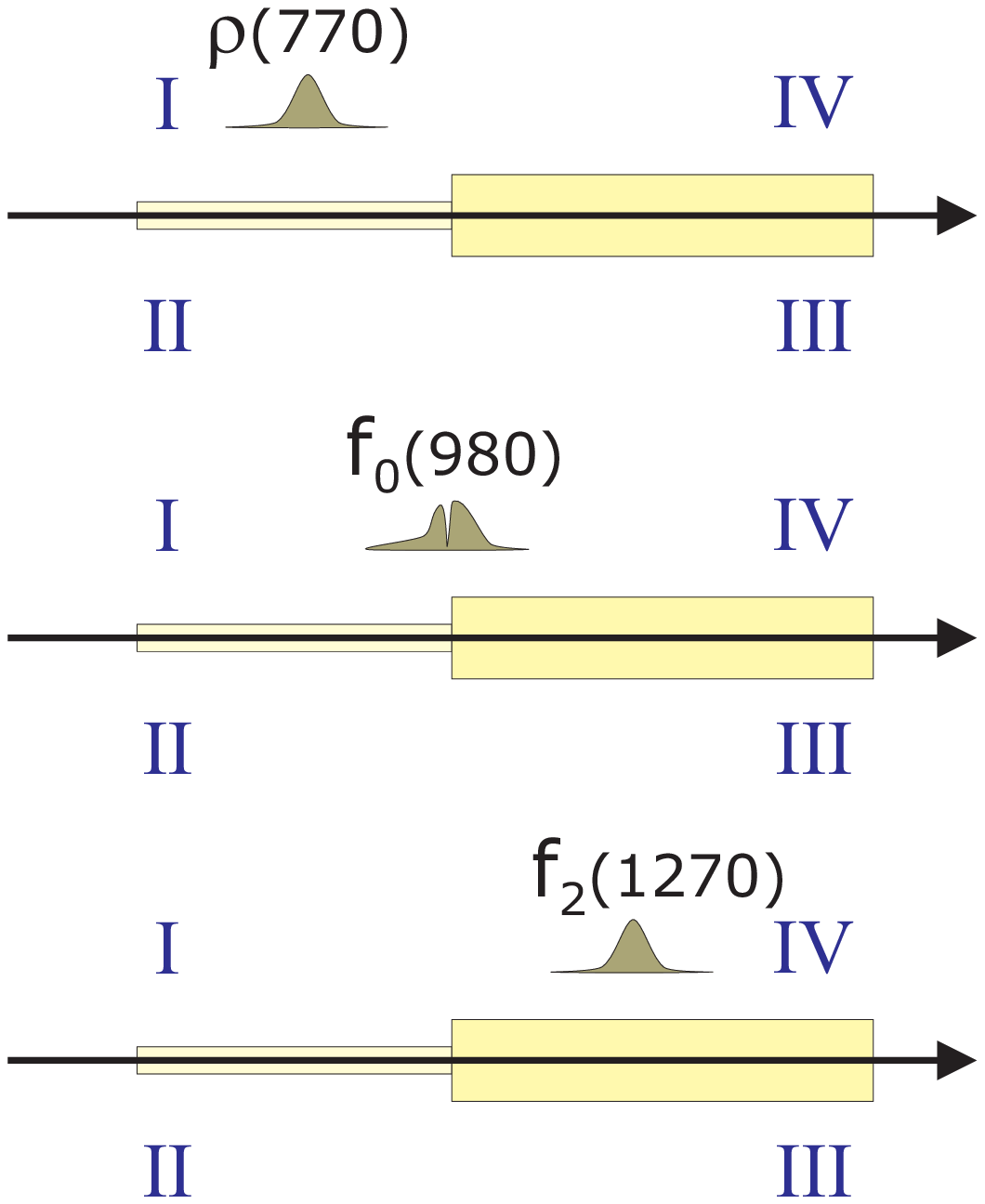,scale=0.5}
\vps(b)\vps\psfig{figure=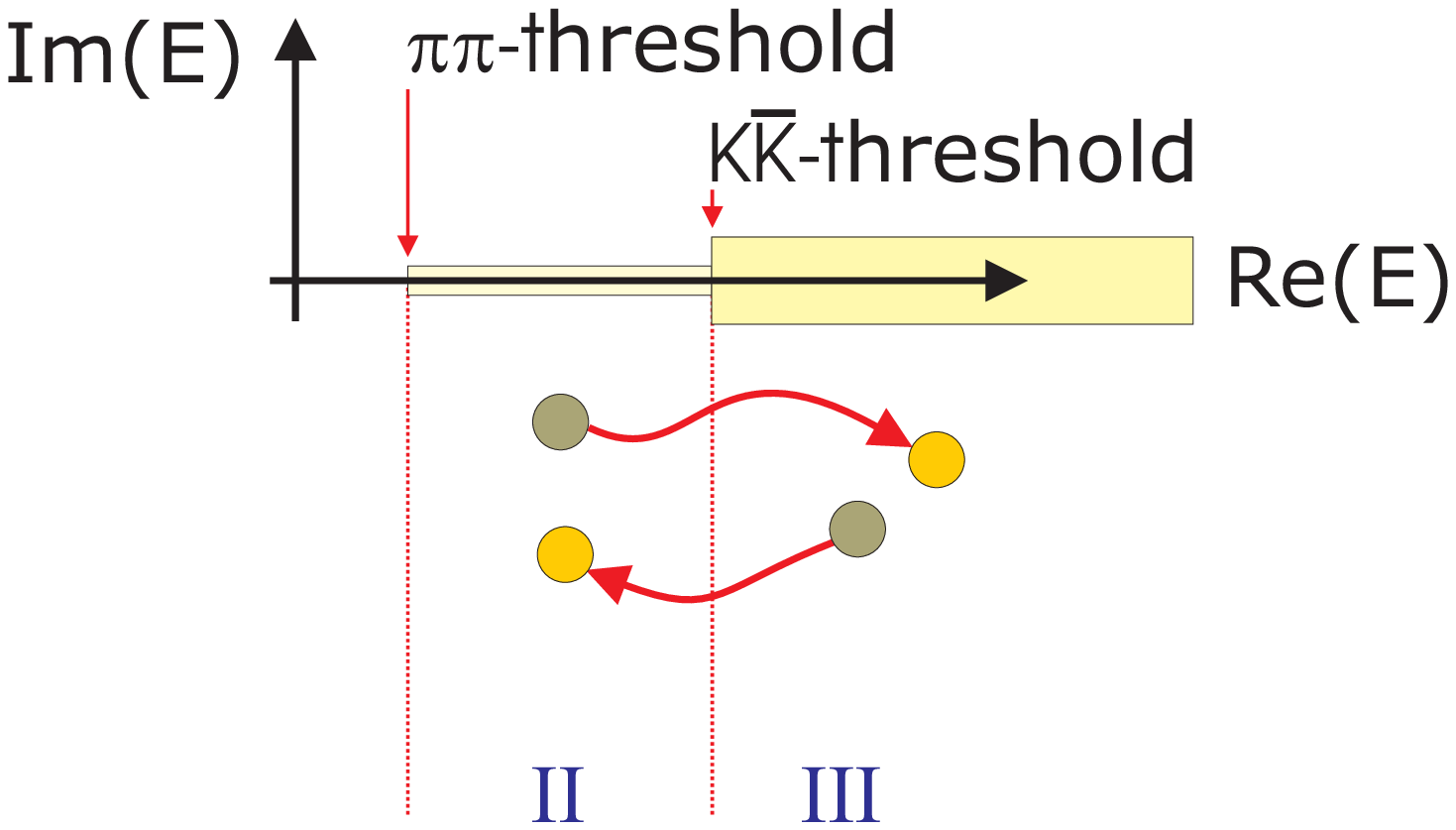,scale=0.5}\ec
\caption[Normal poles and shadow poles at threshold]{Normal poles (a,c) and shadow poles at threshold (b).}
\label{fig:cplx_pole_thres}
\efg
\section*{}
\acknowledgments
I would like to thank T.~Bressani and U.~Wiedner for organizing this hadron school
at such a lovely place leading to so many fruitful discussions.
I also thank S.U.~Chung and M.R.~Pennington for teaching me a lot of the topics discussed
in this primer.\par
In addition I would like to thank bmb+f, GSI and SIF for supporting the preparation of
the lectures.
%
%

%

%

\begin{thebibliography}{0}
%
%
%
\def\etal{{et {\rm{al.}}}}
%
\bibitem{Peters95} \BY{Peters~K., and Klempt E.} Phys. Lett. {\bf B352} (1995) 467.
\bibitem{Jacob59} \BY{Jacob~M., and G.C. Wick} Ann. Phys.(USA) {\bf 7} (1959) 404.
\bibitem{Kotanski66} \BY{Kotanski~A.} Act. Phys. Pol. {\bf 29} (1966) 699; Act. Phys. Pol. {\bf 30} (1966) 629.
\bibitem{Choushirokov58} \BY{Chou~K.C., and Shirokov~M.I.} Sov. Phys. JETP {\bf 7} (1958)851.
\bibitem{Zemach64} Phys. Rev. {\bf B1201} (1964) 133, Phys. Rev.  {\bf B97} (1965) 140, Phys. Rev.  {\bf B109} (1965) 140
%
\bibitem{Day60} \BY{Day~T.B., Snow~G.A and Sucher~J.} Phys. Rev. {\bf 118} (1960) 864.
\bibitem{Batty89} \BY{Batty~C.} Rep. Prog. Phys. {\bf 52} (1989) 1165.
\bibitem{Peters01} \BY{Peters~K.} Nucl Phys. {\bf A692} (2001) 395c
\bibitem{Abele99} \BY{Abele~A. \etal} Eur. J. Phys. {\bf C8} (1999) 67.
%
\bibitem{Martin76} \BY{Martin~B.R., Morgan~D., and Shaw~G.} \TITLE{Pion Pion Interactions in Particle Physics}, Academic Press, 1976.
%
\bibitem{Chung71}\BY{Chung~S.U.} \TITLE{Spin Formalisms}, Yellow Report, CERN 71-8.
\bibitem{Martin70} \BY{Martin A.D., and Spearman~T.D.} \TITLE{Elementary Particle Theory}, North-Holland Publishing Co., Amsterdam 1970.
\bibitem{Burkhardt69} \BY{Burkhardt~H.} \TITLE{Dispersion relation dynamics}, North-Holland, Amsterdam, 1969
%
\bibitem{Wigner46} \BY{Wigner~E.P.} Phys. Rev. {\bf 70} (1946) 15.
\bibitem{Wigner47} \BY{Wigner~E.P., and Eisenbud~L.} Phys. Rev. {\bf 72} (1947) 29.
\bibitem{Dalitz60} \BY{Dalitz~R.H., and Tuan~S.} Ann. Phys. {\bf 10} (1960) 307.
\bibitem{Badalyan80} \BY{Badalyan~A.M., Kok~L.P., Polikarpov~M.I., and  Simonov~Yu.A.} Phys. Rep. {\bf 82} (1982) 31.
\bibitem{Cahn86} \BY{Cahn~R.N.,, and Landshoff~P.V.} Nucl. Phys. {\bf B266} (1986) 451.
\bibitem{Kato65} \BY{Kato~M.} Annals of Physics, {\bf 31} (1965) 130.
\bibitem{Morgan93} \BY{Morgan~D., and Pennington~M.R.} Phys. Rev. {\bf D48} (1993) 1185. 
\bibitem{Martin77} \BY{Martin~A.D., Ozmutlu~E.N. and Squires~E.J.} Nucl. Phys. {\bf B121} (1977) 514.
\bibitem{Hippel72} \BY{Hippel~F.v. and Quigg~C.} Phys. Rev. {\bf 5} (1972) 624.
\bibitem{Flatte76} \BY{Flatt\'e~S.M.} Phys. Lett. {\bf B63} (1976) 224.
%
\bibitem{Au87} \BY{Au~K.L., Morgan~D. and Pennington~M.R.} Phys. Rev. {\bf D35} (1987) 1633.
\bibitem{Amsler95} \BY{Amsler~C. \etal} Phys. Lett. {\bf B355} (1995)425
\bibitem{Rosselet77} \BY{Rosselet~L. \etal} Phys. Rev. {\bf D15} (1977) 574
\bibitem{Grayer74} \BY{Grayer~G. \etal} Nucl. Phys. {\bf B75} (1974) 189
\bibitem{Anisovich02} \BY{Anisovich~V.V., and Sarantsev~A.V.} Eur. Phys. J. {\bf A16} {2003} 229.
%
\bibitem{Aitchison72} \BY{Aitchison~I.J.R.} Nucl. Phys. {\bf A189} (1972) 417.
\bibitem{Watson52} \BY{Watson~K.M.} Phys. Rev. {\bf 88} (1952) 1163.
\bibitem{Longacre82} \BY{Longacre~R.S.} Phys. Rev. {\bf D26} (1982) 82
\bibitem{Longacre86} \BY{Longacre~R.S. \etal} Phys. Lett. {\bf B177} (1986) 223
\bibitem{Lindenbaum92} \BY{Lindenbaum~S.J., and Longacre~R.S.} Phys. Lett. {\bf B274} (1992) 492
\bibitem{Foster68} \BY{Foster~M. \etal} Nucl. Phys. {\bf B6} (1968) 107.
%
\bibitem{Anisovich94} \BY{Anisovich~V.V. \etal} Phys. Lett. {\bf B323} (1994) 233.
\end{thebibliography}
\end{document}